%% file: ms.tex
\documentclass{emulateapj}
\usepackage{rotating}
\usepackage{lscape}

\newcommand{\mycomment}[1]{}
\newcommand{\degs}{\ifmmode ^{\circ}\else$^{\circ}$\fi}
\newcommand{\optebl}{\ifmmode L_{\mathrm 2500\,\AA} \else $~L_{\mathrm 2500\,\AA}$\fi}
\newcommand{\opteml}{\ifmmode l_{\mathrm 2500\,\AA} \else $~l_{\mathrm 2500\,\AA}$\fi}
\newcommand{\lo}{\ifmmode l_o \else $~l_o$\fi}
\newcommand{\xebl}{\ifmmode L_{\mathrm 2\,keV} \else $~L_{\mathrm 2\,keV}$\fi}
\newcommand{\xeml}{\ifmmode l_{\mathrm 2\,keV} \else $~l_{\mathrm 2\,keV}$\fi}
\newcommand{\lx}{\ifmmode l_x \else $~l_x$\fi}
\newcommand{\logz}{\ifmmode{\mathrm log}~z \else log$~z$\fi}
\newcommand{\ax}{\ifmmode{\alpha_x} \else $\alpha_x$\fi} 
\newcommand{\aox}{\ifmmode{\alpha_{\mathrm ox}} \else $\alpha_{\mathrm ox}$\fi} 
\newcommand{\fcgs}{\ifmmode erg~cm^{-2}~s^{-1[B}\else erg~cm$^{-2}$~s$^{-1}$\fi}
\newcommand{\fnucgs}{\ifmmode {\mathrm erg~cm}^{-2}~{\mathrm s}^{-1}~Hz^{-1}\else erg~cm$^{-2}$~s$^{-1}$~Hz$^{-1}$\f
i}
\newcommand{\lnucgs}{\ifmmode erg~s^{-1}~Hz^{-1}\else erg~s$^{-1}$~Hz$^{-1}$\fi}
\newcommand{\lcgs}{\ifmmode erg~~s^{-1}\else erg~s$^{-1}$\fi}
\newcommand{\kms}{\ifmmode~{\mathrm km~s}^{-1}\else ~km~s$^{-1}~$\fi}
\newcommand{\mone}{\ifmmode ^{-1}\else$^{-1}$\fi}
\newcommand{\mtwo}{\ifmmode ^{-2}\else$^{-2}$\fi}
\newcommand{\msun}{\ifmmode {M_{\odot}}\else${M_{\odot}}$\fi}
\newcommand{\lax }{{\lower0.8ex\hbox{$\buildrel <\over\sim$}}}
\newcommand{\gax }{{\lower0.8ex\hbox{$\buildrel >\over\sim$}}}
\newcommand{\nh}{\ifmmode{\mathrm N_{H}} \else N$_{H}$\fi}
\newcommand{\nhgal}{\ifmmode{ N_{H}^{Gal}} \else N$_{H}^{Gal}$\fi}
\newcommand{\nhintr}{\ifmmode{ N_{H}^{intr}} \else N$_{H}^{intr}$\fi}
\newcommand{\nhtot}{\ifmmode{ N_{H}^{tot}} \else N$_{H}^{tot}$\fi}
\newcommand{\atoms}{\ifmmode{\mathrm ~atoms~cm^{-2}} \else ~atoms cm$^{-2}$\fi}
\newcommand{\cmsq}{\ifmmode{\mathrm ~cm^{-2}} \else cm$^{-2}$\fi}
\def\Chandra     {{\em Chandra}}
\newcommand\ha{\ifmmode {\mathrm H}\alpha \else H$\alpha$\fi}
\newcommand\hb{\ifmmode {\mathrm H}\beta \else H$\beta$\fi}
\newcommand{\oi}{\ifmmode{\mathrm [O\,II]} \else [O\,II]\fi}
\newcommand{\oii}{\ifmmode{\mathrm [O\,II]} \else [O\,II]\fi}
\newcommand{\oiii}{\ifmmode{\mathrm [O\,III]} \else [O\,III]\fi}
\newcommand{\ew}{\ifmmode{W_{\lambda}} \else $W_{\lambda}$\fi}
\newcommand{\lognlogs}{{log{\em N}-log{\em S}}}
\newcommand{\XMM}{ XMM-{\em Newton}}
\newcommand{\gmi}{($g-i$) }

\def\plotfiddle#1#2#3#4#5#6#7{\centering \leavevmode
    \vbox to#2{\rule{0pt}{#2}}
    \includegraphics{#1}}

\shorttitle{Properties of SDSS Quasars in the ChaMP} 
\shortauthors{Green et al.}

\received{2008 May 15}
\accepted{2008 August 25}
\begin{document}


\title{A Full Year's Chandra Exposure on
SDSS Quasars from the Chandra Multiwavelength Project} 


\author{Paul J. Green\altaffilmark{1,2}}
\email{pgreen@cfa.harvard.edu}
\author{T. L. Aldcroft\altaffilmark{1},
G. T. Richards\altaffilmark{3},   
W. A. Barkhouse\altaffilmark{4,2}, 
A. Constantin\altaffilmark{1}, 
D. Haggard\altaffilmark{5}, 
M. Karovska\altaffilmark{1}, 
D.-W. Kim\altaffilmark{1}, M. Kim\altaffilmark{6},
A. Vikhlinin\altaffilmark{1}, 
A. Mossman\altaffilmark{1}, 
J. D. Silverman\altaffilmark{7},
S. F. Anderson\altaffilmark{5},
V. Kashyap\altaffilmark{1}, 
B. J. Wilkes\altaffilmark{1}, 
H. Tananbaum\altaffilmark{1}} 

\altaffiltext{1}{Harvard-Smithsonian Center for Astrophysics, 60
  Garden Street, Cambridge, MA 02138} 
\altaffiltext{2}{Visiting Astronomer, Kitt Peak National Observatory
  and Cerro Tololo Inter-American Observatory, National Optical
  Astronomy Observatory, which is operated by the Association of
  Universities for Research in Astronomy, Inc. (AURA) under cooperative
  agreement with the National Science Foundation.}
\altaffiltext{3}{Department of Physics, Drexel University, 3141
  Chestnut Street, Philadelphia, PA 15260, USA}
\altaffiltext{4}{Department of Physics, University of North Dakota,
Grand Forks, ND 58202, USA}
\altaffiltext{5}{Department of Astronomy, University of Washington,
  Seattle, WA, USA} 
\altaffiltext{6}{International Center for Astrophysics,
Korea Astronomy and Space Science Institute,
Daejeon, 305-348, Korea} 
\altaffiltext{7}{Institute for Astronomy, ETH Z\"urich, 8093 Z\"urich, 
  Switzerland}

\begin{abstract}
We study the spectral energy distributions and evolution of a large
sample of optically selected quasars from the Sloan Digital Sky Survey
(SDSS) that were observed in 323 \Chandra\, images analyzed by the
\Chandra\, Multiwavelength Project. Our highest-confidence matched
sample includes 1135 X-ray detected quasars in the redshift range
$0.2<z<5.4$, representing some 36\,Msec of effective exposure.  We
provide catalogs of QSO properties, and describe our novel method of
calculating X-ray flux upper limits and effective sky coverage.
Spectroscopic redshifts are available for about 1/3 of the detected
sample; elsewhere, redshifts are estimated photometrically. We detect
56 QSOs with redshift $z>3$, substantially expanding the known sample.
We find no evidence for evolution out to $z\sim$5 for either the X-ray
photon index $\Gamma$ or for the ratio of optical/UV to X-ray flux
\aox.  About 10\% of detected QSOs show best-fit intrinsic absorbing
columns $>10^{22}\cmsq$, but the fraction might reach $\sim$1/3 if
most non-detections are absorbed. We confirm a significant correlation
between \aox\, and optical luminosity, but it flattens or disappears
for fainter ($M_B\gax -23$) AGN alone.  We report significant
hardening of $\Gamma$ both towards higher X-ray luminosity, and for
relatively X-ray loud quasars. These trends may represent a relative
increase in non-thermal X-ray emission, and our findings thereby
strengthen analogies between Galactic black hole binaries and AGN. For
uniformly-selected subsamples of narrow line Seyfert~1s and narrow
absorption line QSOs, we find no evidence for unusual distributions of
either \aox\, or $\Gamma$.
\end{abstract}


\keywords{galaxies: active -- quasars: general -- quasars: absorption
  lines -- surveys -- X-rays}

\section{Introduction}
\label{intro}

Interest in the properties of active galaxies and their evolution has
recently intensified because of deep connections being revealed between
supermassive black holes (SMBH) and galaxy evolution, such as the
relationship between the mass of galaxy spheroids and the SMBH they
host (the $M_{BH}-\sigma$ connection; \citealt{Ferrarese00,
  Gebhardt00}). A feedback paradigm 
could account for this correlation, whereby winds from active galactic
nuclei (AGN) moderate the SMBH growth by truncating that of their host
galaxies (e.g., \citealt{Granato04}).  Feedback models may explain the
correspondence  
between the local mass density of SMBH and the luminosity density produced
by high redshift quasars \citep{Yu02,Hopkins06b} as well as the
`cosmic downsizing' (decrease in the space density of luminous AGN)
seen in AGN luminosity functions \citep{Barger05,Hasinger05,Scannapieco05}.
If quasar activity is induced by massive mergers (e.g., 
\citealt{Wyithe02, Wyithe05}), then the jigsaw puzzle now assembling
may merge smoothly with cosmological models of hierarchical structure
formation.  

  Many, if not most, of the accreting SMBH in the Universe may be
obscured by gas and dust in the circumnuclear region, or in the
extended host galaxy.  The obscured fraction may depend on both
luminosity and redshift \citep{Ueda03, Brandt05, LaFranca05},
and is indeed likely to evolve on grounds both theoretical (e.g.,
\citealt{Hopkins06a}) and observational
\citep{Treister06,Ballantyne06}.  Such evolution seems to be required
for AGN populations to compose the observed spectrum of the cosmic
X-ray background (CXRB; \citealt{Gilli07}).  However, a full census of
all SMBH remains observationally challenging, since some are heavily
obscured, or accreting at very low rates, below the sensitivity limits
of current telescopes even at low-redshifts.

AGN unification models explain many of the observed differences in the
spectral energy distributions (SEDs) of AGN as being due to the
line-of-sight effects of anisotropic distributions of obscuring
material near the SMBH \citep{Antonucci93}. The intrinsic number ratio
of obscured to unobscured AGN may evolve, and is almost certainly a
function of luminosity. Indeed, the ratio in the Seyfert (low-)
luminosity regime is currently estimated to be $\sim$4, whereas for
the QSO (high-) luminosity regime, it may be closer to unity
\citep{Gilli07}.

Astronomers, like most people, usually look where they can  
see. Type~1 quasars are the easiest AGN to find in large numbers via
either spectroscopic or color selection because of their broad
emission lines (FWHM$\gax$1000\,\kms) and generally blue continuum
slopes and brighter 
magnitudes.  Most of these show few other signs of obscuration such as
infrared excess or weakened X-ray emission.  Large
samples of Type~1 QSOs - as the brightest high redshift objects - have
served to probe intervening galaxies, clusters, and the intergalactic
material (IGM) along the line of sight, right to the epoch of
reionization.  Because observed SEDs are thought to be less affected
by obscuration and therefore more representative of intrinsic
accretion physics, the evolution of this Type~1 sample is of interest
as well. 
 
The SEDs and clustering properties 
of Type~1 QSOs have been studied in increasing detail, probing
farther into the Universe and wider across the sky and the
electromagnetic spectrum.  SEDs including mid-infrared
photometry from {\em Spitzer} were compiled
and characterized recently for 259 quasars from the
Sloan Digital Sky Survey (SDSS) by \citet{Richards06b}.
These studies are most useful for calculating bolometric luminosities
and $K$-corrections towards understanding the energetics of the
accretion process, and its evolution across cosmic time.   
Studies that target SDSS QSO pairs with small separations find a
significant clustering excess on small scales ($\lax 40$\,kpc$h$) of
varying strengths (e.g., \citealt{Hennawi06,Myers07,Myers08}), which
could be due to mutual triggering, or might simply result from the
locally overdense environments in which quasars form \citep{Hopkins08}. 

Optical luminosity function studies of optically selected
quasars (e.g., \citealt{Richards06a,Richards05,Croom04}) date
back decades  (e.g., \citealt{Boyle88}). Accurate luminosity functions
are needed to trace the accretion history of SMBH and to contrast the
buildup of SMBH with the growth of galaxy spheroids.  An
increasing number of optical surveys not only select
AGN photometrically, but also determine fairly reliable
photometric  redshifts for them. These samples stand to vastly improve
the available statistical reliability and the resolution available in
the  luminosity/redshift plane. 

X-ray observations have been found to efficiently select
AGN of many varieties, and at higher surface densities than ever
\citep{Hasinger05}.  Independent of the classical AGN optical emission
line criteria, X-rays are a primary signature of accretion
onto a massive compact object, and the observed X-ray bandpass
corresponds at higher redshifts to restframe energies capable of
penetrating larger intrinsic columns of gas and dust.\footnote{The
observed-frame, effective absorbing column is $N_{\mathrm H}^{\mathrm
  eff}\sim N_{\mathrm  H}/(1+z)^{2.6}$ \citep{Wilman99}.}  Even for
Type~1 QSOs, 
X-ray observations have revealed new connections (e.g.,
between black holes from 10 to 10$^9$ $M_{\odot}$;
\citealt{Maccarone03, Falcke04}) and new physical insights 
such as the possibility of ubiquitous powerful relativistic outflows  
\citep{Middleton07} or of relativistically broadened fluorescent
Fe\,K$\alpha$ emission (see, e.g., the review articles by 
\citealt{Fabian00, Reynolds03}).  Quasars with broad
absorption  lines (BALQSOs) blueward of their UV emission lines turn
out to be highly absorbed in X-rays \citep{Green95, Green96,Green01,
Gallagher02}. Quasars that are radio loud are 2 -- 3 times brighter in
X-rays for the same optical magnitude
\citep{Zamorani81,Worrall87,Shen06}, and also may 
have harder X-ray spectra (e.g., \citealt{Shastri93}, although see 
\citealt{Galbiati05}). 

The high sensitivity and spatial resolution of \Chandra\, and \XMM\,
open other avenues for exploration of quasars and their environments.
Clusters of galaxies have 
been discovered in the vicinity of, or along the sightlines 
to quasars \citep{Green05,Siemiginowska05}.  Lensed quasars
have now been spatially resolved in X-rays, unexpectedly showing
significantly different flux ratios than at other wavebands
\citep{Green02,Blackburne06,Lamer06}.  

The expected evolution in the environment, accretion rates, and masses
of SMBH in AGN should correspond to observable evolution in their SEDs.
The two most common X-ray measurements used are the X-ray power-law
photon index $\Gamma$\footnote{$\Gamma$ is the 
photon number index  of an assumed power-law continuum such that
$N_E(E)=N_{E_0}\,E^{\Gamma}$. In terms of a spectral index $\alpha$ from
$f_{\nu}=f_{\nu_0}\nu^{\alpha}$, we define $\Gamma= (1-\alpha)$.}
and the X-ray-to-optical spectral slope, \aox.\footnote{$\aox$\, is
  the slope of a hypothetical power-law from 2500\,\AA\, to 2~keV;
  $\aox\, = 0.3838~{\mathrm log} (\opteml/\xeml)$}.  
Many of the apparent correlations have been challenged as being
artifacts of selection or the by-products of small, heterogeneous
samples, which impedes progress in our understanding of quasar physics
and evolution.

The archives of current X-ray imaging observatories 
like \Chandra\, and XMM-{\em Newton} are growing rapidly.
and several large efforts for pipeline processing,
source characterization \citep{Ptak03,Aldcroft06}, and catalog
generation are underway.  The 2XMM catalogue \citep{Page06}
is available, and the \Chandra\, source catalog\footnote{The 
  \Chandra\, source catalog webpage is {\tt
    http://cxc.harvard.edu/csc}}.  
is due out in 2008 \citep{Fabbiano07}. 

Serendipitous wide-area surveys with \Chandra\, were
pioneered by the \Chandra\, Multiwavelength Project (ChaMP;
\citealt{Green04,DKim04a}) for high Galactic latitude,
while ChaMPlane \citep{Grindlay05} has
studied the stellar content of Galactic plane fields. The ChaMP,
described in more detail below, also performs multiwavelength source
matching and spectroscopic characterization. These efforts are
greatly augmented by large surveys such as the Sloan
Digital Sky Survey  (SDSS; \citealt{York00}),  
which will obtain spectroscopy of $\sim100,000$ QSOs (e.g.,
\citealt{Schneider07}), and can additionally facilitate the efficient
extrapolation of photometric quasar selection with photometric
redshift estimation almost a magnitude fainter, towards a million QSOs
(e.g., \citealt{Richards08}).   

The current paper studies the X-ray and optical properties of the
subset of these QSOs imaged in X-rays by \Chandra\, as part of the
\Chandra\, Multiwavelength Project \citep{Green04}.

\section{The Quasar Sample}
\label{sample}

 \subsection{The SDSS Quasar Sample}
 \label{dr5qos}

Most large samples of Type~1 QSOs are based on optically selected
quasars confirmed via optical spectroscopy
\citep{Boyle88,Schneider94,Hewett95}.  The largest, most uniform
sample of optically selected quasars by far has been compiled from the
SDSS.  With the completion of the SDSS, we can expect some 100,000
spectroscopically-confirmed quasars.  The SDSS quasars were 
originally identified to $i<19.1$ for spectroscopy by their UV-excess
colors, with later expansion for $z>3$ quasars to $i=20.2$ using
$ugri$ color criteria \citep{Richards02}. The large catalog, the broad 
wavelength optical photometry, and subsequent followup in other
wavebands has meant that research results from the SDSS 
spectroscopic quasar sample better characterize the breadth of the
quasar phenomenon than ever before. However, the limited number of
fibers available, fiber placement conflicts, and above all the bright
magnitude limits of SDSS fiber spectroscopy mean that 10 times as many
quasars have been imaged, and could be efficiently identified from
existing SDSS photometry.   

Using large spectroscopic AGN samples as `training sets' can produce
photometric classification and redshifts of far greater completeness
and depth than spectroscopy.  Without spectroscopic confirmation,
photometric selection criteria strike a quantifiable balance between
completeness and efficiency, 
i.e., a probability can be assigned both to the classification and the
redshift.  Deep photometric redshift surveys like COMBO-17
\citep{Wolf03} have found AGN to $R=24$ and $z=5$ with high 
completeness.  Efficient photometric selection of quasars in the SDSS
using a nonparametric Bayesian classification based on kernel density
estimation is described in \citet{Richards04} for SDSS point sources
with $i<21$.  An empirical algorithm to determine photometric
redshifts for such quasars is described in \citet{Weinstein04}.  The
spectroscopic training samples for these methods now include far more
high redshift quasars, and so the algorithms have been re-trained to
include objects redder than ({\em u-g})=1.0 to classify high-z
quasars, and applied to the much larger SDSS Data Release 6 (DR6).
This large catalog of $\sim$1 million photometrically-identified QSOs
and their photometric redshifts is described in \citet{Richards08}.
They estimate the overall efficiency of the catalog to be better than
72\%, with subsamples (e.g., X-ray detected objects) being as
efficient as 97\%.  These estimates are based on an analysis of the
autoclustering of the objects in the catalog \citep{Myers06},
which is very sensitive to stellar interlopers. However, at the faint
limit of the catalog some additional galaxy contamination is expected.

For luminosity and distance calculations, we adopt 
a $H_0 = 70$~km\,s$^{-1}$\,Mpc$^{-1}$, $\Omega_{\Lambda} = 0.7$, 
and $\Omega_{M} = 0.3$ cosmology throughout.  We assume $\alpha=-0.5$
for the optical continuum power-law slope ($v\propto \nu^{\alpha}$,
where $\nu$ is the emission frequency),
and derive the rest-frame, monochromatic optical luminosity at
2500\AA\, ($l_{2500\AA}$; units erg~s$^{-1}$Hz$^{-1}$) using the SDSS
dereddened magnitude with central wavelength closest to $(1+z)\times\,
2500$\AA.  

\subsection{The Extended \Chandra\, Multiwavelength Project (ChaMP)}
\label{champx}

The \Chandra\, Multiwavelength Project (ChaMP) is a wide-area
serendipitous X-ray survey based on archival X-ray images of the
($|b|>20\deg$) sky observed with the AXAF CCD Imaging Spectrometer
(ACIS) onboard \Chandra.  The full 130-field Cycle\,1--2 X-ray catalog
is public \citep{MKim07a}, and the most comprehensive X-ray number
counts (\lognlogs) to date have been produced, thanks to 6600 sources
and massive source-retrieval simulations \citep{MKim07b}.  We have also
published early results of our deep ($r$$\sim$25) NOAO/MOSAIC optical
imaging campaign \citep{Green04}, now extended to 67 fields
(Barkhouse et al 2008, in preparation).  ChaMP results and data can be
found at  {\tt http://hea-www.harvard.edu/CHAMP}.

To improve statistics and encompass a wider range of source types, we
have recently expanded our X-ray analysis to cover a total of 392
fields through \Chandra\, Cycle~6.  We chose only
fields overlapping with SDSS DR5 imaging.  To ease analysis and
minimize bookkeeping problems, the new list of \Chandra\, pointings
(observation IDs; obsids hereafter) avoids any overlapping
observations by eliminating the observation with the shorter exposure
time. As described in \citet{Green04}, we also avoid 
fields with large ($\gax$3\arcmin) extended sources in either optical
or X-rays (e.g., nearby galaxies M101, NGC4725, NGC4457, or clusters
of galaxies MKW8, or Abell 1240).  Spurious X-ray sources (due to
e.g., hot pixel, bad bias, bright source readout streaks, etc.) have
been flagged and removed as described in \citet{MKim07a}.   
Of the 392 ChaMP obsids, 323 overlap the SDSS DR5 footprint. 

\begin{figure}
% \plotfiddle{PSFILE}{VSIZE}{ROTANG}{HSCALE}{VSCALE}{HTRANS}{VTRANS}
\plotone{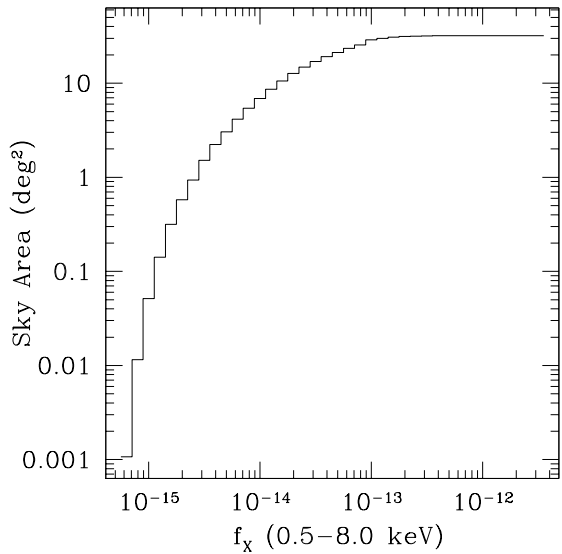}
%\vspace*{-1.25cm}
\caption{
Sky area vs $B$ band (0.5-8\,keV) flux limit for the 323
obsids included in our ChaMP/SDSS field sample. Flux limit
is defined here as the number of counts detectable
in 90\% of simulation trials, converted to flux
assuming a power-law $\Gamma=1.7$ at $z$=0 and the Galactic $N_H$
appropriate to each obsid.  Chip S4 (CCD ID \#8) is excluded
throughout. The area covered at the brightest fluxes is 32\,deg$^2$. }
%\vskip-0.2cm
\label{xskycover}
\end{figure}

  The ChaMP has also developed and implemented an {\tt xskycover}
pipeline which creates sensitivity maps for all ChaMP sky regions
imaged by ACIS.  This  allows (1) identification of
imaged-but-undetected objects (2) counts limits for 50\% and 90\%
detection completeness and (3) corresponding flux upper-limits at any sky
position as well as (4) flux sensitivity vs. sky coverage for any
subset of obsids, as needed for \lognlogs\, and luminosity function
calculations.  Our method is decribed in the Appendix, and has been
verified recently by \citet{Aldcroft08} using the Chandra Deep Field
South (CDF-S).  
The final sky ChaMP/SDSS coverage area (in deg$^2$) for the 323
overlapping fields as a function of broad band ($B$ band 0.5-8\,keV)
flux limit is shown in Figure~\ref{xskycover} (see caption for
specific definition of this limit).  The area covered at
the brightest fluxes is 32\,deg.$^2$  On average 5 CCDs are
activated per obsid. 

We have downloaded into the ChaMP database all the SDSS photometry,
and the list of photo-$z$ quasar candidates within 20\arcmin\, of the 
\Chandra\, aimpoint for each such obsid.\footnote{For
14 obsids, we extended to 28\arcmin\, radius, to achieve full coverage
of the \Chandra\, footprint. For other obsids, the SDSS imaging strips
do not completely cover the \Chandra\, field of view.}  Because the
\Chandra\, point spread function (PSF) increases with off-axis angle (OAA),
comparatively few sources are detected beyond this radius, and source
centroids also tend to be highly uncertain.  Of X-ray
detected candidates, we will show in Section~\ref{redshifts} that
98\% of these candidates with spectra are indeed QSOs. 
% (10,378 QSOs total)
% A final revision of the
% DR5 QSO catalog to exclude more contaminants reduced the parent QSO
% candidate sample by 11\%; of those objects, 90\% were non-detections.

Next we describe the identification of high-confidence ChaMP X-ray
counterparts to SDSS QSOs in \S~\ref{match}.  We then discuss 
in \S~\ref{redshifts} spectroscopic
identifications for these objects.  Section~\ref{detfrac} then
describes results for several interesting QSO subsamples, including
our treatment of SDSS quasars that were not X-ray detected.

 \subsection{X-ray/Optical Matching}
 \label{match}

\begin{figure}
% \plotfiddle{PSFILE}{VSIZE}{ROTANG}{HSCALE}{VSCALE}{HTRANS}{VTRANS}
%\epsscale{1.6}
\plottwo{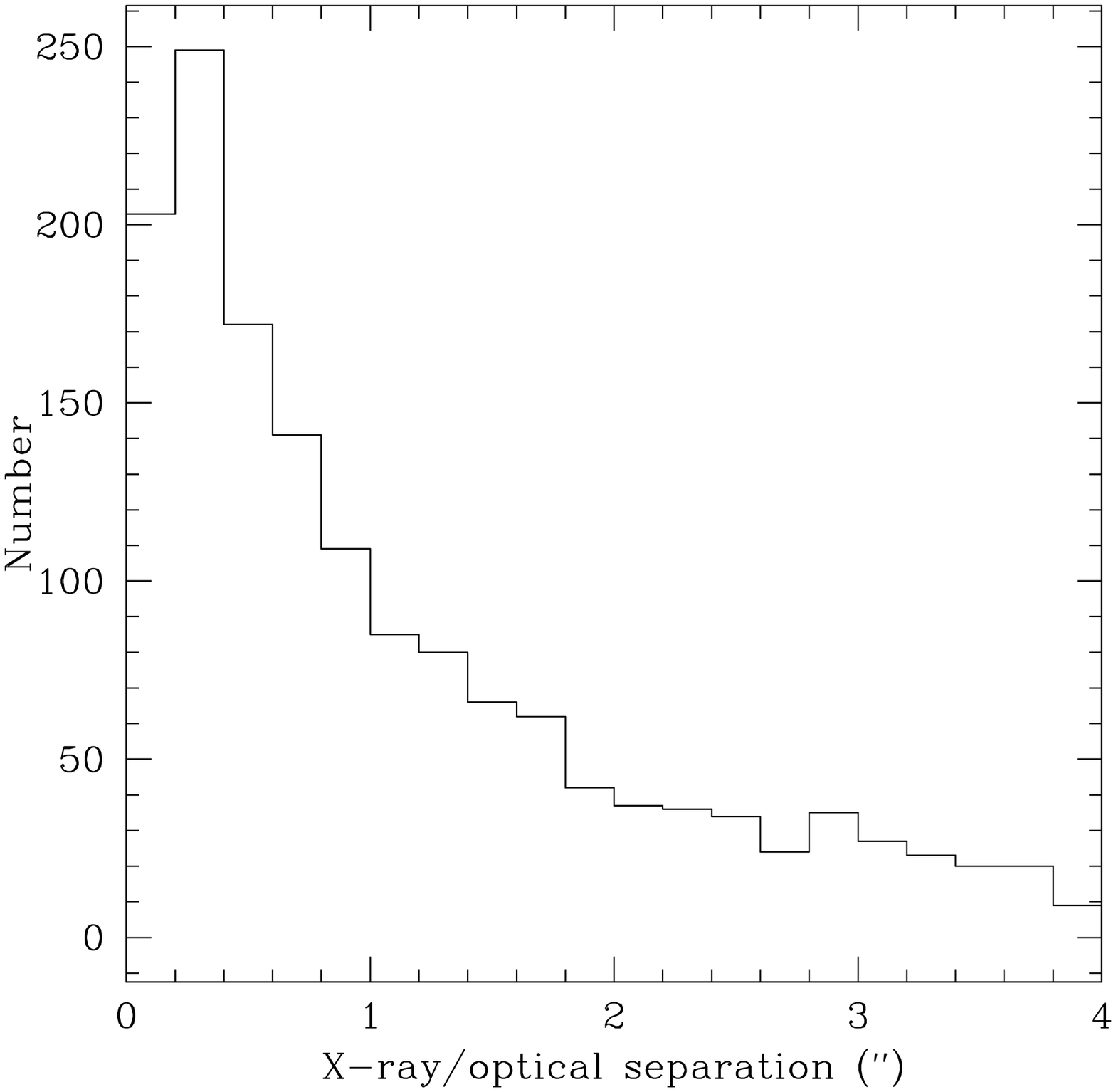}{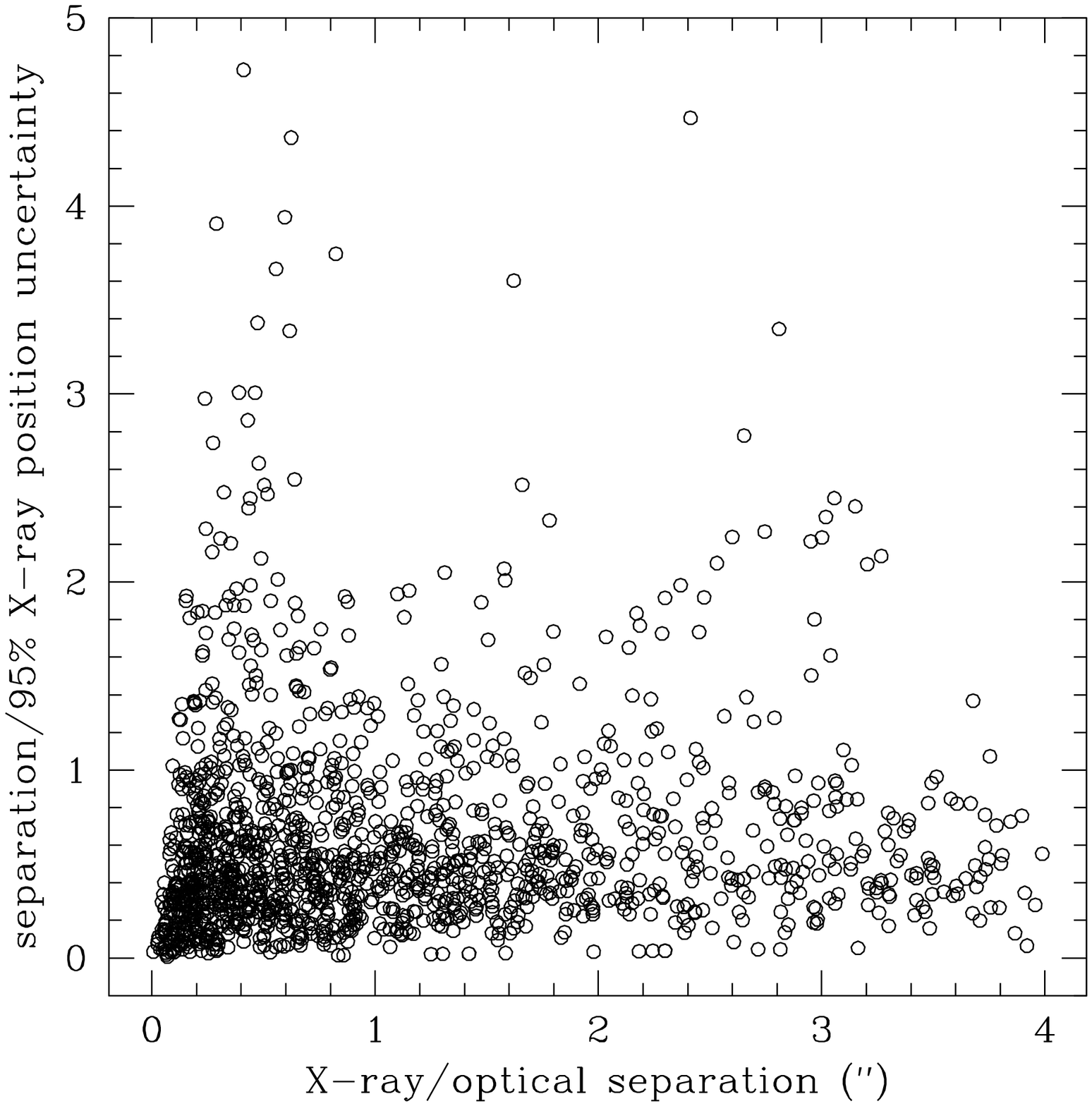}
%\vspace*{-1.25cm}
\caption{
Positional offsets of matched sample.  
LEFT: The histogram of X-ray/optical centroid separations in the
full matched sample of 1376.  The mode is 0.3\arcsec,
the median is 0.76\arcsec, and the mean is 1.1\arcsec.
RIGHT: The ratio of the X-ray/optical centroid separation to the 95\%
X-ray positional uncertainty vs. separation. The fraction of
objects with separations larger than the 95\% uncertainty 
(i.e., with ratio$>$1) remains relatively constant, and virtually no
separations wider than 3\arcsec\, are larger than the
positional uncertainty $\sigma_{XP95}$. }
%\vskip-0.2cm
\label{dxohist}
\end{figure}

The positional uncertainty of ChaMP X-ray source centroids has been
carefully analyzed via X-ray simulations by \citet{MKim07b} and
depends strongly on both the number of source counts and the OAA.
Cross-correlating ChaMP X-ray source centroids with QSOs, we find
that 95\% of the X-ray/optical separations $d_{XO}$ of matched QSOs
are smaller than the 95\% X-ray positional uncertainty
$\sigma_{XP95}$. We first perform an automated matching procedure
between each optical QSO position and the ChaMP's X-ray source
catalog.  We adopt 4\arcsec\, as our matching radius criterion. 
Figure~\ref{dxohist} shows that 95\% of the matched sample has an
X-ray/optical position difference of $<$3\arcsec, and expansion to
include match radii above 4\arcsec\, would not substantially increase
the sample size.  Searching the ChaMP catalog for X-ray sources within
4\arcsec\, of the optical SDSS quasar coordinate, yields 1376
unique matches in the ``Matched'' sample.   

Although the overall efficiency of the SDSS photometric QSO catalog is
only expected to be 72\% \citep{Richards08}, the rarity of luminous
Type~1 quasars and X-ray sources means that matched objects should be
quite clean. By repeatedly offsetting the SDSS coordinates of each QSO by
36\arcsec\, and rematching to the ChaMP X-ray catalog, we derive a
spurious match rate of just 0.7\%.\footnote{The ratio of the number of
X-ray matches to the optical control sample to the number of matches
in the actual sample is 0.0067.}  This excellent result is thanks to
\Chandra's $\sim$arcsec spatial resolution, which at SDSS depths allows
for unambiguous counterpart identification, given that both Type~1
QSOs and X-ray sources are relatively sparse on the sky.

We  identify and remove a variety of objects with potentially 
poor data, including overlapping multiple sources (some of which are
targeted lenses), bright X-ray sources suffering from pile-up, optical
sources with photometry contaminated by close brighter sources or
within large extended galaxies or stellar diffraction spikes, etc.  

 In addition to the automated matching procedure, we 
also perform visual inspection (VI) of both X-ray and optical images,
overplotting the centroids and their associated position errors.  
We retain only the highest confidence matches ({\tt matchconf}=3).  Most
of the 105 objects we thereby eliminate have large ratios of
$d_{XO}$/$\sigma_{XP95}$, or multiple candidate optical counterparts. 
We note that some of the most interesting celestial systems may be
found among sources with {\tt matchconf}$<$3.  For example, these
might include QSOs that are lensed, have bright jets, or are 
associated with host or foreground optical clusters or galaxies.  
Systems that are poorly-matched, multiply-matched, or 
photometrically contaminated may account for up to $\sim$10\% of the
full X-ray-selected sample.  We therefore caution against blind
cross-correlation of large  
source catalogs (e.g., the Chandra Source Catalog)\footnote{The Chandra
Source Catalog, available at {\tt http://cxc.cfa.harvard.edu/csc},
contains all X-ray sources detected by the Chandra X-ray Observatory.}
without such detailed quality control and visual examination of
images.  However, since we seek here to analyze the multiwavelength
properties of a large clean sample of QSOs, and since most of these
more complicated systems require significant further analysis or
observation, we defer their consideration to future studies.

%> most matchconf=3 should
%> be within sqrt(poserr_95**2 + sdss_err**2), where sdss_err is
%> an estimate of the total (random + systematic) astrometric errors
%> in the SDSS.    Pier, J.~R., et al. 2003, AJ, 125, 1559 put this at about
%> 0.15arcsec rms total (conservatively, for r<20).

\subsection{Spectroscopic Redshift Information}
\label{redshifts}

\begin{figure}
%\vspace*{-1cm}
%\epsscale{1.1}
\plotone{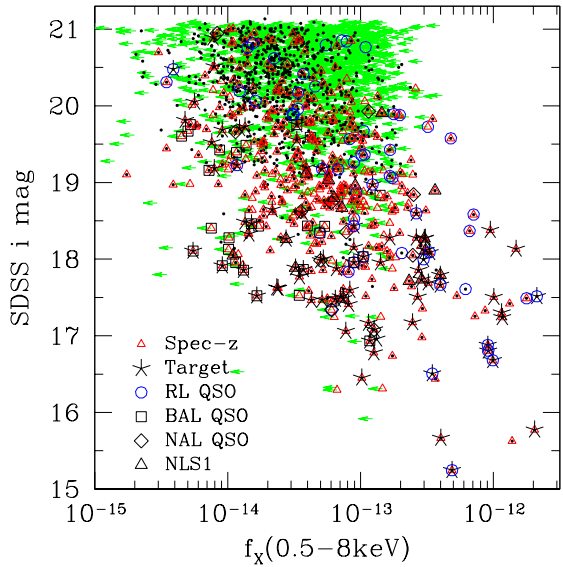}
%\vspace*{-4cm}
\caption{SDSS $i$ magnitude vs. broadband (0.5-8\,keV)
X-ray flux for the SDSS/ChaMP QSO the Main~sample.  The
flux shown is based on the best-fit PL-free {\tt yaxx} model, with the
effect of Galactic absorption removed from both X-ray and optical.
X-ray detections are marked by black dots, and flux upper limits by
green arrows. Open red triangles show QSOs with existing spectroscopic
redshifts, clearly biased towards brighter optical mags.  Radio-loud
quasars (large open blue circles), BAL QSOs (open black squares), NAL
QSOs (open black diamonds) and NLS1s (open black triangles),
and \Chandra\, PI targets (large asterisks) are also indicated as
shown in the legend. } 
\label{imagx}
\end{figure}

\begin{figure}
% \plotfiddle{PSFILE}{VSIZE}{ROTANG}{HSCALE}{VSCALE}{HTRANS}{VTRANS}
%\epsscale{2}
\plottwo{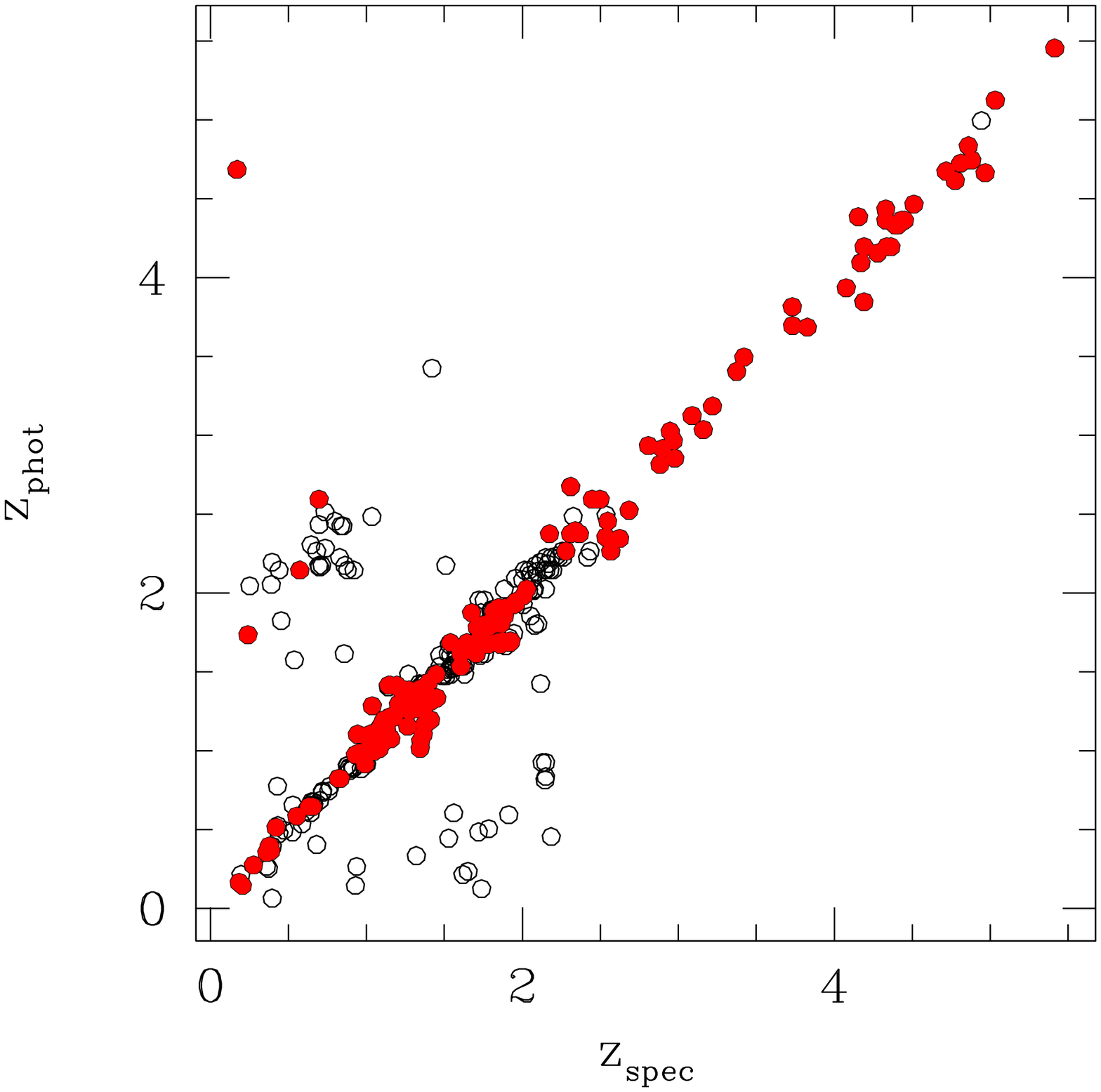}{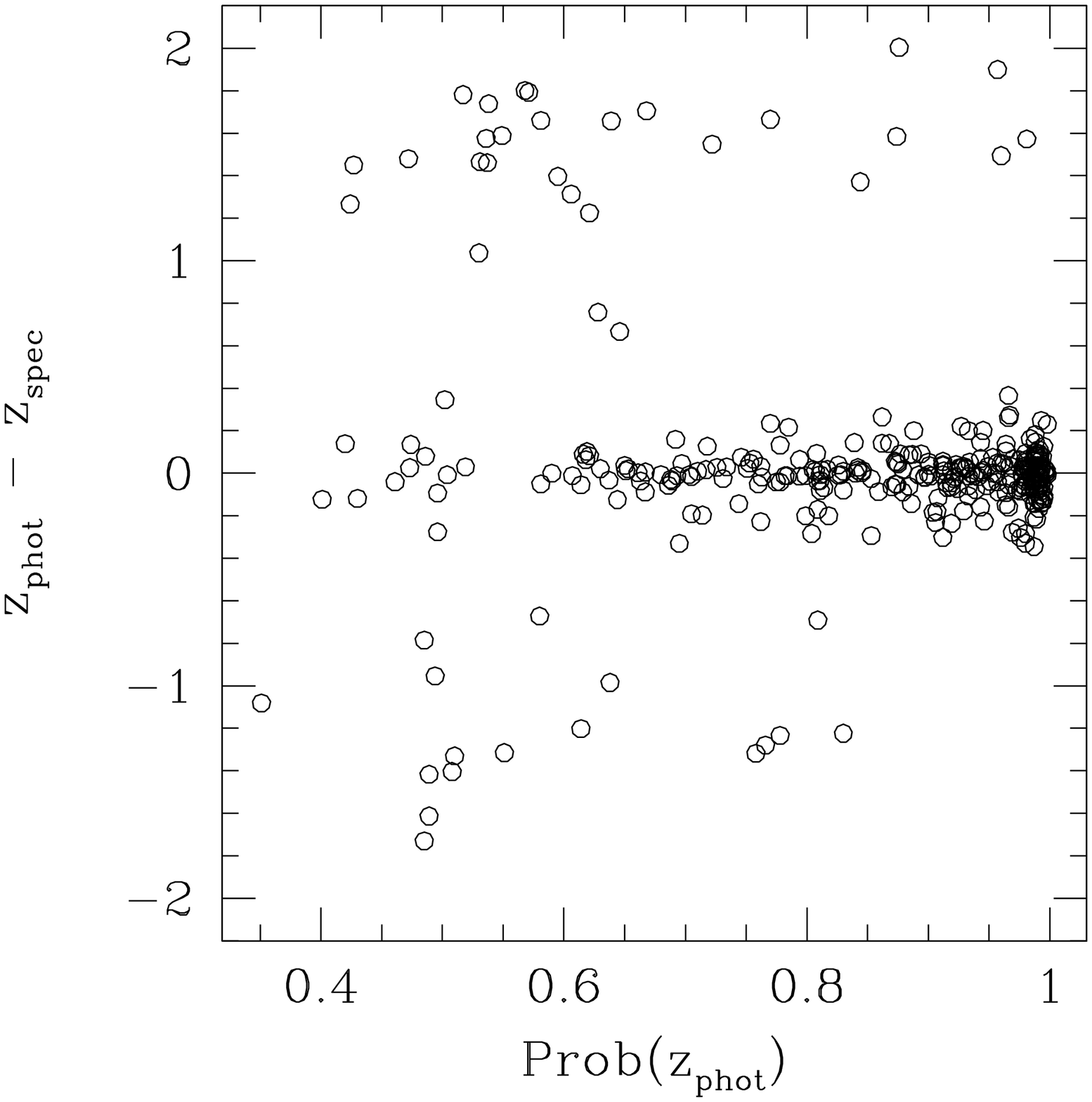}
\caption{LEFT: 
Photometric redshift vs. spectroscopic redshift for 
QSOs detected in ChaMP fields.  Filled (red) circles 
show QSOs for which the formal photometric redshift probability 
is $>$95\%.  A fraction of objects have large errors in their
photometric redshifts. About 18\% of QSOs with $z_{spec}$$<$1
have $z_{phot}>2$. This drops to 13\% using only Prob($z_{phot}$)$>$0.5.
RIGHT: 
Difference between the photometric and true (spectroscopic) 
redshift for QSOs in our sample, plotted against 
photometric redshift probability, again illustrates the
reliability of these probabilities.
}
%\vskip-0.2cm
\label{zcompare}
\end{figure}

After cross-correlation with the X-ray catalog, we sought
spectroscopic redshifts for any objects in the
photometric QSO catalog.  For this purpose, we obtained redshifts
from existing ChaMP spectroscopy, from the SDSS (DR6) database itself,
and then finally we searched the literature by cross-correlating
optical positions with the NASA Extragalactic Database (NED), using a
2\arcsec\, match 
radius.  Of 1376 matched objects, we found high confidence
spectroscopic redshifts for 407, of which 43 spectra
were observed by the ChaMP. In striking testimony to the quality of
the quasar selection algorithm (especially once candidates have been
matched to X-ray sources) only 8 of these (2\%) are not
broad line AGN.  Three are narrow emission line galaxies, 2 are
absorption line galaxies (no emission lines of equivalent width
$W_{\lambda}>5$\AA), and 3 are known BL~Lac objects. We exclude these
objects from all samples described below.  Because the magnitude
distribution (and therefore the photometric color errors) for objects
lacking spectroscopic 
classifications is fainter (see Fig~\ref{imagx}), we expect that the
overall fraction of misclassified photo-z QSOs is larger than 2\%, but
it would be difficult to estimate it without deeper spectroscopic
samples.  A plot of the photometric vs. spectroscopic redshifts of
these quasars is shown in Fig~\ref{zcompare}.  

\section{Samples, Subsamples, and Detection Fractions}
\label{detfrac}

The definition of our main sample and a variety of subsamples are
described in subsequent sections and summarized in
Table~\ref{tsamples}.  The size of our Main~sample
allows us to investigate the effects of luminosity or redshift limits,
X-ray non-detections, PI target bias, strong radio-related emission (RL
QSOs), broad and narrow line absorption (BAL and NAL QSOs), and
Narrow-Line Seyfert 1s (NLS1s).  In Table~\ref{tsamples}, samples to which we
refer frequently are arrayed under `Primary Samples' in decreasing
order of the number of detections.  Samples mentioned only
once or twice in this paper are listed in similar order
under `Other Samples'.  Tables listing bivariate statistical results  
(Tables~\ref{treg_gamma} -- \ref{treg_oaox}) later in the paper list 
samples in this same order for reference.  

We define the Main~sample to be the 2308  SDSS QSOs that fall on
\Chandra\, ACIS chips in a region of effective exposure $>$1200\,sec
(excluding CCD \#8; see below), regardless of \Chandra\, X-ray
detection status.   Our cleaned matched sample (the MainDet~sample)
of X-ray/optical matched QSOs contains  1135 distinct X-ray sources
with high optical counterpart match confidence, where we have removed
all sources (1) with significant contamination by nearby bright optical sources,   
%%%%%%% 51
(2) significant overlap  with other X-ray sources, 
%%%%%%% 20
(3) detected on ACIS-S chip S4 ({\tt CCD}\,\#8), because of its high
background and streaking,  
%%%%%%% 71
(4) dithering across chips (which renders unreliable the {\tt yaxx}
X-ray spectral fitting described below) 
%%%%%%% 21 
or (5) with spectroscopy indicating that the object is not a Type~1
QSO.  
%%%%%%% 10
% These cuts eliminate 2 otherwise valid serendipitously observed BALQSOs and 2 RL QSOs.   

\begin{figure}
\plottwo{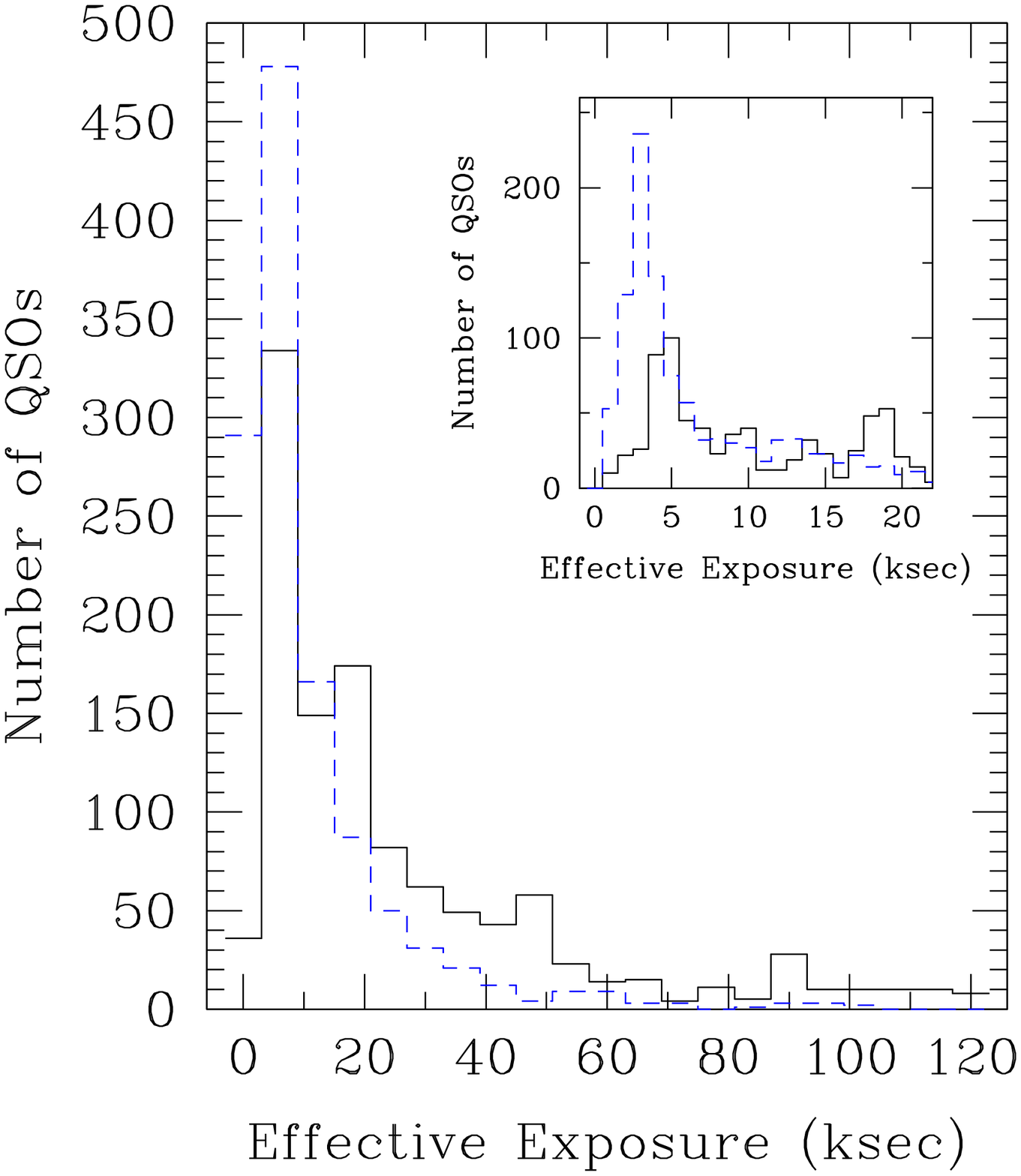}{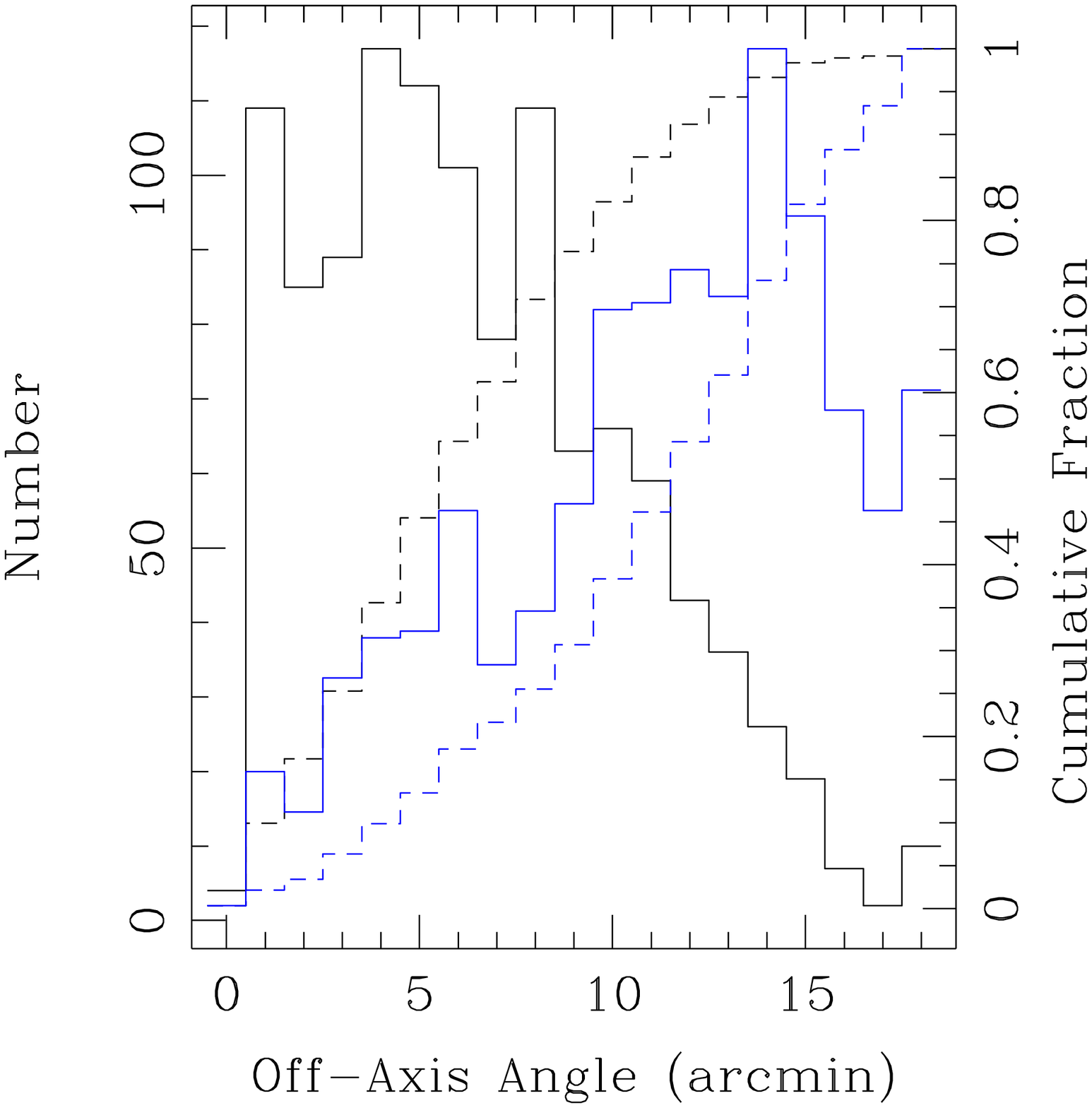}
%\vspace*{-1.25cm}
%\epsscale{1.4}
\caption{LEFT:
Histogram of effective \Chandra\, exposure times for QSOs in
the Main~sample, in units of ksec, with bin size 6\,ksec for detections
(solid black line) and limits (dashed blue line).  For detections,
the mean and median exposure times are 25.9 and 17.6\,ksec,
respectively.  Inset shows detail at low exposure times using
bin size 1\,ksec.  
RIGHT: Histogram of \Chandra\, off-axis angles for the Main~sample, in units
of arcmin.  Black solid histogram shows how detections trend towards
small OAA. Mean and median for detections are 6.4\arcmin\, and
5.8\arcmin, respectively. Blue histogram shows that limits
trend towards large OAA. Dashed histograms 
show the corresponding cumulative fractions. About 90\% of the 
detections (compared to $\sim50\%$ of the limits) are at
OAA$<$12\arcmin\, off-axis.}  
%\vskip-0.2cm
\label{thisto}
\end{figure}

The mean exposure time for the MainDet~sample is 25.9\,ksec per QSO, with an
average of 3.6 QSOs detected on each  \Chandra\, field.\footnote{For
  14 of 323 \Chandra\, fields, no QSOs are detected.}  For the
1173  non-detections in the Main~sample, the mean exposure time is
17.6\,ksec.  A histogram of exposure times for all QSOs in the
Main~sample is shown in Fig~\ref{thisto}.   

\begin{figure*}
\plottwo{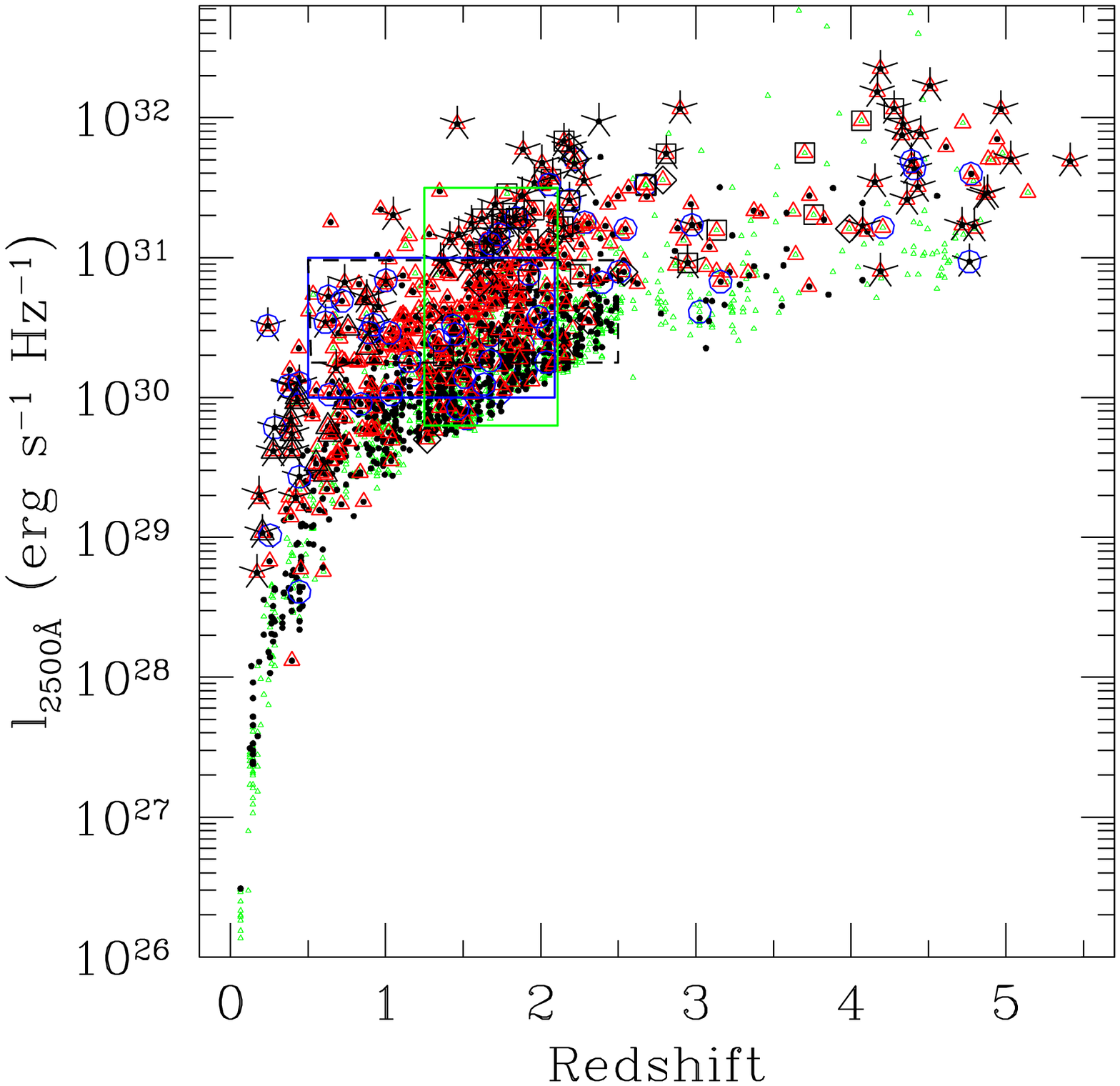}{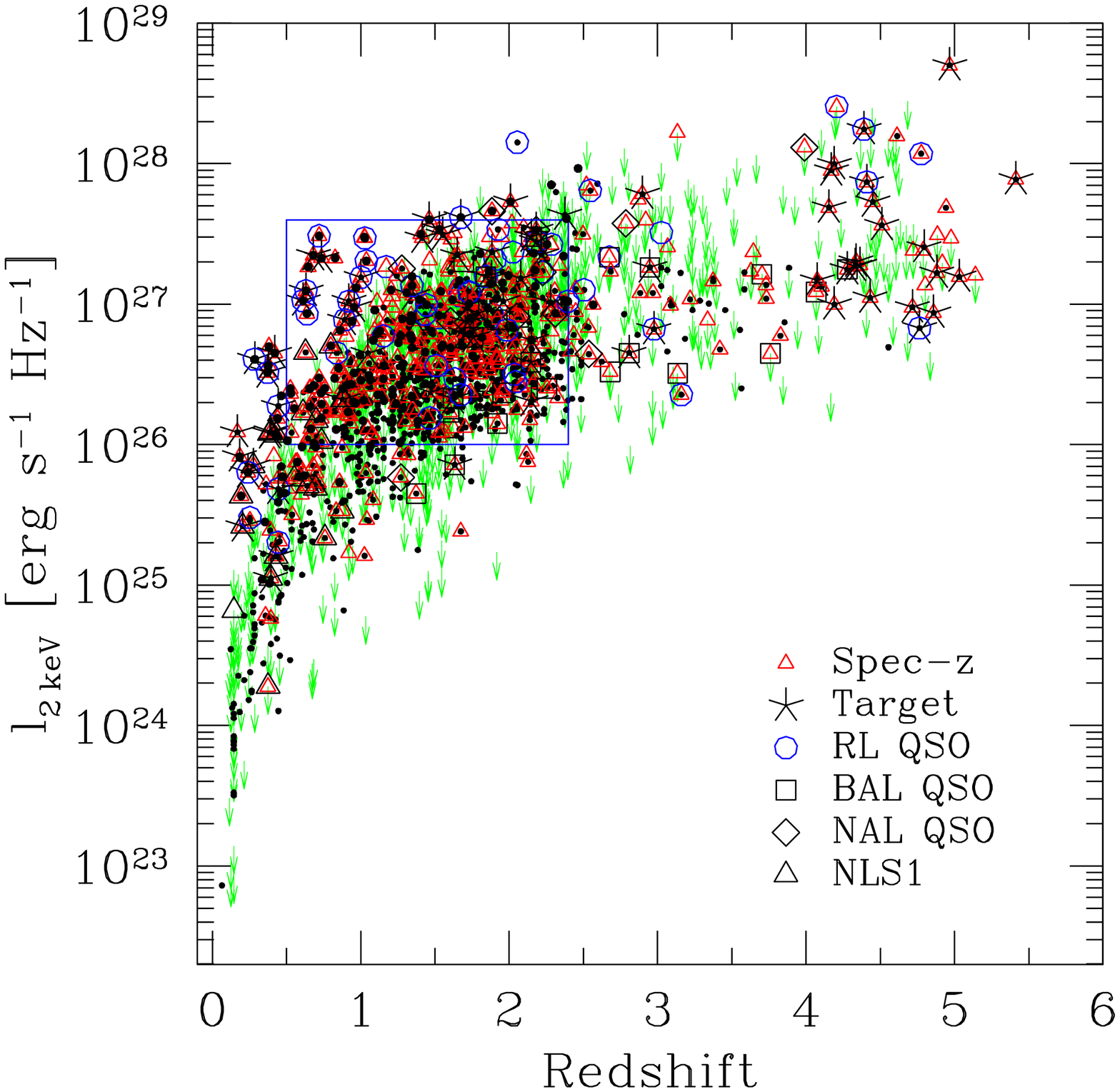}
\caption{Luminosity at 2500\AA\, (LEFT) and 2\,keV (RIGHT)
vs. redshift for the SDSS/ChaMP the Main~sample.  Above $z\sim 2.5$, the
number of QSOs declines steeply, due to the SDSS magnitude limit
and to the decreased efficiency of the photometric selection algorithm
as it crosses the stellar color locus (see Fig~\ref{rawdethistos} and
\citealt{Richards02}). These plots show 56 QSOs with $z>3$, of which 34
are new serendipitous detections. X-ray upper limits are shown as
small green triangles here.  See Fig~\ref{imagx} for symbol types.
The dashed black rectangle surrounds the zBoxDet~sample, a portion of
the \opteml$-z$  plane chosen to test for redshift dependence. The
green rectangle ``LoptBox'' surrounds  the LoBox~sample, used to test for
dependence on \opteml.  The blue rectangle surrounds the
zLxBox~sample,  chosen to avoid X-ray flux limit bias.  }     
%\vskip-0.2cm
\label{lz}
\end{figure*}

\begin{figure*}
\plottwo{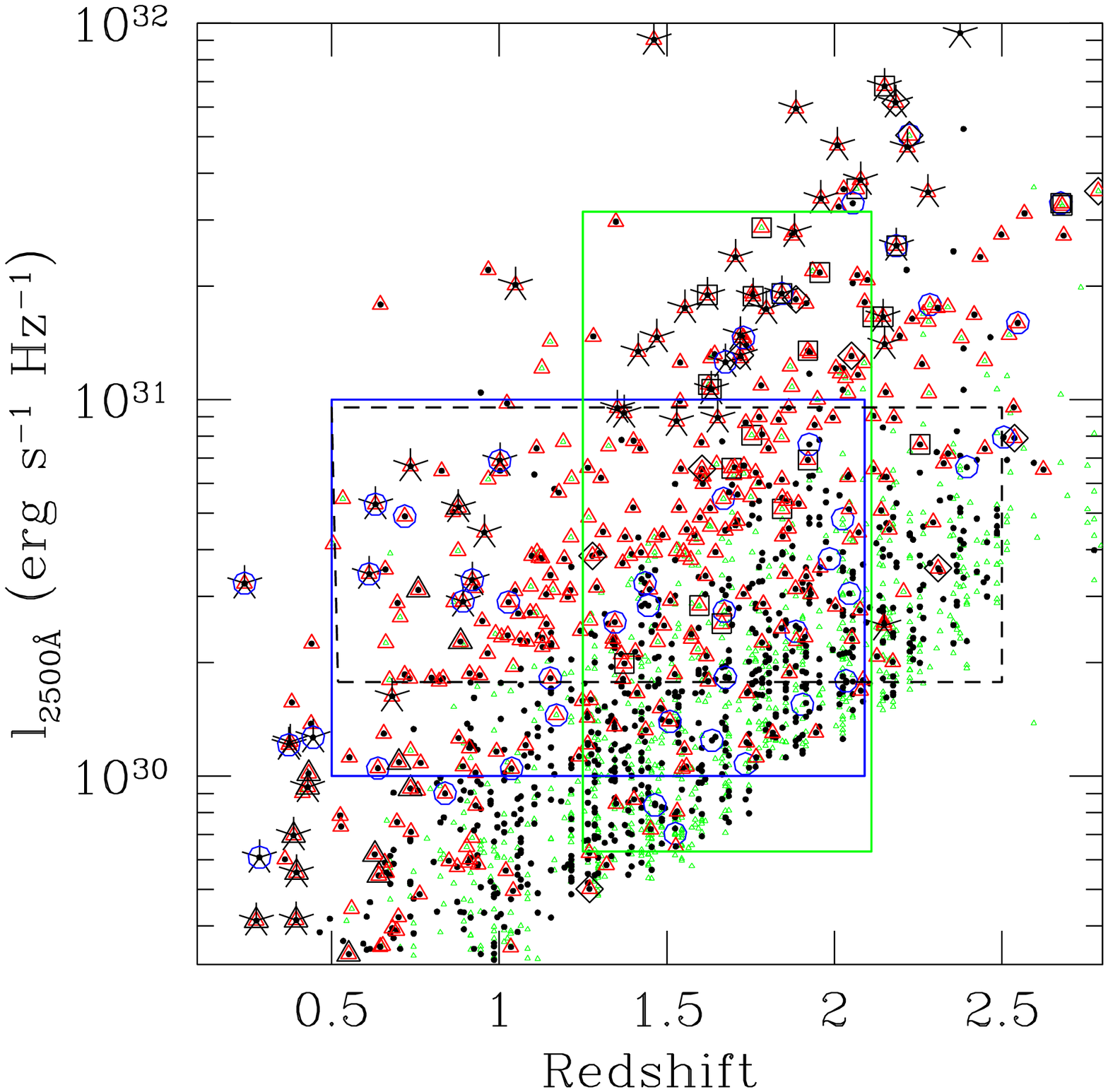}{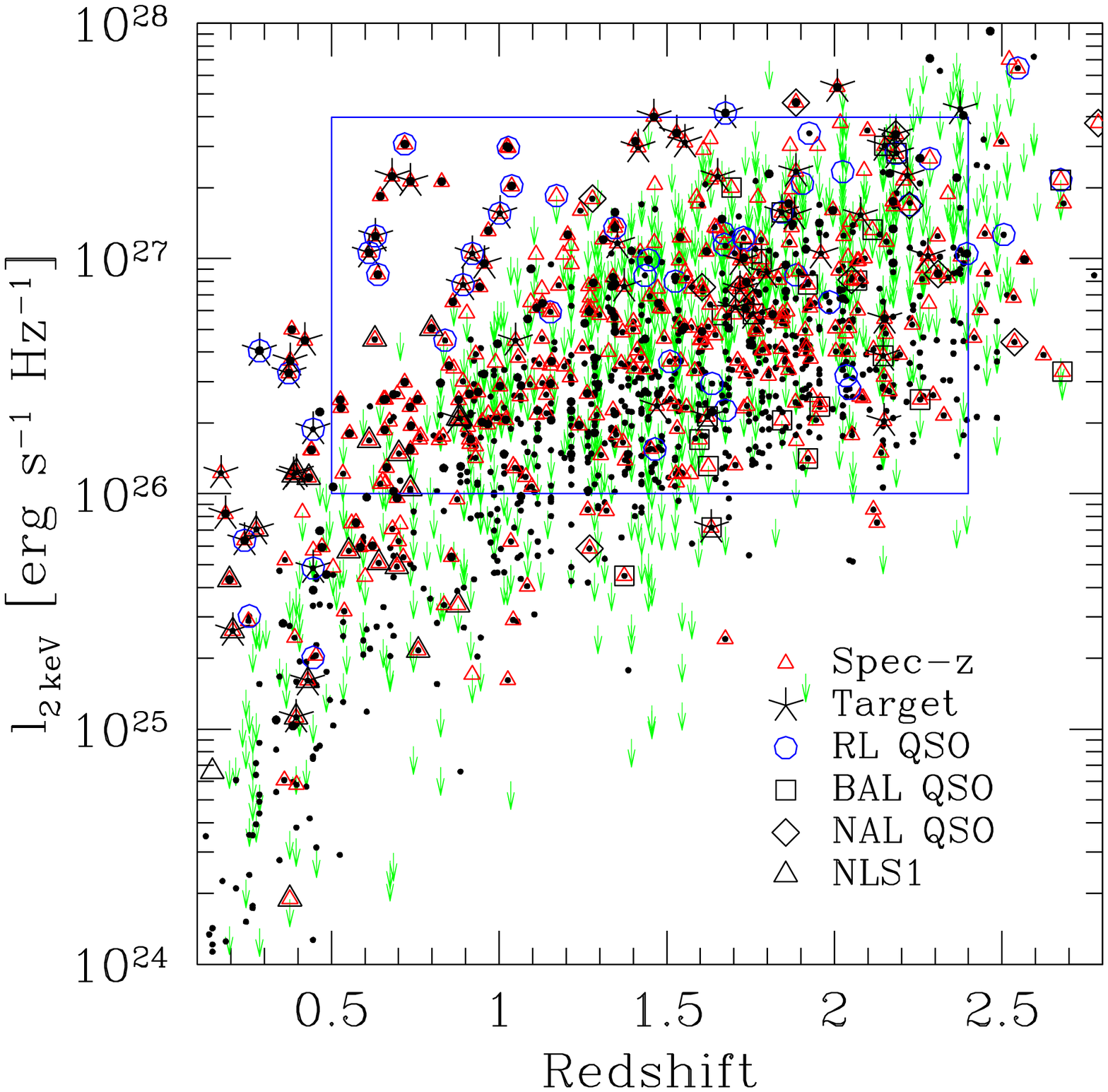}
\caption{LEFT: Zoom-in of 2500\AA\, luminosity vs. redshift.
Objects with spectroscopic redshifts (open red triangles) tend to be
at high optical luminosities by selection.  Detectably
radio-loud QSOs are shown with open blue circles.  
See Fig~\ref{imagx} for symbol types. BALQSOs (open black
squares) are mostly detected at $z>1.6$ where the CIV region enters the
optical bandpass. There appears to be no preference of BALQSOs for
high optical luminosity, apart from the bias caused by \Chandra\, target
selection. 
RIGHT: Zoom-in of 2\,keV luminosity vs. redshift. Here the RL QSOs
clearly populate the upper luminosity envelope.  BALQSOs are
preferentially X-ray quiet, unlike the QSOs with NALs only (open
diamonds).  } 
%\vskip-0.2cm
\label{lz_zoom}
\end{figure*}

We publish key data for 1135  QSOs in the MainDet~sample  in
Table~\ref{tdets}, marking 82  sources that are the intended
\Chandra\, principal investigator (PI) targets. X-ray sources in plots  
include only the MainDet~sample or subsets of it.  Fig~\ref{lz} shows
luminosity vs. redshift for the MainDet~sample. A large fraction of 
the $z>4$ objects are \Chandra\, targets (large black stars).  Strong
redshift-luminosity trends are seen both in optical and X-ray,
as is expected from any flux-limited survey.  However, the 
factor $\sim$30--50 range in luminosity is unusual for a single
sample; such breadth is usually only achieved using sample
compilations encompassing diverse selection techniques. 
In Fig~\ref{lz}, the large number of objects in our sample makes it
difficult to distinguish the point-types presenting object class
information, so Fig~\ref{lz_zoom} shows a zoom-in on the
most densely-populated regions of the $L-z$ plane.

Since we start with Type~1 SDSS QSOs, we are studying an
optically selected sample, and the selection function 
is complex \citep{Richards06a}. If we limit the analysis to
detections-only, then the sample is both optically- and
X-ray-selected, and the selection function becomes increasingly 
complex.  If instead we include all X-ray upper limits
in the analysis, the sample remains fundamentally optically selected,
but then statistical analyses must incorporate the
non-detections (see \S~\ref{stats}). 

\begin{figure}
%\epsscale{1.9}
\plottwo{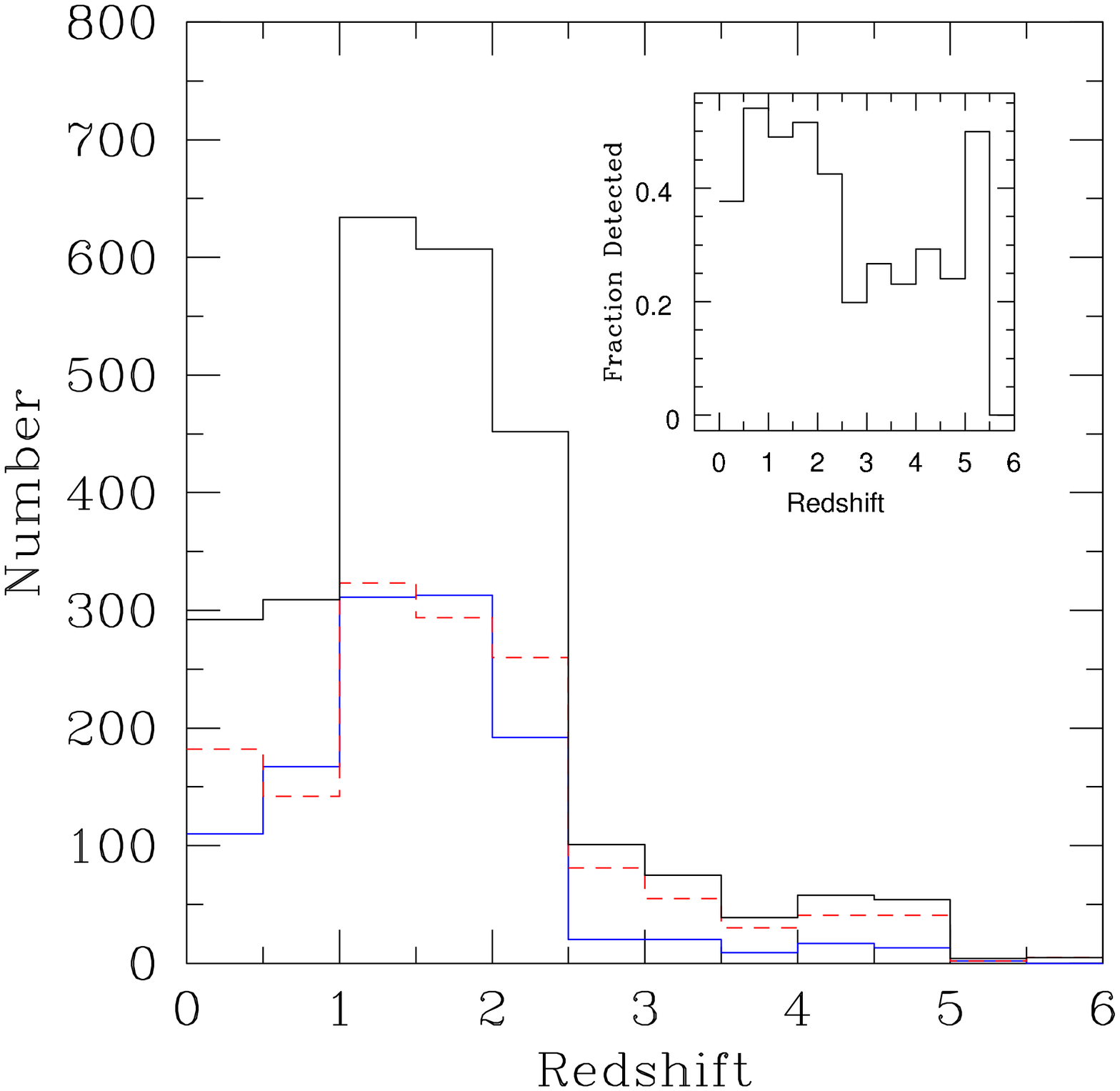}{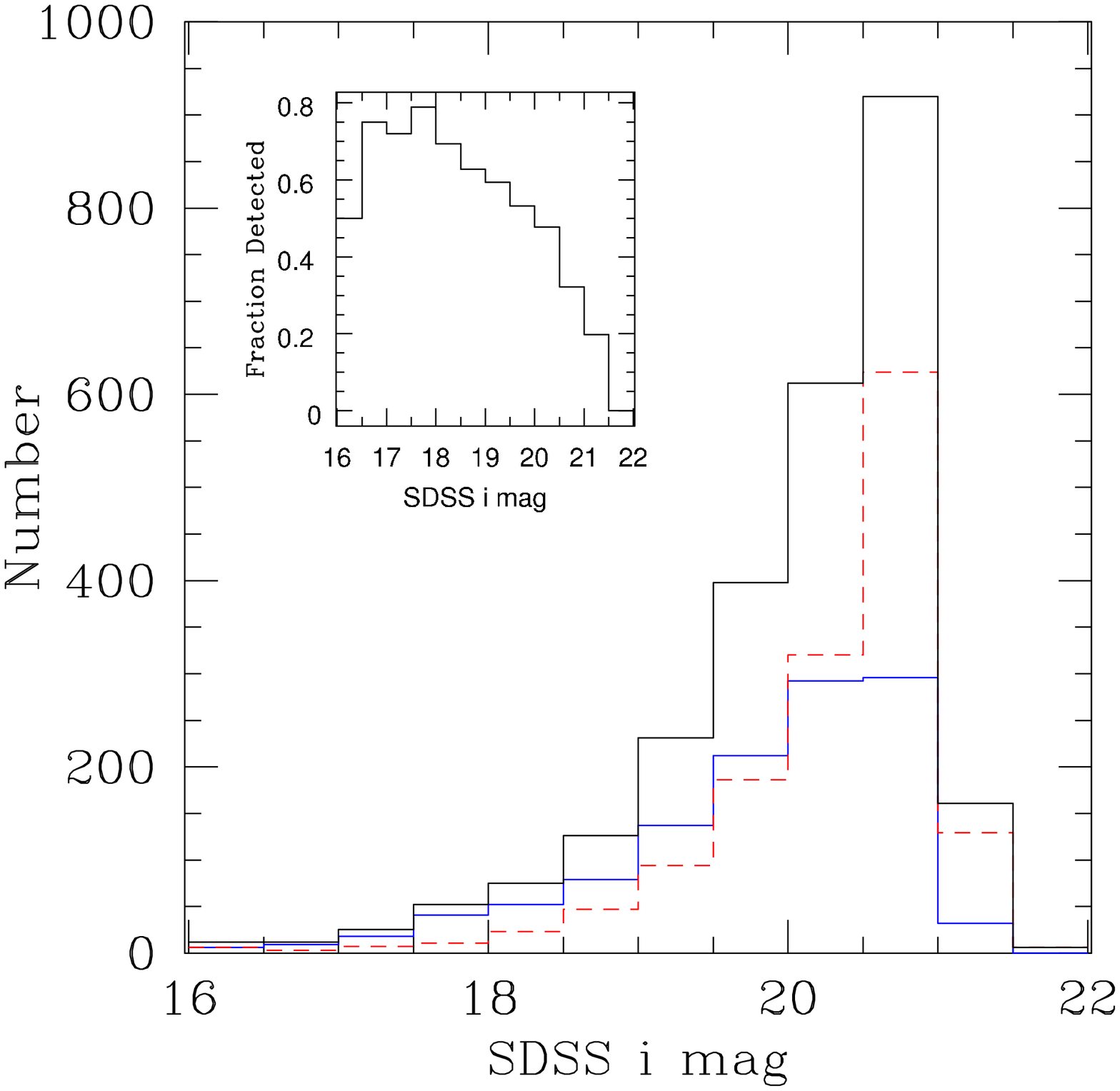}
\caption{
LEFT:
Histogram of redshifts for detected (solid blue), non-detected (red
dashed) and all (black solid) QSOs in the the MainDet~sample.  
We tally the `best' redshift for each object (i.e., spectroscopic
redshifts are always used when available). Above $z\sim 2.5$, the
number of QSOs declines steeply, due to the  decreased efficiency of
the  photometric selection algorithm as it crosses the stellar color
locus \citep{Richards02}.  The inset shows the detected fraction as a
function of redshift.  
RIGHT:
Histogram of SDSS $i$ mag for detected (solid blue), non-detected (red
dashed) and all (black solid) QSOs.  The inset shows the detected
fraction as a function of magnitude.}
%\vskip-0.2cm
\label{rawdethistos}
\end{figure}

\begin{figure}
%\epsscale{1.9}
\plottwo{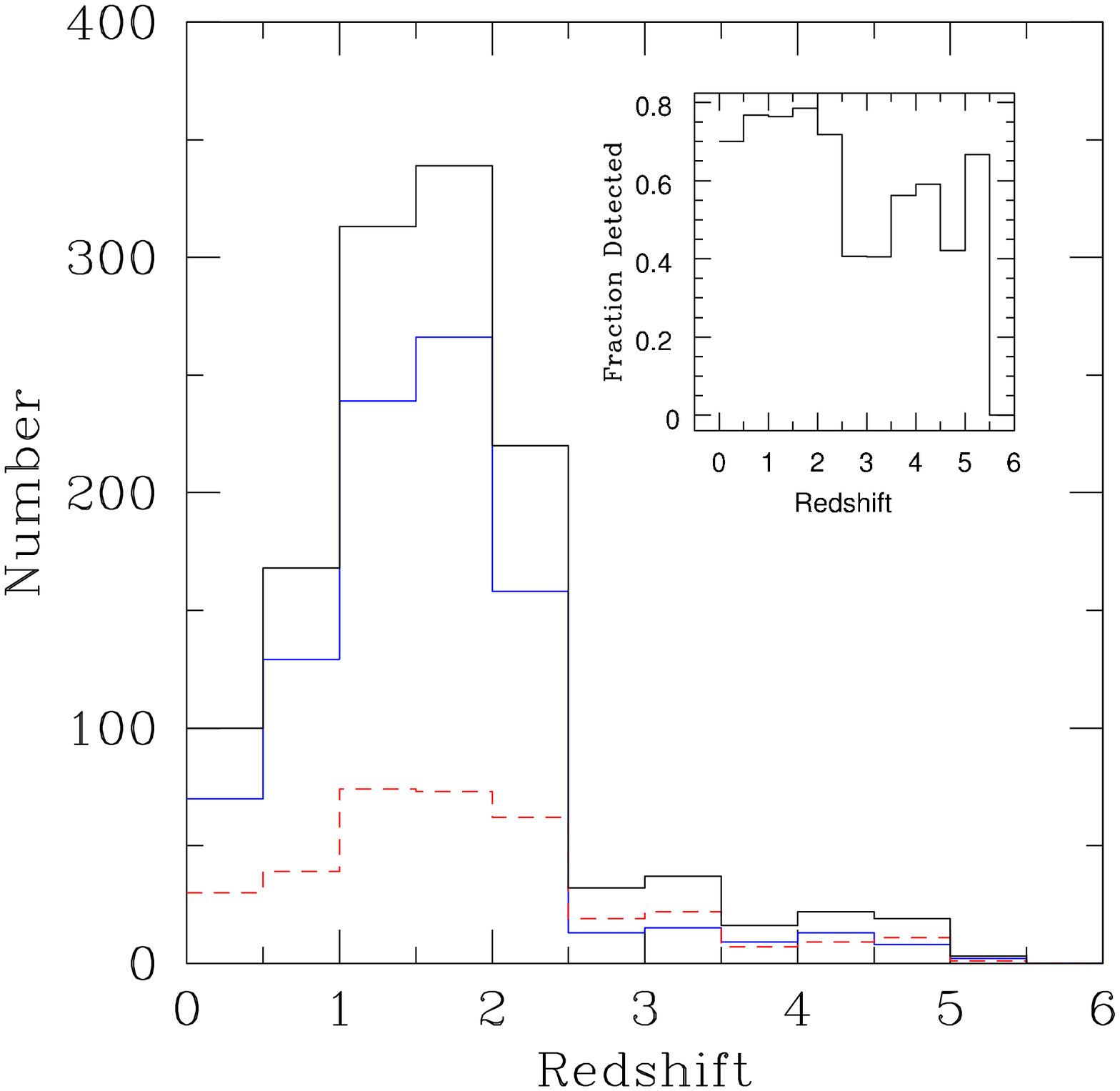}{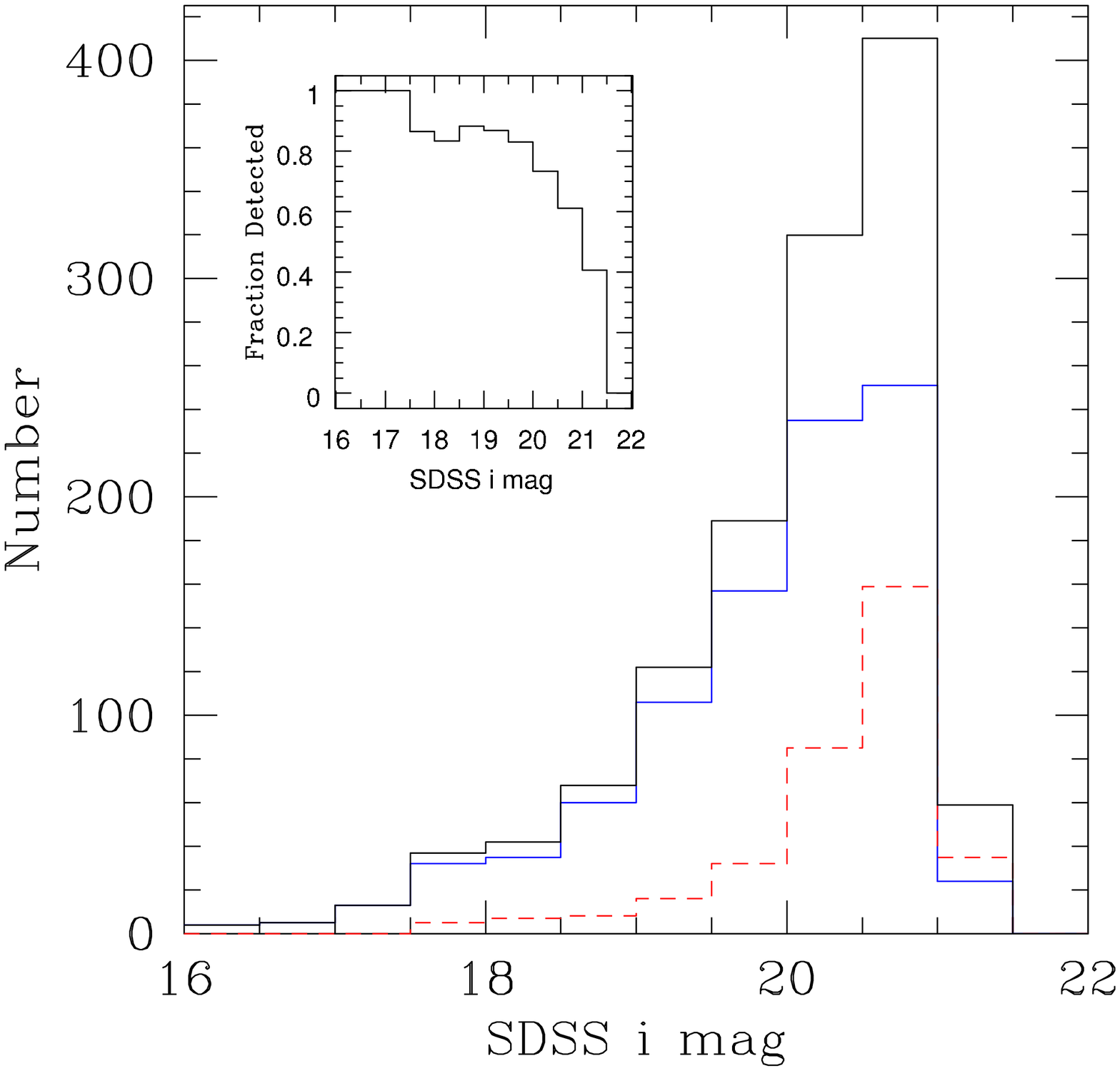}
\caption{LEFT:
Histogram of redshifts for detected (solid blue), non-detected (red
dashed) and all (black solid) QSOs, after restriction to obsids with
$T>4$\,ksec and QSOs with OAA$<$12\arcmin\, (the D2L~sample). The inset shows
the detected fraction as a function of redshift.
RIGHT:
Histogram of SDSS $i$ mag for detected (solid blue), non-detected (red
dashed) and all (black solid) QSOs.  The inset shows the detected
fraction as a function of magnitude.}
%\vskip-0.2cm
\label{hidethistos}
\end{figure}

The ChaMP's {\tt xskycover} pipeline allows us to investigate the
detection fraction for the full SDSS QSO sample, shown in
Fig~\ref{rawdethistos}. Of  2308  SDSS QSOs that fall on an ACIS
chip (the Main~sample) in our 323  ChaMP fields, 1135  (49\%) are
detected in the MainDet~sample.  Detection fractions as a function
of SDSS QSO mag and redshift for 
this sample are shown in Fig~\ref{hidethistos}.  To minimize
sample biases, we can also examine detection fractions as a function of
X-ray observing parameters like exposure time and OAA. To
simultaneously optimize detected sample size and detection fraction, we 
simply maximize $N_{det}^2/N_{lim}$, where $N_{det}$ and $N_{lim}$
are the number of X-ray detections and non-detections (flux upper
limits), respectively.  We find that an X-ray-unbiased
subsample with a significantly higher detection rate is
achieved by limiting consideration to the 1269  QSOs with
OAA$<12\arcmin$\, and exposure time $T>4$\,ksec, of which 922 (72\%)
are detections.  This high detection fraction sample is called
the D2L~sample (Table~\ref{tsamples}). 

Data for X-ray non-detections is available in 
Table~\ref{tlims}. We include in Table~\ref{tlims} only 
the 347 limits in the D2L~sample, where the flux limits are sensitive
enough to be interesting.  
By ``interesting'', we mean that the limits are close to or brighter
than the faint envelope of detections.  A flux limit several times
brighter than that envelope provides no statistical contraints
whatsoever on the derived distributions or regressions. For
X-ray non-detections, data are more sparse all around for several
reasons.  QSOs with limits are optically fainter (mean and median
$i$=20.4\,mag for the D2L~sample limits 
compared to $i=19.9$ for detections).  Being fainter, fewer have SDSS
spectroscopy.  Also, as non-detections, none have been targeted for
spectra by the ChaMP, so globally only about 10\% of non-detected QSOs
have optical spectra. The fraction of radio detections is also smaller
(1.2\% vs 4.8\%).  None are Chandra PI targets.  Finally, X-ray
non-detections lacking optical spectroscopy are somewhat less likely
to be QSOs.  The selection efficiency (fraction of QSO candidates that
are actual QSOs) between about $0.8<z<2.4$ is $\sim$95\%
\citep{Richards04,Myers06}, but \citet{Richards08} estimate that near
the faint limit of $i\sim$20.4\,mag, the overall QSO selection
efficiency is $\sim 80\%$. Particular attention must be paid to
possible galaxy contamination at the faint end as the autoclustering
estimates of the efficiency do not include galaxy interlopers at faint
limits where SDSS star-galaxy separation begins to break down.
However, many of these 'spurious' cross-matches may turn out to be
(e.g., low-luminosity) AGN.  In any case, the increased level of
contamination by non-QSOs is another rationale for limiting the number
of non-detections to those with sensitive X-ray limits. The nature of
the statistical analysis (discussed in \S~\ref{stats}) is such that
non-detections are included, but are assumed to follow the
distribution of detections, and so effectively have a lower weight in
the results.

% See Fig~\ref{imagx}

% select srcid from dr5qso_xomatch4as d, ospec s where d.srcid=s.xsrcid and s.typconf=3;

Amongst the undetected QSOs in the Main~sample, 165  have SDSS
spectroscopy, of which 144 are high confidence spectroscopic QSOs.  
% Judging from the SDSS specClass parameter, 
The 21 non-QSOs comprise 16 stars, 5 galaxies. The higher (13\%) 
rate of non-star spectroscopic classifications amongst undetected QSOs 
is not surprising, since X-ray detection greatly increases the
probability that an optical AGN candidate is indeed an AGN.
From the upper limit QSO sample, we remove the 22 non-QSOs, and
use the SDSS spectroscopic redshifts instead of the photometric
redshifts wherever applicable.

\vskip1cm
\subsection{Targets}
\label{targets}

The ``Non-target'' detected sample (the noTDet~sample) of 1053 QSOs 
further eliminates 82 objects ($7.2\%$ of the MainDet~sample)
that are the intended targets of the \Chandra\, observation
wherein they are found.  Targets are on average brighter than most of the
QSO sample (see Fig~\ref{imagx}), but more 
importantly were chosen for observation for a variety of reasons
unrelated to this study.  In particular, targets tend to be more
luminous than serendipitous QSOs (Figs~\ref{lz},~\ref{lz_zoom}), and
several are known lenses (e.g., HS\,0818+1227, PG\, 1115+080,
UM\,425=QSO\,1120+019).  The bias in sample characteristics is
largely mitigated in the subsamples excluding targets (see
Table~\ref{tsamples}). 
However, the exclusion of targets also produces a (much smaller) bias
because some objects with similar characteristics would have been
included (at a lower rate) were the \Chandra\, pointings all truly
random.  Because many of the targets are indeed of interest (e.g.,
high-z QSOs), we include them in most discussions, but always check
that results are consistent without them. We also note that some
target bias probably affects the X-ray sample even after the exclusion
of PI target QSOs, because PI targets may cluster with other
categories of X-ray sources such as other AGN, galaxies or clusters. 
Overall, a comparison of regression
results\footnote{Table~\ref{treg_ox}-\ref{treg_oaox} shows similar
  results comparing 
e.g., the MainDet~sample and the noTDet~sample, or the D2LNoRB~sample and the D2LNoTRB~sample.}
for several of our subsamples that differ only in target exclusion
does not indicate a significant target bias, due at least in part to
our large sample sizes. 

\subsection{Radio Loudness}
\label{rlqs}

Quasars with strong radio emission are observed to be more X-ray
luminous (e.g., \citealt{Green95,Shen06}).  At least some of the
additional X-ray luminosity is likely 
to originate in physical processes related to the radio jet rather
than to the accretion disk, so it may be important to recognize
those objects that are particularly radio loud.  

The Faint Images of the Radio Sky at Twenty-Centimeters (FIRST) 
survey \citep{Becker95} from the NRAO Very Large Array (VLA)
has a typical (5$\sigma$) sensitivity of $\sim$1\,mJy, and
covers most of the SDSS footprint on the sky.  Following
\citet{Ivezic02}, we adopt a positional matching radius of 1.5\arcsec,
which should result in about 85\% completeness for core-dominated
sources, with a contamination of $\sim$3\%.  We thereby match  69
sources to the Matched sample.  \citet{Jiang07} matched the FIRST to SDSS
spectroscopic quasars and found that about 6\% matched within 5\arcsec.
% 2566/31835 total matches, but 1944 <5arcsec, and 622 are FRIIs <30arcsec
We might expect a lower matched fraction because our optical photometric
sample extends 1-2\,mag fainter.  On the other hand, we are looking at
X-ray-detected  quasars, so the actual matched fraction of $\sim5\%$
is similar.  We also matched all the quasar optical positions
to the FIRST within 30\arcsec, and visually examined all the FIRST
images to look for multiple matches and/or lobe-dominated quasars.
There are 26 sources we judged to have reliable morphological
complexity that are resolved by FIRST into multiple sources.
The NRAO VLA Sky Survey (NVSS; \citealt{Condon98}), lists detections
for 19 of these.  Comparing NVSS to summed FIRST fluxes, we found 
NVSS fluxes slightly larger - $<2\%$ difference in the mean
($\sim20\%$ max).  Since the NVSS beam is larger
($45\arcsec$) than FIRST ($5\arcsec$), FIRST detection algorithms
may exclude some of the extended source flux as background, so we
include the NVSS fluxes for these 19 objects, and summed FIRST
fluxes for those remaining.

% See Double-Lobed Radio Quasars from the Sloan Digital Sky Survey
% \bibitem[de Vries et al.(2006)]{2006AJ....131..666D} de Vries, W.~H., 
% Becker, R.~H., \& White, R.~L.\ 2006, \aj, 131, 666 
% Note that from among X-ray NON-DETECTIONS (based on SDSS DR5 QSOs 
% that fall on ACIS chips in dr5qsos_skyout), only 19 new FIRST sources 
% are found above those with X-ray detections... not worth the work!

Following \citet{Ivezic02}, we adopt a radio loudness parameter $R$ as 
the logarithm of the ratio of the radio to optical monochromatic flux: 
$R=\log(F_{\mathrm 20\,cm}/F_i) = 0.4(i-m_{\mathrm 20\,cm})$, where
$m_{\mathrm 20\,cm}$ is the radio AB magnitude \citep{Oke83},
$m_{\mathrm 20\,cm}=-2.5\log(F_{\mathrm 20\,cm}/3631\textrm{~Jy})$
calculated from the integrated radio flux density, and 
$i$ is the SDSS $i$-band magnitude, corrected for Galactic
extinction. We adopt a radio-loudness threshold $R=1.6$.
Thus there are 72 QSOs in the Main~sample with radio detections, of
which 57 (79\%) are radio-loud.  For the MainDet~sample (detections
only), there are 55 radio-detected QSOs, of which 43 (78\%) are radio-loud.  

Given the ($\sim$1\,mJy) source detection limit of the FIRST Survey,
all RL QSOs will be detected to about $i\sim$20.4\,mag.  
For the magnitude range 17$<i<$20 where the statistics are good
and the FIRST is sensitive to all RL QSOs, we find 41 of
529 (8$\pm$1\%) such QSOs from the MainDet~sample are detected by the FIRST,
with 29 (5.4\%) that are radio loud.  Since the 34\% of our full 
the MainDet~sample that is fainter than $i$=20.4 suffers from incomplete radio
loudness  measurements, some 2\% may be unidentified RL QSOs. 
A similar fraction pertains if we count X-ray non-detections as well
(the Main~sample).

\begin{figure}
\plotone{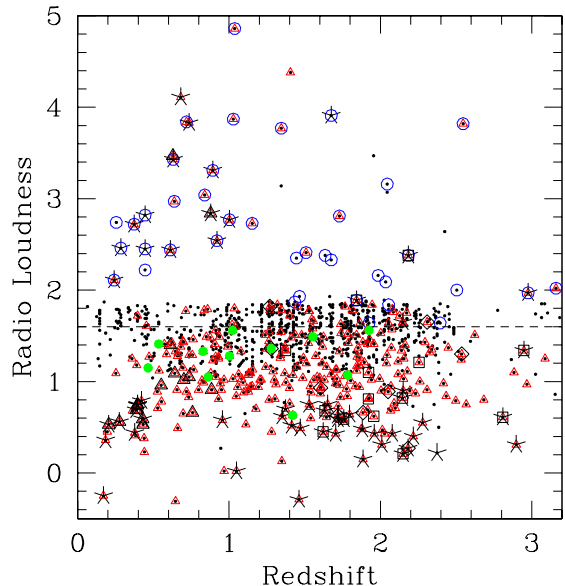}
%\vspace*{-1.25cm}
\caption{Radio-loudness vs. redshift for the MainDet~sample
(detections). Radio-loud objects $R>1.6$  are shown with open blue
circles.  Radio-quiet but FIRST radio-detected objects are shown as
filled green circles. All other symbols (described in Fig~\ref{imagx})
have radio flux upper limits only.  Note that most \Chandra\, targets
are distinctly either loud or quiet, highlighting a bias in the target
subsamples.   
}
\label{z_RL}
\end{figure}

\subsection{Broad and Narrow Absorption Line Quasars}
\label{bals}

We identified QSOs with broad absorption lines (BALs) and narrow
absorption lines (NALs) directly by visual inspection of
QSOs with spectroscopy, finding 16 BAL and 11 NAL QSOs   in
the the MainDet~sample. Ten (two)  of the BAL (NAL) QSOs were the
\Chandra\, PI targets. 

The best estimates to date of the raw BALQSO fraction among
optically selected quasars range from about 13 -- 20\%
\citep{Reichard03,Hewett03}. From the SDSS DR3 sample
of \citet{Trump06}, \citet{Knigge08} carefully define what is a BALQSO
and correct for a variety of selection effects to derive an estimate
of the {\em intrinsic} BALQSO fraction of 17\%$\pm$3.
The vast majority of
BALQSOs in the SDSS are above redshift 1.6 because
only then does the CIV absorption enter the spectroscopic
bandpass.\footnote{A much smaller number of the rare low-ionization
BALQSOs (with BALs just blueward of MgII) are found at 
lower redshifts.}  If we determine our BALQSO fraction in the MainDet~sample
only counting the serendipitous (non-Target) QSOs with $z>1.6$ 
and spectroscopic redshifts, we find just 4 out of 119  QSOs with
BALs.  Even with sensitive X-ray observations such as these, X-ray
selection is strongly biased against the high ionized absorbing
columns along the line of sight towards the X-ray emitting regions of
BALQSOs.   Of the 24  absorbed (BAL or NAL) QSOs two are detectably
radio loud; SDSS\,J171419.24+611944.5 - a BAL QSO - and
SDSS J171535.96+632336.0 - a NAL QSO - are targets selected (by
\Chandra\, PI Richards) as reddened QSOs.  

\subsection{Narrow Line Seyfert 1s}
\label{nls1s}

X-rays from NLS1s are of particular interest because they were thought
to show marked variability and strong soft X-ray excesses
(e.g., \citep{Green93, Boller96}).  NLS1s are proposed
to be at one extreme of the so-called \citep{BG92} ``eigenvector 1'',
which have been suggested to correspond to low SMBH masses 
\citep{Grupe04} and/or high (near-Eddington) accretion rates
\citep{Boroson02}.   

For objects with SDSS spectra encompassing H$\beta$ ($z<0.9$),
we identify as Narrow Line Seyfert 1s (\citealt{Osterbrock85}; 
NLS1s hereafter) those objects with FWHM(H$\beta$)$<$2000\kms
and line flux ratio \oiii/\hb $<3$.  The FWHM measurements are
obtained  via  FWHM=$2.35\,c\,\sigma/\lambda_0/(1+z)$, where 
$\sigma^2$ is the variance of the  Gaussian curve that fits the 
$\hb$ emission line, $z$ is the redshift, and $\lambda_0$ is the
rest-frame wavelength of the \hb\, line (4863 \AA). We extract
measurements of $\sigma$ (in \AA) and line fluxes from
the SDSS DR6 SpecLine table. 

Our MainDet~sample contains at least 19 NLS1s (those with spectroscopy in
the redshift range to include \hb).  Among the non-detections, we
identify 3 more.  Six of these 22 objects are targets described in
\citep{Williams04}, and SDSS\,J125140.33+000210.8 was a target
selected (by \Chandra\, PI Richards) as a dust-reddened QSO.  
The \citet{Williams04} sample consisted of 17 SDSS NLS1s selected from
the RASS to be X-ray-weak. Their study confirmed earlier suggestions
that strong, ultrasoft X-ray emission is not a universal
characteristic of NLS1s.  

We therefore present here new results for a sample of 15 SDSS NLS1s
observed by \Chandra.  This admittedly small sample is nevertheless
the largest published sample of optically-selected NLS1s with unbiased
X-ray observations.  (The sample of \citet{Williams04} was selected
to be X-ray weak, while the \citet{Grupe04} ROSAT sample was selected
to have strong soft X-ray emission.)  We find no evidence for unusual
SEDs from the distributions of either \aox\, or $\Gamma$. 

\section{Optical Colors and Reddening}
\label{reddening}

\begin{figure*}
% \plotfiddle{PSFILE}{VSIZE}{ROTANG}{HSCALE}{VSCALE}{HTRANS}{VTRANS}
\plottwo{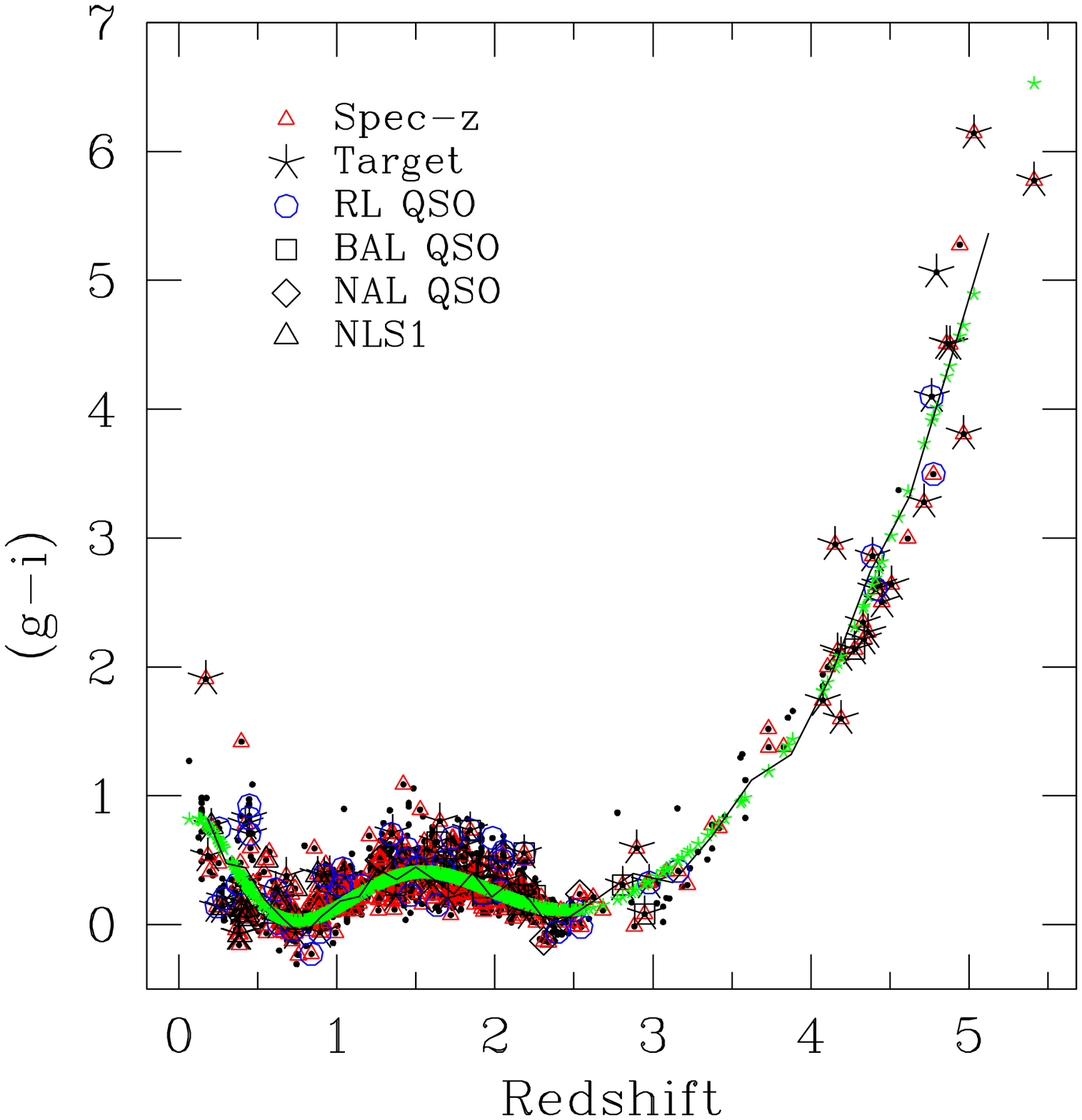}{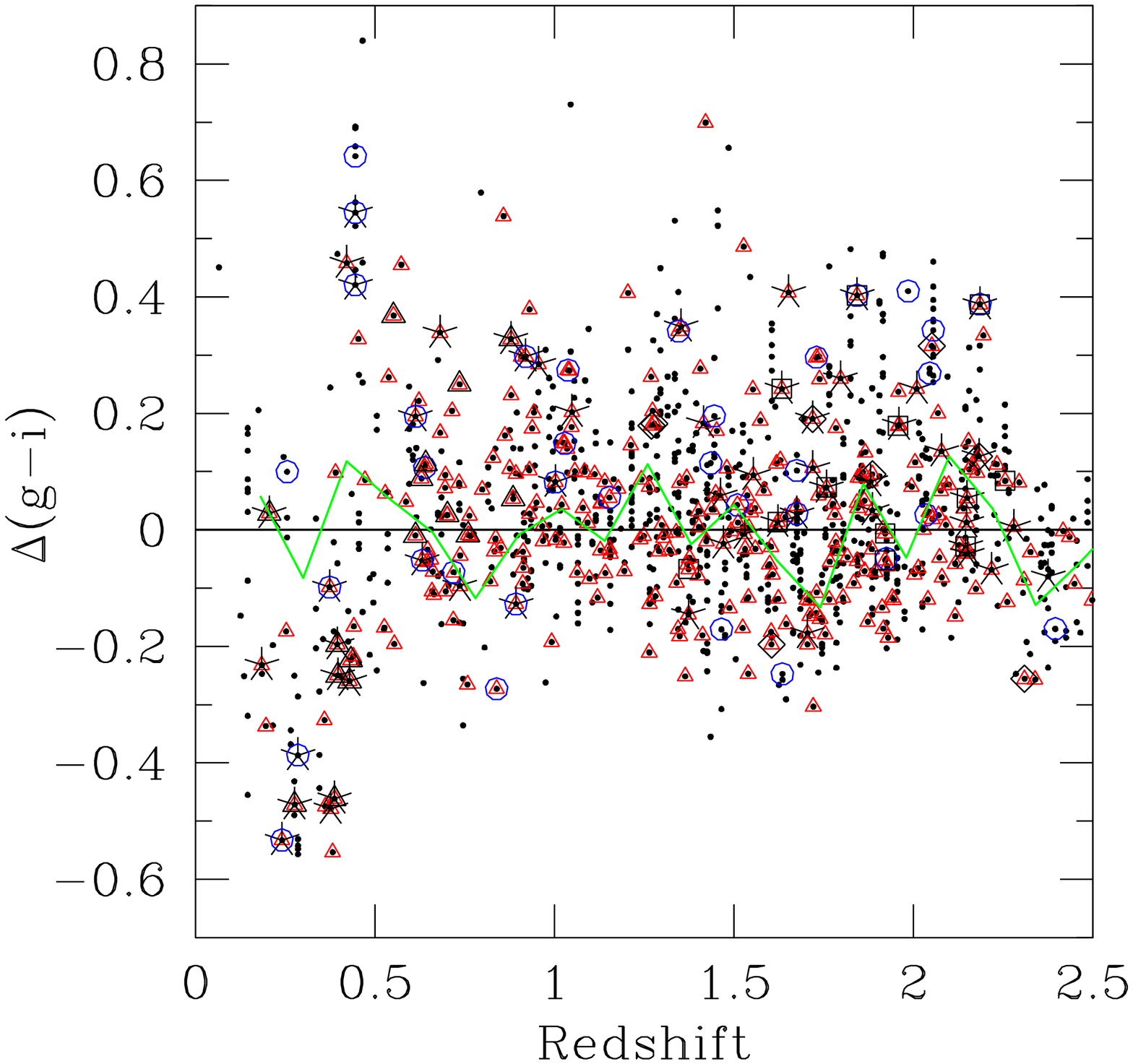}
\caption{LEFT: SDSS \gmi\, color vs. best redshift.
Symbols show individual QSOs in the SDSS/ChaMP the MainDet~sample.  The
black line shows the mean color at each redshift (in redshift bins of
0.12 for $z<$2.5 and 0.25 for higher redshifts) for the full optical
QSO sample (regardless of \Chandra\, imaging).  We derive a smooth
(9th order) polynomial fit to those means, whose value is 
plotted as the ``expected'' \gmi\, with a green asterisk (at the
redshift of each actual QSO). 
RIGHT:  The color excess  $\Delta$\gmi (the difference between the
actual and ``expected'' \gmi\, color) is plotted against redshift for
a limited redshift range, to show highlights.  The residuals of the
polynomial fit to the mean binned \gmi\, of the full optical sample are
shown connected by a solid green line. See Fig~\ref{imagx} for symbol
types.  Most targets, BALQSOs, and RL QSOs are redder than average. 
}
\label{zcolors}
\end{figure*}

In Fig~\ref{zcolors}, we plot the \gmi\, colors of the
matched SDSS/ChaMP QSO sample as a function of redshift, and compare 
to the optical-only sample.\footnote{The optical-only sample
refers to all SDSS QSOs
within $\sim$20\arcmin\, of all \Chandra\, pointings, regardless
of whether its position falls on an ACIS CCD. We use a 9th-order polynomial
fit to the optical-only sample with the following coefficients:
0.698892, 3.011733, -20.358267, 37.850353, -33.617121, 16.652032, -4.861889,
0.833027,-0.077526, 0.003026}.  The \Chandra-detected sample does not 
show significantly different colors from the full optical sample.
This likely attests to (1) the sensitivity of the \Chandra\, imaging
relative to the magnitude limit of the optical sample and (2) the
fact that Type~1 QSOs are largely unabsorbed in both the optical and 
X-ray regimes.  

The right panel of  Fig~\ref{zcolors} shows that 10 of 14 BALQSOs
are above the $\Delta\,\gmi$=0 line.  This reflects that SDSS BALQSOs tend to
be redder than average \citep{Reichard03,Dai07}. Most 
of the RL QSOs are also redder than average.  \citet{Richards01}
found a higher fraction of intrinsically reddened quasars among those
with FIRST detections. \citet{Ivezic02} found that RL QSOs are redder 
than the mean (at any given redshift) in \gmi\, by
0.09$\pm$0.02\,mag. Fig~\ref{zcolors} confirms a similar trend in  
the X-ray detected SDSS/ChaMP sample.  At the same time, a small
number of RL QSOs are found on the blue extreme of the color-excess
distribution.  These trends are fully consistent with the detailed
results found by stacking FIRST images of SDSS quasars
\citep{White07}, independent of X-ray properties.

\section{X-ray Spectral Fitting with {\tt yaxx}}
\label{yaxx}

Besides comparing the broadband multiwavelength properties of QSOs,
\Chandra\, imaging provides X-ray spectral resolution capable of yielding
significant constraints on the properties of emission
arising nearest the supermassive black hole. While the ChaMP
calculates Hardness Ratios (HR) and appropriate errors for every source,
these can be difficult to interpret, since HR convolves the intrinsic
quasar SED with telescope and instrument response, and does not
take redshift or Galactic column $\nhgal$ into account.  A direct
spectral fit of the counts distribution using the full instrument
calibration and known redshift and $\nhgal$ provides a much more
direct measurement of quasar properties.  Note that even
in the low-counts regime one can obtain robust estimates of fit
parameter uncertainties using the \citet{Cash79} fit statistic.

We use an automated procedure to extract the spectrum and fit up to
three models to the data.  For all objects in the Matched sample,
we first define a circular source region
centered on the X-ray source which contains 95\% of 1.5\,keV photons
at the given off-axis angle.  An annular background region is also
centered on the source with a width of 20\arcsec. We exclude any
nearby sources from both the source and background
regions.  We then use CIAO\footnote{http://cxc.harvard.edu/ciao} tool
{\tt psextract} to create a PHA(pulse height amplitude) spectrum
covering the energy range 0.4-8\,keV.  

Spectral fitting is done using the
CIAO {\it Sherpa}\footnote{http://cxc.harvard.edu/sherpa} tool in an automated
script known as {\tt yaxx}\footnote{http://cxc.harvard.edu/contrib/yaxx}
\citep{Aldcroft06}.  All of the
spectral models contain an appropriate Galactic neutral absorber.  For
all sources we first fit two power-law models which include a Galactic
absorption component frozen at the 21\,cm value:\footnote{Neutral
Galactic column density $\nhgal$ taken from \citet{Dickey90} for the
\Chandra\, aimpoint position on the sky.} (1) fitting photon index
$\Gamma$, with no intrinsic absorption component (model ``{\tt PL}'')
and (2) fitting an intrinsic absorber with neutral column $\nhintr$ at the
source redshift, with photon index frozen at $\Gamma=1.9$ (model
``{\tt PLfix}'').  Allowed fit ranges are $-1.5<\Gamma<3.5$ for {\tt
PL} and $10^{18}<\nhintr<10^{25}$ for {\tt PLfix}.  These fits use the
Powell optimization method, and provide a robust and reliable
one-parameter characterization of the spectral shape for any number of
counts.  Spectra with less than 100 net counts\footnote{Source counts
derived from {\tt yaxx}  may differ at the $\sim$1\% level from those
derived by ChaMP XPIPE photometry, due to slightly different
background region conventions.}  were fit using the ungrouped data
with Cash statistics \citep{Cash79}.  Spectra with more than 100
counts were grouped to a minimum of 16 counts per bin and fit using
the $\chi^2$ statistic with variance computed from the data.  

Finally, X-ray spectra with over 200 counts were also fit with a
two-parameter absorbed power-law where both $\Gamma$ and the $\nhgal$
were free to vary within the above ranges (model ``{\tt PL\_abs}''). 

\subsection{X-ray Spectral Continuum Measurements}
\label{gamma}

We compile ``best-PL'' measurements, where for fewer than 200 counts,
we use $\Gamma$ from the {\tt PL} ($\nhintr$ fixed at zero) fits and
for higher count sources we use {\tt PL\_abs} (both $\Gamma$ and
$\nhintr$ free).  In the MainDet~sample there are 156 sources  with 200
counts or more. High-count objects are found scattered at all luminosities
below $z\sim$2.5. QSOs with $>$200 counts (0.5-8\,keV) with both
$\Gamma$ and $\nhintr$ fits in {\tt yaxx}, are well-distributed in \xeml\,
amongst the detections, due to the wide range of exposure times. 

The mean $\Gamma$ for all the 1135 QSOs in the MainDet~sample is
1.94$\pm$0.02 with median 1.93.  Means and medians for the
MainDet~sample and the subsamples discussed in this section are listed
in Table~\ref{tuni}. 
    % from means.for   1.94$\pm$0.55 is sigma
The typical (median) error in $\Gamma$, $\Delta\Gamma\sim 0.5$,
is similar to the dispersion 0.54 in best-fit values of $\Gamma$.
If we limit the sample to the 314 sources with $>$100\,counts, the
typical error is 0.32, with no change in mean or median $\Gamma$.
The $\Gamma$ distribution we find is similar to that
found recently for smaller samples of BLAGN. \citet{Just07} studied a
sample of luminous optically-selected quasars observed by {\em
Chandra,\,ROSAT,} and {\em XMM-Newton}, and found
$<$$\Gamma$$>$=1.92$\pm0.09$ for 42 QSOs.  
\citep{Mainieri07} studied a sample of 58 {\em X-ray selected} BLAGN
in the {\em XMM-COSMOS} fields, and found  $<$$\Gamma$$>$=2.09 with a
dispersion of $\sim$0.26. \citet{Page06} found
 $<$$\Gamma$$>$=2.0$\pm0.1$ with a dispersion of $\sim$0.36
for 50 X-ray selected BLAGN in the  13$^H$ {\em XMM-Newton/Chandra}
deep field.

\begin{figure*}
% \plotfiddle{PSFILE}{VSIZE}{ROTANG}{HSCALE}{VSCALE}{HTRANS}{VTRANS}
\plottwo{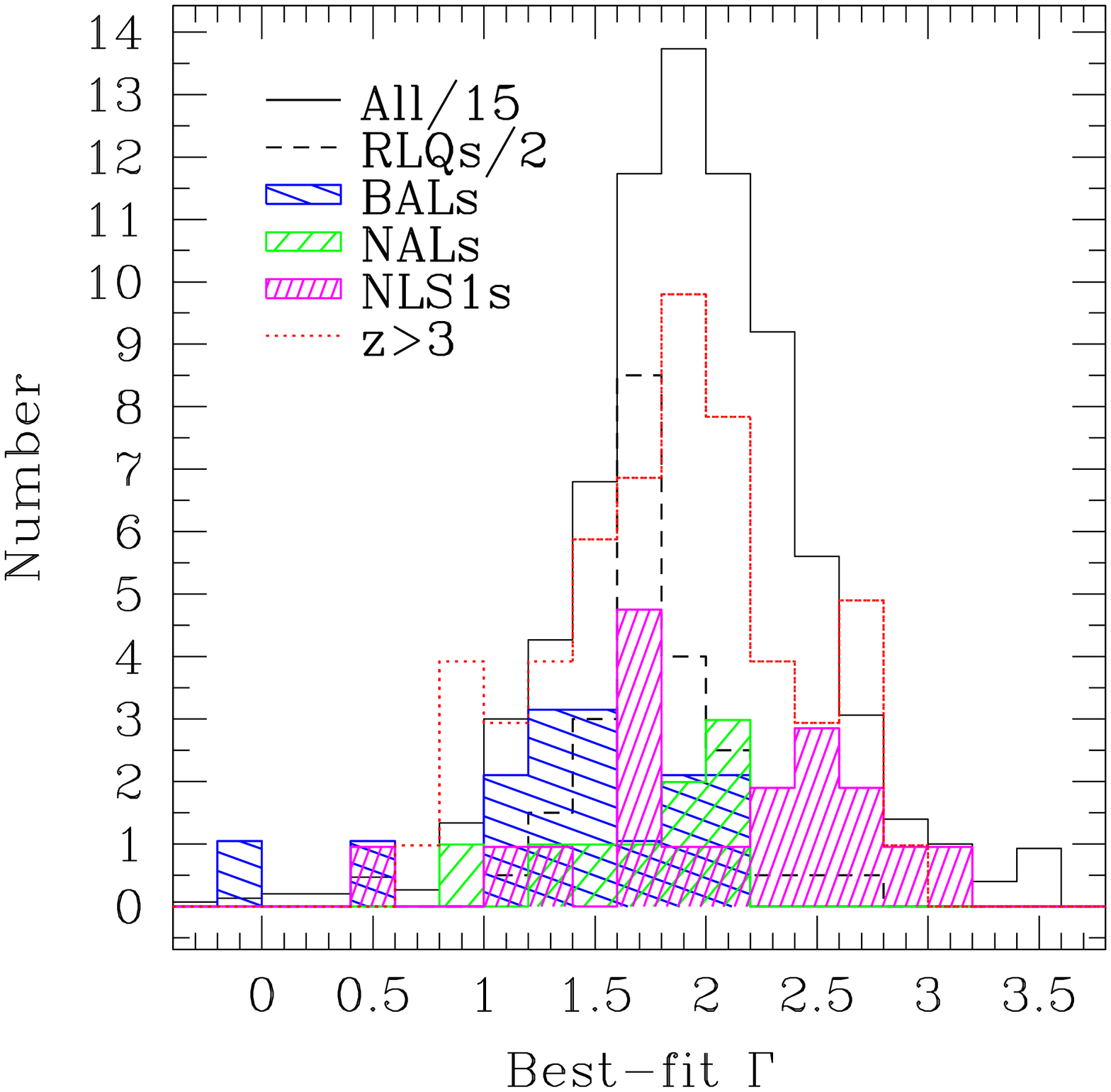}{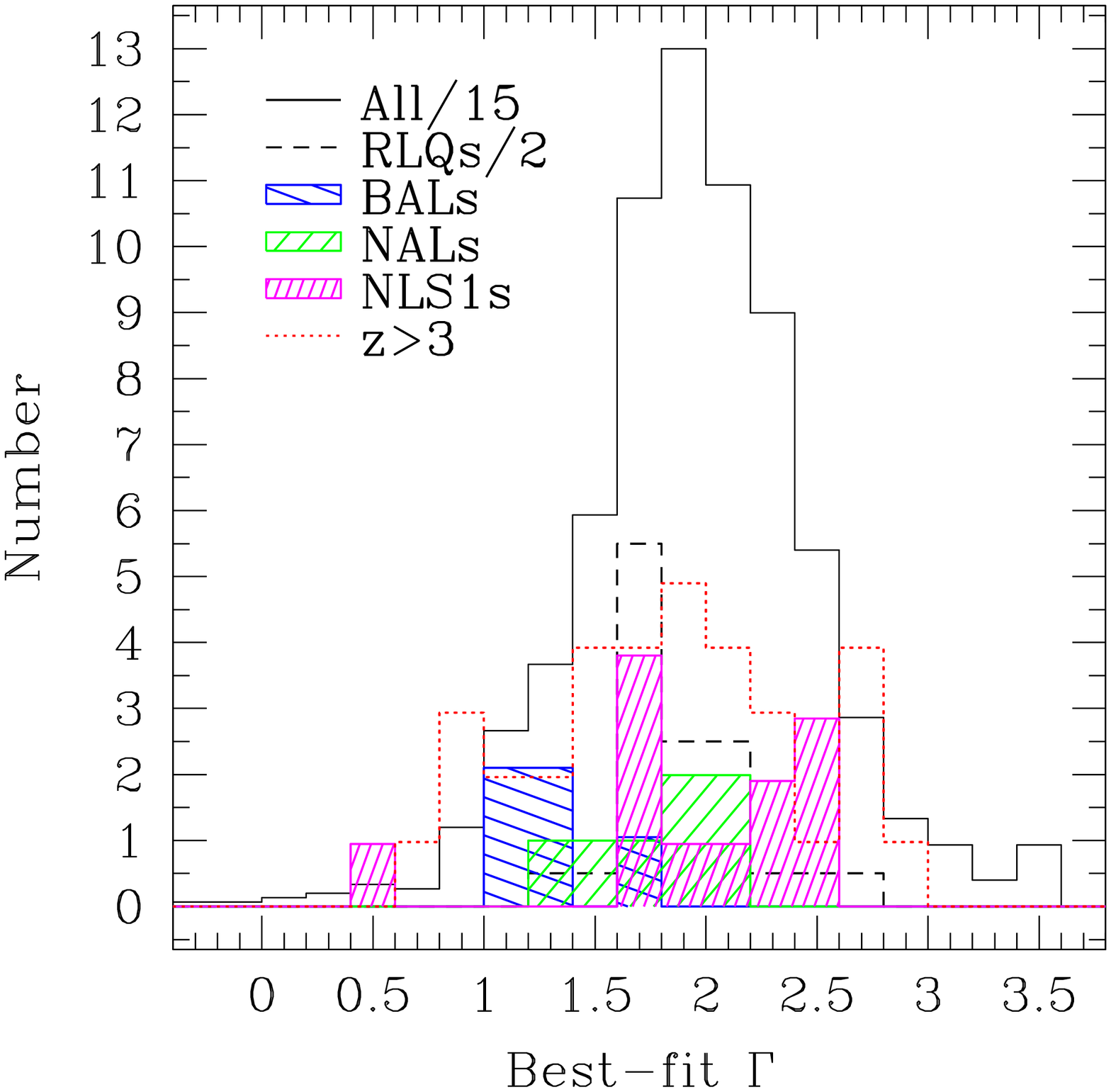}
\caption{LEFT: Histogram of best-fit X-ray spectral power-law
slope $\Gamma$ (best-PL) for the MainDet~sample. The full sample
(black solid) histogram has been divided by 15, and the RL QSO
(long-dashed black) histogram by 2, for ease of comparison with
smaller subsamples. BALQSOs (blue down-slash shading) show very
flat slopes, due to strong intrinsic absorption.  The NAL
distribution (green up-slash shading) suggests a possible bimodality.
Neither the NAL nor the NLS1 (magenta dense shading) nor the
hig redshift (short-dashed red) histograms are significantly different
by K-S  test from the full sample distribution.  RIGHT: Same plot, but
for the noTDet~sample which omits PI targets.  The RL QSO distribution
is less distinct here, and the softest (largest $\Gamma$) NLS1s
disappear. }    
%\vskip-0.2cm
\label{gammahisto}
\end{figure*}

Figure~\ref{gammahisto} shows a histogram of best-fit power-law 
slopes for several interesting subsamples of QSOs from both the 
MainDet~sample and from the noTDet~sample which omits PI targets.
The mean and median values for these subsamples are listed in
Table~\ref{tuni}.  We do not separately plot the radio-quiet (RQ) QSO
sample, since it follows quite closely the shape of the full sample
histogram. 

For 43 detected RLQSOs in the MainDet~sample, the nominal mean slope is 
$<$$\Gamma$$>$=1.73$\pm$0.05 with median 1.65, with a distribution
significantly flatter than for the 704 definitively RQ QSOs 
($<$$\Gamma$$>$=1.91$\pm$0.02) in the
MainDet~sample, using the two-sample tests described in \S~\ref{uni} below.
RL QSOs are known to have flatter high energy continuua from
previous work (e.g., \citealt{Reeves00}). 
   % from means.for $<$$\Gamma$$>$=1.80$\pm$0.36 with median 1.74.

For the 15  known BALQSOs in the MainDet~sample, the nominal mean slope is 
$<$$\Gamma$$>$=1.35$\pm$0.15 with median 1.3, and the distribution 
is significantly different than for the full MainDet~sample (minus
BALs, NALs, and NLS1), using the two-sample tests described in
\S~\ref{uni} below. The difference is even more significant
($P_{max}<0.01\%$) when comparing only to the 667 definitively RQ QSOs
($RL<1.6$). Since the mean (median) number of (0.5-8\,keV) counts for
BALQSOs is just 27 (15), the nominally lower $\Gamma$ likely reflects
undetected absorption \citep{Green01,Gallagher02}. 
    % from means.for $<$$\Gamma$$>$=1.37$\pm$0.61 with median 1.43.

The mean slope $<$$\Gamma$$>$=1.68$\pm$0.13 with median 1.75 for the
9 known NALs in the MainDet~sample is indistinguishable from the full
MainDet~sample (minus BALs, NALs, and NLS1), probably but the
NAL statistics are poor.  On the other 
hand, a smaller, non-overlapping sample of NALs observed by \Chandra\, 
published by \citet{Misawa08} agrees that the X-ray properties of
intrinsic NAL quasars are indistinguishable from those of the
larger quasar population.

For the 19  known NLS1s in the MainDet~sample, the nominal mean slope is 
$<$$\Gamma$$>$=2.01$\pm$0.15 with median 1.95, indistinguishable
from the comparison sample (MainDet~sample minus BALs, NALs, and NLS1).

\subsubsection{X-ray Spectral Evolution}
\label{gammaz}

No signs of evolution have been detected for the intrinsic power-law
slope $\Gamma$ of QSOs:  $z$$>$4 samples
\citep{Vignali05,Shemmer06} show $\Gamma$$\sim$2, just like those at  
lower redshifts \citep{Reeves00}.
In a recent small sample of high (optical) luminosity QSOs
\citep{Just07} also found no trend of $\Gamma$ with redshift.
A larger compilation also shows at best marginal signs of
evolution, and only for well-chosen redshift ranges \citep{Saez08}.

\begin{figure}
% \plotfiddle{PSFILE}{VSIZE}{ROTANG}{HSCALE}{VSCALE}{HTRANS}{VTRANS}
\plotone{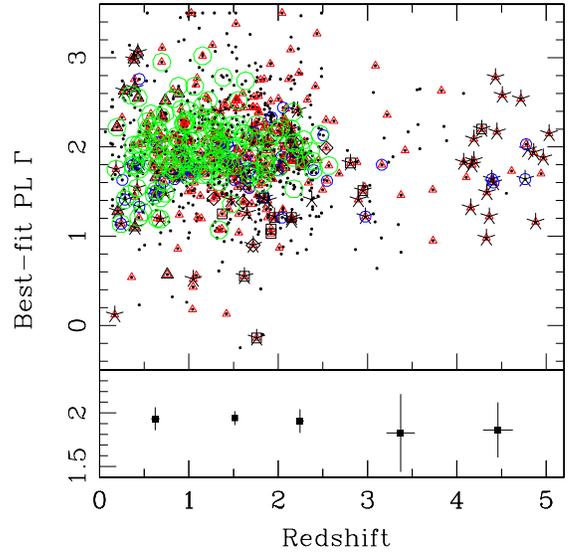}
\caption{TOP: Best-fit X-ray spectral power-law slope $\Gamma$
vs. redshift for the MainDet~sample.  Large open green circles show QSOs
with more than 200 (0.5-8\,keV) counts and simultaneous $\Gamma$-$\nhintr$
fits.  No evidence is seen for evolution in $\Gamma$. 
See Fig~\ref{imagx} for symbol types.  Dots to distinguish 
detections from limits in other plots are omitted here, since
by definition all objects are detected.  Radio-loud QSOs
(open blue circles) tend to have flatter slopes.  Some of the flattest slopes
seen are for BALQSOs (open black squares), due to their strong
intrinsic absorption. Typical mean errors on $\Gamma$ are $\sim 0.5$ 
below $z=2$, rising to 0.8 at the highest redshifts.
BOTTOM: Mean $\Gamma$ values are shown at the mean redshift for QSOs
in 5 redshift bins of width $\Delta\,z=1$ for the
MainDet~sample. Errorbars show the error in the means for both axes. }
%\vskip-0.2cm
\label{gamma_z}
\end{figure}

Fig~\ref{gamma_z} shows $\Gamma$ vs. redshift both for the individual
QSOs (top), and for subsamples (bottom) in fixed bins of $\Delta\,z$.
We also try another binning, where we grow redshift bins from $z=0$
until each bin has between 100 and 150 objects (except for the high
redshift bin $z>$3 which has just 58).  The largest difference between
any of the bins is 0.17, whereas the typical rms dispersion in all
bins is $\sim$0.5.  

The noTDet~sample and the NoRBDet~sample remove contributions from targets, and 
from RL or absorbed QSOs, respectively. the HiCt~sample is the subset of
the MainDet~sample with counts$>$200, where fits for $\nhintr$ are also
performed.  This decreases the effect of QSOs that may  
appear to have hard $\Gamma$ but are actually absorbed.
the HiCtNoTRB~sample also has only counts$>$200, but further removes
targets, RL, BAL, and NAL QSOs.  We find no evidence for
evolution; the null hypothesis (no correlation) cannot be rejected
with $P<8\%$  for any of samples 0, 1, 2, 20, or 21, which test
possible contamination or biases from targets, RL, and/or absorbed
QSOs.

Given that the typical uncertainty in $\Gamma$ for a single QSO
($\sim$0.5) is comparable to the sample dispersion, simultaneous
spectral fitting of subsamples of detected QSOs in bins of redshift (al\`a
\citealt{Green01}) might improve constraints, but such a project is
both beyond the scope of this paper and probably unwarranted given the
scarce evidence for evolution to date.  

\subsubsection{X-ray Spectral Trend with X-ray Luminosity}
\label{gammal}

There is conflicting evidence as to whether $\Gamma$ correlates with
X-ray luminosity $L_X$ \citep{Dai04,Page04}, or perhaps with 
Eddington ratio \citep{Porquet04,Kelly07b,Shemmer08,Kelly08}.  
\citet{Saez08} find significant softening (increase) of $\Gamma$ with
luminosity in selected redshift ranges, for a combination of Type~1
and Type~2 X-ray selected AGN from the CDFs.  Type~2 objects
dominate numerically, especially at low $L_X$, and show larger scatter
in $\Gamma$ at all luminosities, so Type~2 objects dominate the trends
that they observe.  Our sample is complementary in that it treats only
Type~1 QSOs, and provides the largest, most uniform such spectral
study to date.  

%\footnote{Simulations in \citet{Saez08}
%which demonstrate that undetected $\nhintr$ cannot account for the
%trends they see are performed assuming 550 counts, whereas the sample
%objects have a minimum of 170 counts.}

\begin{figure}
% \plotfiddle{PSFILE}{VSIZE}{ROTANG}{HSCALE}{VSCALE}{HTRANS}{VTRANS}
\plotone{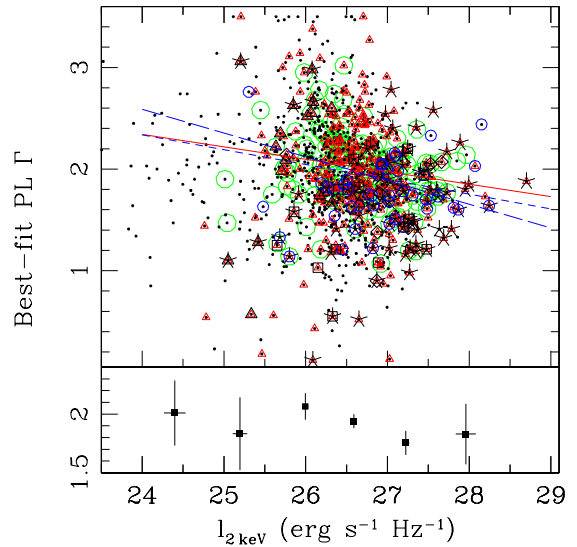}
\caption{TOP: Best-fit X-ray spectral power-law slope $\Gamma$
vs. X-ray luminosity for the MainDet~sample. See Fig~\ref{imagx} for symbol
types. Large open green circles show QSOs
with more than 200 (0.5-8\,keV) counts and simultaneous $\Gamma$-$\nhintr$
fits.  A significant but shallow trend towards harder spectra 
(smaller $\Gamma$) is evident from the best-fit regression lines shown:
the MainDet~sample (red solid), the HiCtNoTRB~sample with
counts$>$200 (blue short-dashed) or HiLoLx~sample with log\,\opteml$>29.8$ and
$\xeml>26$ (blue long-dashed).  Fit parameters are shown in
Table~\ref{treg_gamma}. BOTTOM: Mean $\Gamma$ values are shown at the
mean luminosity for QSOs in 6 bins of width $\Delta\,l_X=0.75$ for 
the MainDet~sample. Errorbars show the error in the means for both axes. }  
%\vskip-0.2cm
\label{gamma_xeml}
\end{figure}

We detect a significant anticorrelation between $\Gamma$ and $\xeml$
across a several subsamples (see Table~\ref{treg_gamma} and
Fig~\ref{gamma_xeml}), but none between $\Gamma$ and either redshift
or \opteml.  The best-fit continuum slope $\Gamma$
hardens (decreases) with increasing luminosity for the MainDet~sample.
Since $\Gamma$ is itself used to calculate \xeml\, via the 
standard X-ray power-law $k$-correction, there might be some danger
that the apparent anticorrelation is induced.  We investigate this in two
ways. First we examine the range in the ratio of \xeml\, calculated
assuming our best-fit $\Gamma$ $k$-correction to that calculated with
a fixed $\Gamma=1.9$.  Across the full X-ray luminosity range, the
different assumptions induce a change of $\sim$0.1\,dex ($\sim25\%$),
insufficient to account for the observed trend.  Second, we examine
whether $\Gamma$ correlates significantly with \opteml, which is
clearly unaffected by the X-ray $k$-correction.  There is indeed a
significant trend in the MainDet~sample  (see Table~\ref{treg_gamma}).

One possibility is that the observed trend of $\Gamma$ with
luminosity might be due to an undetected increase in $\nhintr$.
To test this, we examined the relationship in the HiCtNoTRB~sample,
the subset of the MainDet~sample with counts$>$200, where fits for
$\nhintr$ are also performed. We do not claim that all absorption is
detected in the HiCtNoTRB~sample.  However, if indeed undetected
absorption caused the trend in the MainDet~sample, we would expect the
anticorrelation to weaken or flatten in the HiCtNoTRB~sample.
Instead, the anticorrelation is still significant, and the slope is
actually steeper (see Table~\ref{treg_gamma}).

It is also conceivable that soft X-ray emission associated with
starformation activity might contaminate the sample at
low luminosities.  To test this, we create the HiLoLx~sample with
definitively QSO-like luminosities in both wavebands, i.e.,
log\,\opteml$>29.8$ and $\xeml>26$. For this sample, the anticorrelation
remains strong and and steep (see Table~\ref{treg_gamma}). 

We speculate that the balance between thermal (accretion disk) and
non-thermal X-ray emission may shift towards higher luminosities.
If at high \xeml, non-thermal emission represents a larger
fraction of the emitted X-rays, we would expect a hardening (decrease) 
of $\Gamma$ with luminosity, as is seen.  We would also
expect to find more RL QSOs at high \xeml, and we do.
Since quasars such as RL QSOs with a larger fraction of non-thermal
emission also show stronger X-ray emission relative to optical
(smaller \aox), we might expect a correlation between
$\Gamma$ and \aox\, such that as the spectrum hardens, \aox\, 
decreases.  This is shown to be the case in \S~\ref{aox_gamma}
below.  The idea of a significant non-thermal, probably jet-related
emission component even in RQ QSOs is not new.  \citet{Blundell98}
found strong evidence from Very Long Baseline Array (VLBA)
observations  for jet-producing central engines in 8 of the 12 RQ QSOs 
in their sample.  \citet{Barvainis05} find similarities between
RL and RQ quasars spanning a host of radio indicators:
variability, radio spectral index, and VLBI-scale components, 
suggesting that the physics of radio emission in the inner regions of
all quasars is essentially the same, involving a compact, partially
opaque core together with a beamed jet. \citet{Czerny08} similarly
find evidence for a blazar component in RQ QSOs by modeling
their variability.

More sensitive empirical tests of whether the observed trend is due to
substantial (but not directly detectable) absorption depressing the
observed soft X-ray continuum at high luminosities, or to an
increasing thermal fraction at lower luminosities could be performed
by stacking counts in narrow energy bands from \Chandra\, images of
all QSOs (detected or not).  Stacking has been used effectively this
way in the CDFs (e.g., \citealt{Steffen07,Lehmer07}), but the task is
more daunting for the ChaMP, where care must be taken to properly
account for the effects of a much wider range of $\nhgal$, CCD quantum
efficiency values, and exposure times.

A study of 35 Type~1 QSOs by \citet{Shemmer08} finds that $\Gamma$
increases (softens) with $L/L_{Edd}$, the latter derived from
FWHM(H$\beta$) and $\nu L_{\nu}$(5100\AA).  \citet{Kelly08} examine 
a larger sample, and find complicated relationships
between $M_{BH}, L_{UV}/L_{Edd},$ and $L_X/L_{Edd}$ that
change direction depending on the emission line used to estimate
$M_{BH}$.  In a subsequent paper, we are extending our 
current analysis to include estimates for $M_{BH}$ and  $L/L_{Edd}$
for the spectroscopic subsample of the current paper.

\subsection{X-ray Intrinsic Absorption Measurements}
\label{nh}

\begin{figure}
\plotone{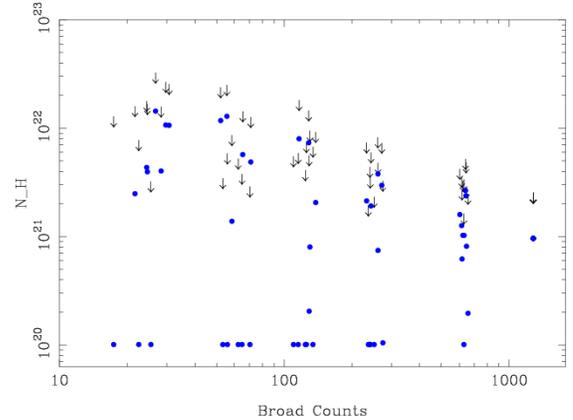}
\caption{
Best-fit log intrinsic absorption vs. (0.5-8\,keV) counts
for 10 random exposure-time subsamples scaled to achieve
2\%, 5\%, 10\%, 20\%, and 50\%  of the full original exposure 
time (100\%, at the far right) for one bright X-ray source
observed by \Chandra.  Best-fit values of \nhintr\, are shown as blue
dots, with the corresponding 90\% upper confidence limit shown as
black arrows. This plot shows a clear trend, reflecting the skewed
(one-sided positive definite and logarithmic) nature of the $\nhintr$
parameter. } 
%\vskip-0.2cm
\label{cts_nhsim}
\end{figure}

The quality of X-ray measurements of intrinsic absorbing columns 
depends strongly on the number of counts in the quasar,
but also on the redshift of the object, as illustrated
in a simple test in Fig~\ref{cts_nhsim}.  

\begin{figure}
% \plotfiddle{PSFILE}{VSIZE}{ROTANG}{HSCALE}{VSCALE}{HTRANS}{VTRANS}
%\plotfiddle{comparespec.eps}{3in}{0.}{400}{400}{-260}{30} 
%\epsscale{1.9}
\plottwo{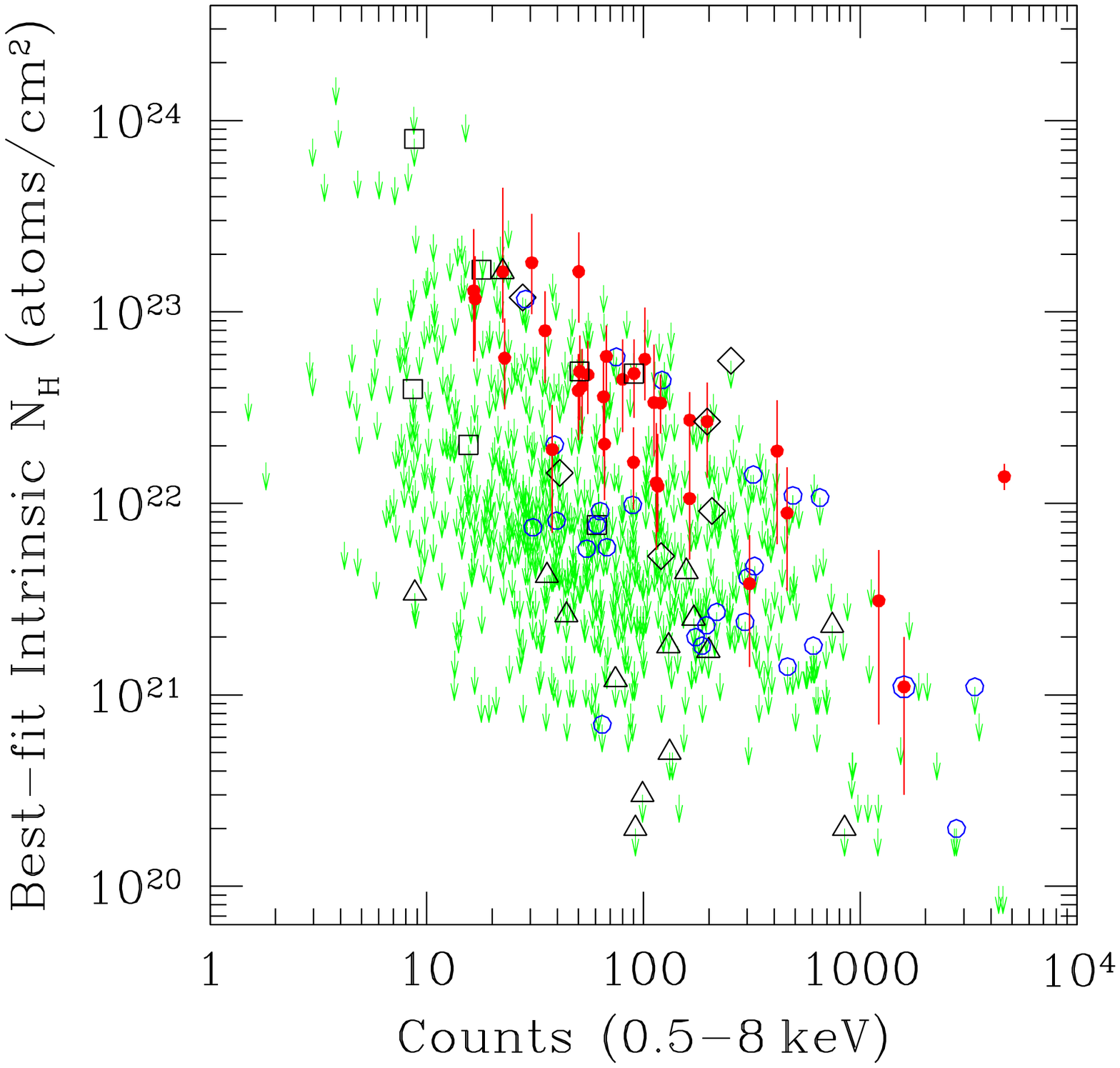}{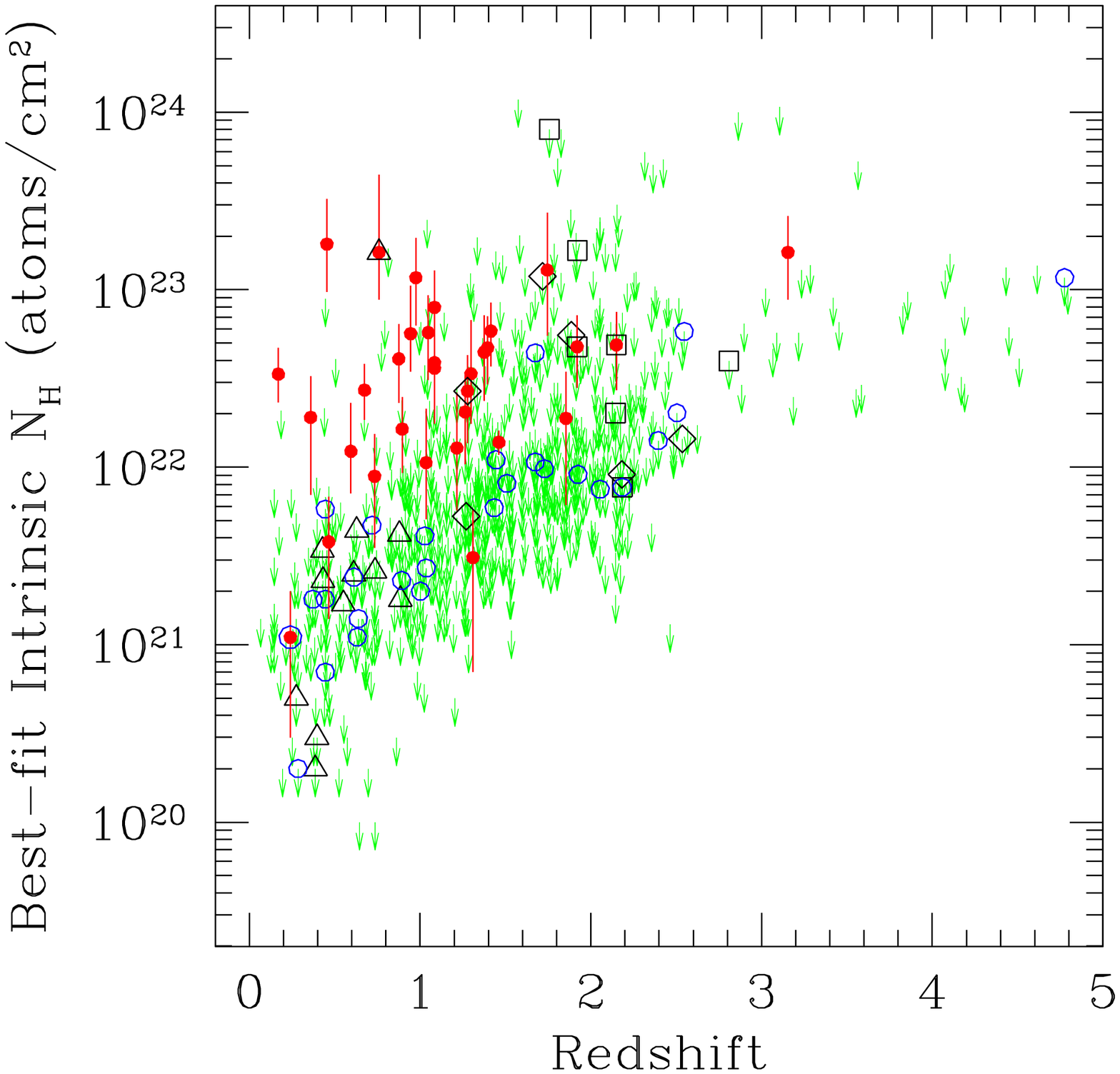}
\caption{
LEFT: Best-fit intrinsic absorption vs. (0.5-8\,keV) counts for
detected QSOs.   Object type symbols are as described in
Fig~\ref{imagx}. Green arrows indicate upper limits, while detections
(where the 90\% lower limit exceeds log\,\nhintr =20) are red dots 
with 90\% errorbars.  The strong decrease in the upper limit envelope
traces the increased sensitivity as a function of detected counts.
Only a handful of objects show significant absorption
above 200 counts where simultaneous fits to $\Gamma$ and \nhintr\, are
performed.  
RIGHT: Best-fit intrinsic absorption vs. redshift.
The strong increase in the upper limit envelope reflects both
the decrease in counts and the decreased sensitivity as a function of
redshift, due to the restframe absorbed region below $\sim$2\,keV
redshifting out of the \Chandra\, bandpass.} 
\vskip-0.2cm
\label{nhfig}
\end{figure}

Fig~\ref{nhfig} shows our best-fit intrinsic absorption column
measurements, which are overwhelmingly upper limits, either when
assuming (as for counts$<$200) that $\Gamma=1.9$ or when fitting
both $\Gamma$ and \nhintr\, (for QSOs with $>$200 counts).
Just one of the RL QSOs has detectable \nhintr, which is nevertheless
not large (10$^{21}$\cmsq). Jet-related X-ray emission may reduce
any absorption signatures.

Strong intrinsic absorption is relatively rare in radio-quiet Type~1
QSOs.  In the strictest interpretation of AGN unification models, none
of these broad line AGN should be significantly X-ray-absorbed.  
Small obscured fractions might be expected by selection, which
requires that the broad line region (BLR) not be heavily
dust-reddened, i.e. our view of the BLR is unobscured. 
X-ray absorbed BLAGN have therefore sometimes been called ``anomalous''.
The obscured fraction of BLAGN from the literature spans a wide
variety of samples and analysis methods, but most define
``obscured'' as \nhintr$>10^{22}$\cmsq and find fractions of
about 10\% or less \citep{Perola04, Page06, Mainieri07}.
By contrast, selecting optically unobscured AGN only from optical/IR 
photometry, and using X-ray hardness ratios from XMM, \citet{Tajer07}
find 30\%.

To estimate the obscured fraction of SDSS QSOs from the ChaMP,
we first limit to the D2L~sample and $z<2$ to maximize the fraction 
that are X-ray detected (see Fig~\ref{hidethistos}).  We further limit to
QSO-like optical luminosities log\,\opteml$>29.8$, yieleding 630 QSOs,
for which 498 (79\%) are X-ray detections.  If we rather stringently
require that the 90\% lower bound to the best-fit log\,$\nhintr$
exceeds 22, then 50 of the 498 (10\%) of X-ray detected QSOs are
obscured.  If instead we simply use the best-fit $\nhintr$, 98 of 498
(20\%) of detected QSOs are obscured.  However, if we make the
unlikely assumption that all X-ray-undetected QSOs are missed due to
heavy intrinsic absorption, the obscured fraction could be as high as
29\% and 36\% for these two different methods, respectively.   

We note that the correlation of dust to total column of X-ray absorbing
metals is not strict.  For instance, we know that a significant
fraction of optically selected QSOs (13 -- 20\%; \citealt{Reichard03,
Hewett03, Knigge08})  appear as BALQSOs, which are thought to be seen along a
sightline piercing warm (ionized) absorbers.  BALQSOs are highly
absorbed in X-rays (Green et al. 1995, 1996, 2001, Gallagher et
al. 2002).  All the BALQSOs here have \nhintr\, detections or upper limits
that are $>3\times 10^{22}$\cmsq.  An even larger fraction may harbor
undetected warm absorbing material that is too smooth in its velocity
and/or geometric distribution to show distinct absorbing troughs
\citep{Gierlinski04,Green06}. The frequency of warm X-ray absorbers
has been shown to be about 50\% \citep{Porquet04} in low-redshift (PG)
QSOs. Again, the bias of (color- plus broad-line-based) optical
selection decreases their number here.  The fraction of
detectably-large $\nhintr$ in the fully X-ray-selected ChaMP is
larger, but nearly all such examples are found in objects which appear
optically as narrow-line AGN or absorption line galaxies (XBONGs), as
shown by \citet{Silverman05a}. 

\section{X-ray to Optical SEDs}
\label{seds}

Beyond the desire to know of real trends in the SEDs of QSOs for the
sake of understanding accretion physics, such trends influence other
scientifically important research.  Corrections to derive bolometric
luminosities ($L_{bol}$) from measured $L_X$ depend strongly on SED
trends.  $L_{bol}$ in turn is key to constraining such fundamental
parameters as accretion efficiency and/or SMBH densities in the
Universe (e.g., \citealt{Marconi04}), or active accretion lifetimes or
duty cycles (e.g., \citealt{Hopkins05, Adelberger05}).  As another
example, to compare the AGN space densities found via optical
vs. X-ray selection, \citet{Silverman05b,Silverman08} and others use
the measured trend in \aox\, to convert the X-ray luminosity function
(LF) to an optical LF.

Numerous studies have debated the strength and origin of trends in the
X-ray-to-optical luminosity ratio of optically selected quasars.  This
ratio is herein (and typically) characterized by the spectral slope \aox.
Most studies include large samples mixing both targeted and
serendipitous X-ray observations (e.g.,
\citealt{Avni82,Wilkes94,Vignali03,Strateva05,Steffen06,Kelly07a}).  
All these studies, using widely varying techniques and sample
compilations, have found a strong correlation between \aox\, and
optical luminosity, and several have also suggested a residual
evolution of \aox\, with redshift or lookback time, even after the
luminosity correlation is accounted for.

While large sample sizes and appropriate statistical techniques 
(see \S~\ref{stats}) have been sought in these studies, the impact of
sample selection effects combined with large intrinsic sample
dispersions in luminosity 
may still dominate the observed correlations.  These effects have been
modeled and simulated by \citet{Yuan98} and \citet{Tang07} among
others, who retrieve seemingly statistically significant but
artificially-induced correlations.  They highlight the need for large,
well-defined samples to alleviate this problem, and they also note the
importance of minimizing the effects of a strong $L-z$ correlation
induced by a single sample flux limit. The ChaMP/SDSS sample is a
large step towards alleviating these problems.

Contamination by unrelated physical processes should be eliminated
wherever feasible. As mentioned above, it is well-known that RL QSOs
tend to be X-ray 
bright, and there is evidence that a distinct,  jet-related physical
process produces that extra X-ray emission.  BALQSOs, on the other hand, 
tend to be X-ray weak, because of intrinsic absorption from the
warm (ionized) winds \citep{Green01,Gallagher02}.   When investigating
distributions  and correlations amongst \aox, \lo, and \lx, most previous
authors have chosen to eliminate both BAL and RL QSO, to 
minimize contamination of (presumably) pure accretion-dominated
X-ray emission.  It is important to note that this precaution is
neither complete, nor perhaps correct.  For example, the intrinsic
radio-loudness distribution of quasars does appear to be well-modeled
by a quasi-normal distribution with a 5\% tail of RL objects
\citep{Cirasuolo03}, but the relationship of that distribution
to X-ray emission is not well-characterized. Even after removal of
quasars with a radio/optical flux ratio above some limit,
radio-related X-ray emission may pervade quasar samples, and affect
the distributions we study.  Similarly, the best quantitative measures of
absorption profiles via e.g., `balnicity' \citep{Weymann91}
indicate that the fraction of quasars with intrinsic
outflows may be significantly underestimated \citep{Reichard03}, with
classic BALQSOs just the tip of the iceberg.  Equally important may be 
that BALs to date are mainly detected only beyond $z\sim1.4$ in
ground-based optical spectroscopic quasar samples ($z\sim1.5$ in SDSS 
spectra), so the low redshift, low luminosity end of the quasar  
distribution may harbor significant undetected warm absorption.
We therefore examine a variety of subsamples within and without these
classes.  When we compare bivariate regression results 
\opteml, \xeml, and \aox, directly to Steffen et al. (2006; S06
hereafter), we follow their convention of excluding known RL QSOs and
BAL QSOs, and expand somewhat to also eliminate QSOs with evident
NALs, and NLS1s as well.   We note that their \aox\, definition is the
negative of the more conventional one we adopt here. 

\subsection{Statistical Tools and Methods}
\label{stats}

 Many previous studies have emphasized the importance of
including X-ray upper limits in the samples, which require appropriate
statistical treatment using e.g., survival analysis.  
Results from survival analysis can depend
strongly on the number of detections, and therefore on the detection
threshold, since in flux-limited samples, most detections are near
that threshold.   The \Chandra\, fields used here span a wide range of
effective exposure times, substantially alleviating the powerful
effects that a single survey flux limit can have (e.g., by creating a 
very tight $L_x-z$ correlation).  Previous studies
have managed to limit the fraction of upper limits by using mostly targeted
X-ray observations (e.g., \citealt{Wilkes94}), where exposure times are
often chosen based on optical flux, or by judicious choice of
subsamples (e.g., \citealt{Strateva05,Steffen06}) designed to fill the
luminosity-redshift plane.  For example, \citet{Strateva05} uses SDSS
spectroscopic AGN from the DR2, observed by ROSAT PSPC for more than
11\,ksec.  For this purpose, the DR2 AGN sample is already biased,
since many spectroscopic targets are chosen as X-ray ROSAT or FIRST
radio detections. Other AGNs they include to fill in
the $L-z$ plane were specifically targeted for \Chandra\, or \XMM\,
observations for a variety of reasons. A slightly larger study by
\citet{Steffen06} added objects using mostly photometric AGN
classifications and redshifts from COMBO-17, together with the
Extended \Chandra\, Deep Field (CDF) South (ECDF-S), but also several
other small bright samples for the low $L-z$ region.  Even after
selecting and combining various samples with high detection fractions,
the use of Survival Analysis techniques to incorporate limits must
sometimes be abandoned.\footnote{For example, \citet{Steffen06} 
simply drop the X-ray limits when regressing with \xeml\, as the
independent variable.  We choose instead to incorporate the limits
using the two-dimensional Kaplan-Meier (2KM) test \citep{Schmitt85}.}
In summary, if significant selection biases may affect either the
constituent subsamples or their ensemble, neither the inclusion of
upper limits, nor the use of complex statistical analysis methods
should convey the impression that statistical results are as robust as
those from a uniform, complete, and well-characterized sample.     

It is also worthwhile to consider the fact that some previous
studies have appended low-luminosity subsamples (Seyferts~1s).
Our SDSS sample excludes some of these objects because they
would be spatially resolved. This may bias our sample in the
sense that for low-redshift AGN (e.g., 107 objects in the MainDet~sample with 
$z\leq0.55$) we include only those that are compact (optical
light distribution consistent with the expected SDSS PSF).  Most 
previous studies have instead attempted to include such objects, 
but then they correct for the inclusion of substantial host galaxy
emission via e.g., spectroscopic template fitting \citep{Strateva05}.
This may bias those samples in a different way by excluding the host
contribution only for nearby objects.

\subsubsection{Univariate Analyses}
\label{uni}

To compare two sample distributions, we test the `null' hypothesis
that two independent random samples subject to censoring (e.g.,
$\Gamma$ for RL and RQ QSO) are randomly drawn from the same parent
population. The programs for two sample tests that we use are the
Gehan, or generalized Wilcoxon test, and the logrank test in ASURV.
Again, we require $P_{max}$$<$0.05 to call the distinction
significant. For  $P_{max}$$<$0.10, we call the difference
``marginal''.  The Kaplan-Meier method we employ to estimate the mean
of a distribution allows for the inclusion of censored values (upper
or lower limits).

\subsubsection{Bivariate Regressions}
\label{biv}

Except where noted, all the correlations studied between X-ray and
optical luminosity, and between \aox\, and \lo\, are highly
significant: $P<10^{-4}$ by Cox Proportional Hazard, Kendall's $\tau$
or Spearman's $\rho$ tests, as implemented in the ASURV (Survival
Analysis for Astronomy) package \citep{LaValley92}.  We deem a
correlation significant if the maximum probability $P_{max}$ from all
three tests is 0.05 or less.  We perform bivariate linear regressions
using the two-dimensional Kaplan-Meier (2KM) test \citep{Schmitt85} as
implemented in ASURV, which permits linear regression with limits in
either axis.\footnote{Our results from the other 2 bivariate
regression algorithms in ASURV (the Buckley-James method and the
parametric EM Algorithm) are quite consistent.}  We use 20 bootstrap
iterations for error analysis, (more than sufficient given the large
sample size here) and 20 bins in each axis, with origins 27.3 (23.0,
1.0) for \opteml\, (\xeml, \aox), except where samples have been
explicitly restricted in luminosity, or by object type to have
$N<200$, whereupon we use 10 bins.

For bivariate regressions between X-ray and optical luminosity,
we make no assumptions about which luminosity constitutes the
dependent or independent variable, and calculate the mean 
(bisector) of the ordinary least squares (OLS) regression lines
which minimize residuals for $Y(X)$, $X(Y)$, respectively.\footnote{In
the absence of limits, these results reduce reliably to  the bisectors
found by the SLOPES program \citep{Feigelson92,Babu92,Isobe90}.}
While the intrinsic dependence of \aox(\lo) or \aox(\lx) is unknown,
most studies assume \aox\, to be the dependent variable, and quote
slopes  accordingly, so we will follow suit largely for purposes of
comparison.   We further caution that in samples with large
dispersions, different regression methods can yield different results
(see a recent mathematical review of related issues by
\citealt{Kelly07b}).  

For completeness, we provide statistical results in table format
for a variety of subsamples which may be of interest, even beyond
those discussed in the text.  For ease of reference, tables listing
bivariate statistical results  (Tables~\ref{treg_gamma} --
\ref{treg_oaox}) list samples in the same order as the table defining
samples (Table~\ref{tsamples}. 

\subsection{\aox\, for QSO Subsamples}
\label{aox}

Here we use the D2L~sample, with high detection fraction, to examine
the \aox\, distributions of several subsamples, including targets.
The mean \aox\, for 1269 QSOs in the D2L~sample is 1.429$\pm$0.005 with
median 1.370.  Means and medians for the D2L~sample and for subsamples 
discussed in this section are listed in Table~\ref{tuni}, as are
result for two-sample tests.

For 31  RLQSOs in the D2L~sample, the mean \aox=1.377$\pm$0.028 with
median 1.392.  Using the two-sample tests described in \S~\ref{uni}
above, this distribution is only marginally different than for the
1238 non-RLQSOs in the D2L~sample.  RL QSOs are thought to be more
X-ray loud than RQ QSOs.

For the 23  known BALQSOs in the D2L~sample, the  mean \aox\, is 
1.717$\pm$0.028 with median 1.66, and the distribution 
is significantly different than for the non-BAL the D2L~sample.
The apparent X-ray weakness has been shown to be consistent
with intrinsic absorption of a normal underlying X-ray continuum
\citep{Green01,Gallagher02}. 

For the 8  known NALs in the D2L~sample, the mean \aox is 
1.463$\pm$0.056 with median 1.5, but the poor statistics render the
distribution indistinguishable from the overall D2L~sample.    

For the 19  known NLS1s in the D2L~sample, the mean \aox is 
1.54$\pm$0.08 with median 1.43, again indistinguishable
from the overall D2L~sample.

\subsection{X-ray luminosity \xeml\, vs. optical \opteml}
\label{optx}

\begin{figure}
% \plotfiddle{PSFILE}{VSIZE}{ROTANG}{HSCALE}{VSCALE}{HTRANS}{VTRANS}
%\plotfiddle{opteml_xeml0.eps}{3in}{0.}{400}{400}{60}{-260} 
%\epsscale{1.2}
\plotone{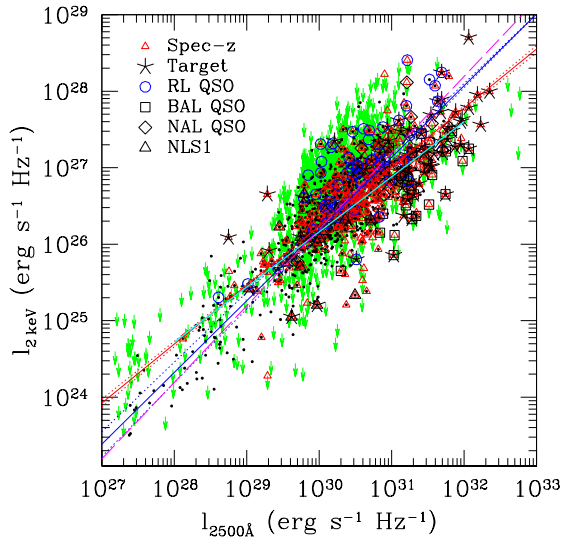}
\caption{X-ray (2\,keV) vs. optical (2500\AA) luminosity
for the the Main~sample. See Fig~\ref{imagx} for symbol types.
The long-dashed magenta line is of slope unity, normalized
to the sample means. The flattest fit is the best-fit (OLS bisector)
relation from \citet{Steffen06} (their Eq.\,1c), shown as a solid
cyan line, spanning the luminosity range of their compilation.  The
best-fit OLS bisector regression line to the SDSS/ChaMP the Main~sample
(including non-detections) is shown in blue, spanning the full
plot. Short-dashed lines plotted using  1-$\sigma$ statistical errors
to the ChaMP fits are so close to the best-fit as to be barely
discernible on the plot.  The red line is a $Y(X)$ regression on the
same data, illustrating the sensitivity of fits to regression method
in high dispersion data. }     
%\vskip-0.2cm
\label{fopteml_xeml}
\end{figure}

Fig~\ref{fopteml_xeml} shows a highly significant correlation
of \xeml\, with \opteml, and plots our best-fit regression lines.
The bisector regression relationship for the SDSS/ChaMP 
sample (the Main~sample,  limits included, 2308  QSOs), is  
$${\mathrm log}(l_{\mathrm 2\,keV}) = 
	(1.117\pm 0.017)\,{\mathrm log}(l_{\mathrm 2500\AA})
	- (7.59\pm 0.64) $$

\begin{figure}[t!]
% \plotfiddle{PSFILE}{VSIZE}{ROTANG}{HSCALE}{VSCALE}{HTRANS}{VTRANS}
%\epsscale{1.2}
\plotone{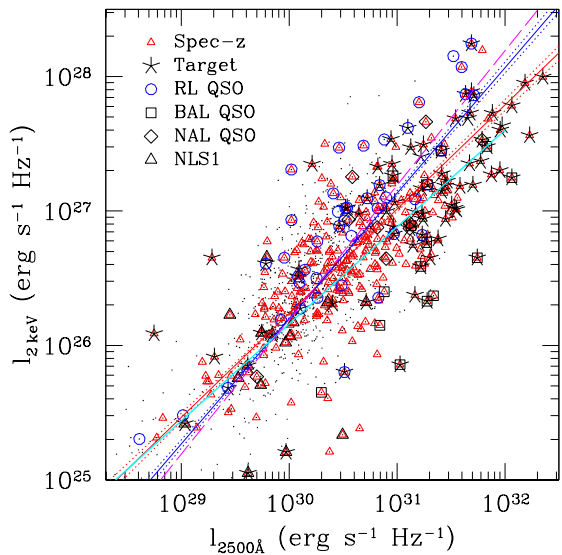}
\caption{Zoom-in of the X-ray (2\,keV) vs. optical (2500\AA)
luminosity plot for the MainDet~sample.  See Fig~\ref{imagx} for symbol types.
Here we show a smaller luminosity
range, and omit limits to highlight the classes indicated in the key.
RL QSOs clearly populate the upper end of the distribution in X-ray
luminosity, while BALQSOs are underluminous in X-rays.  The fits shown here 
are performed on the MainDet~sample with detections only.
Line types are the same as in Fig~\ref{fopteml_xeml}.} 
%\vskip-0.2cm
\label{fopteml_xemlzoom}
\end{figure}

\noindent This slope is close to linear, and significantly different
than the bisector slope $\beta=0.72\pm0.01$ derived by Steffen et
al. (2006; S06 hereafter) from their smaller, more diverse sample.
The subsample philosophically closest to that of S06 is the NoRB~sample,
or its high detection fraction version the D2LNoRB~sample, since they exclude
known RL and BAL QSOs (as well as NALs and NLS2s), 
but include targets.  The D2LNoRB~sample bisector slope is $\beta=1.115\pm
0.015$.  Our bisector slope results in Table~\ref{treg_ox} are closer
to linear than S06 for all samples tested across the full luminosity
range, including those that omit limits
altogether. Fig~\ref{fopteml_xemlzoom} shows the MainDet~sample
(detections only) across a smaller luminosity range, to highlight the
different QSO types.  

A nearly linear result was also found by \citet{Hasinger05}
for a sample of Type~1 QSOs that spanned a similarly large range of
luminosities as our own.  That work used an {\em X-ray selected
sample} with a high ($\sim$95\%) completeness for spectroscopic
identifications, and concluded that the non-linear trends seen
in optically selected samples probably result from selection 
effects.  

\subsection{\aox\, vs. Optical Luminosity}
\label{aox_lo}

\begin{figure*}
% \plotfiddle{PSFILE}{VSIZE}{ROTANG}{HSCALE}{VSCALE}{HTRANS}{VTRANS}
\plottwo{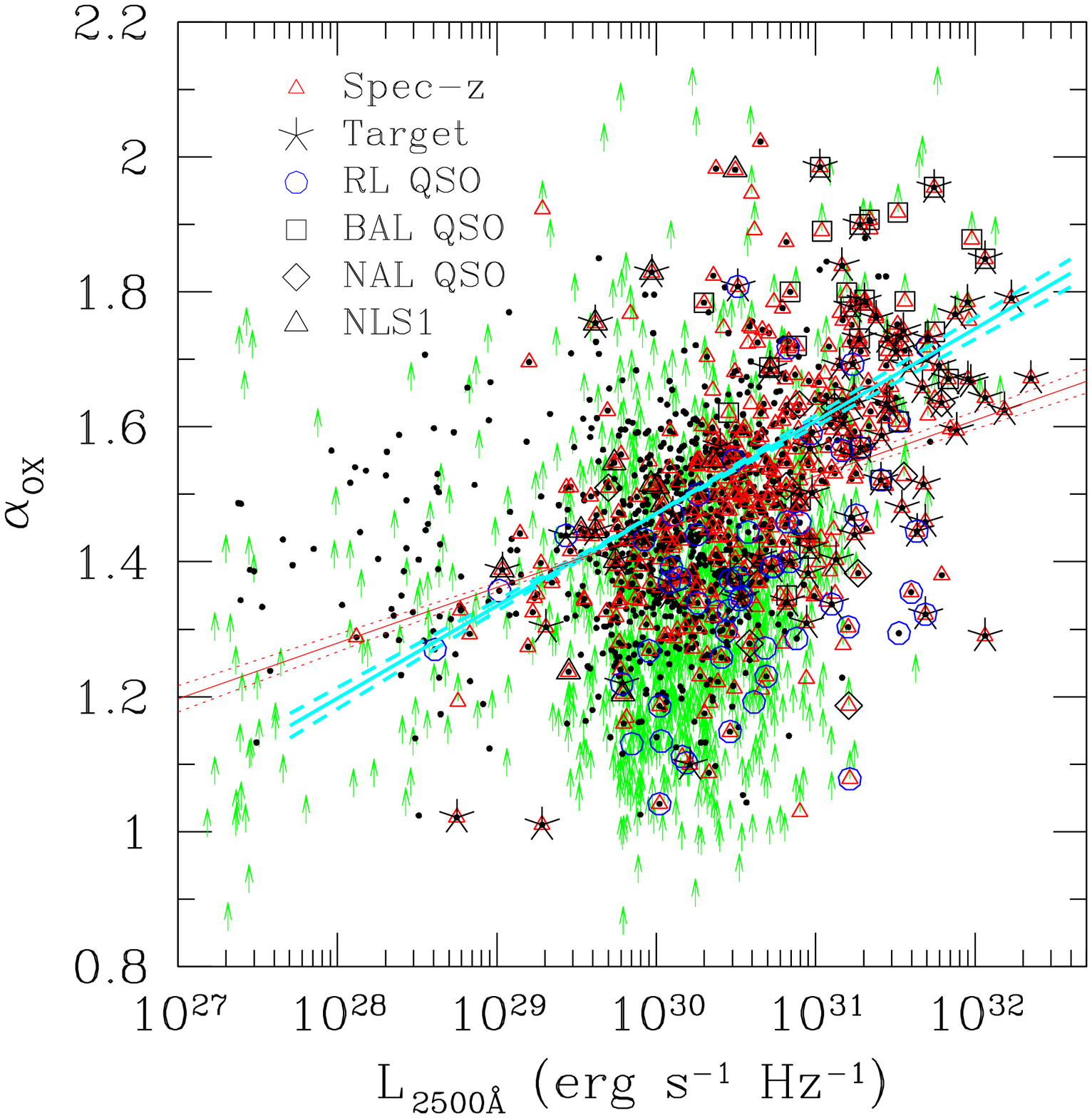}{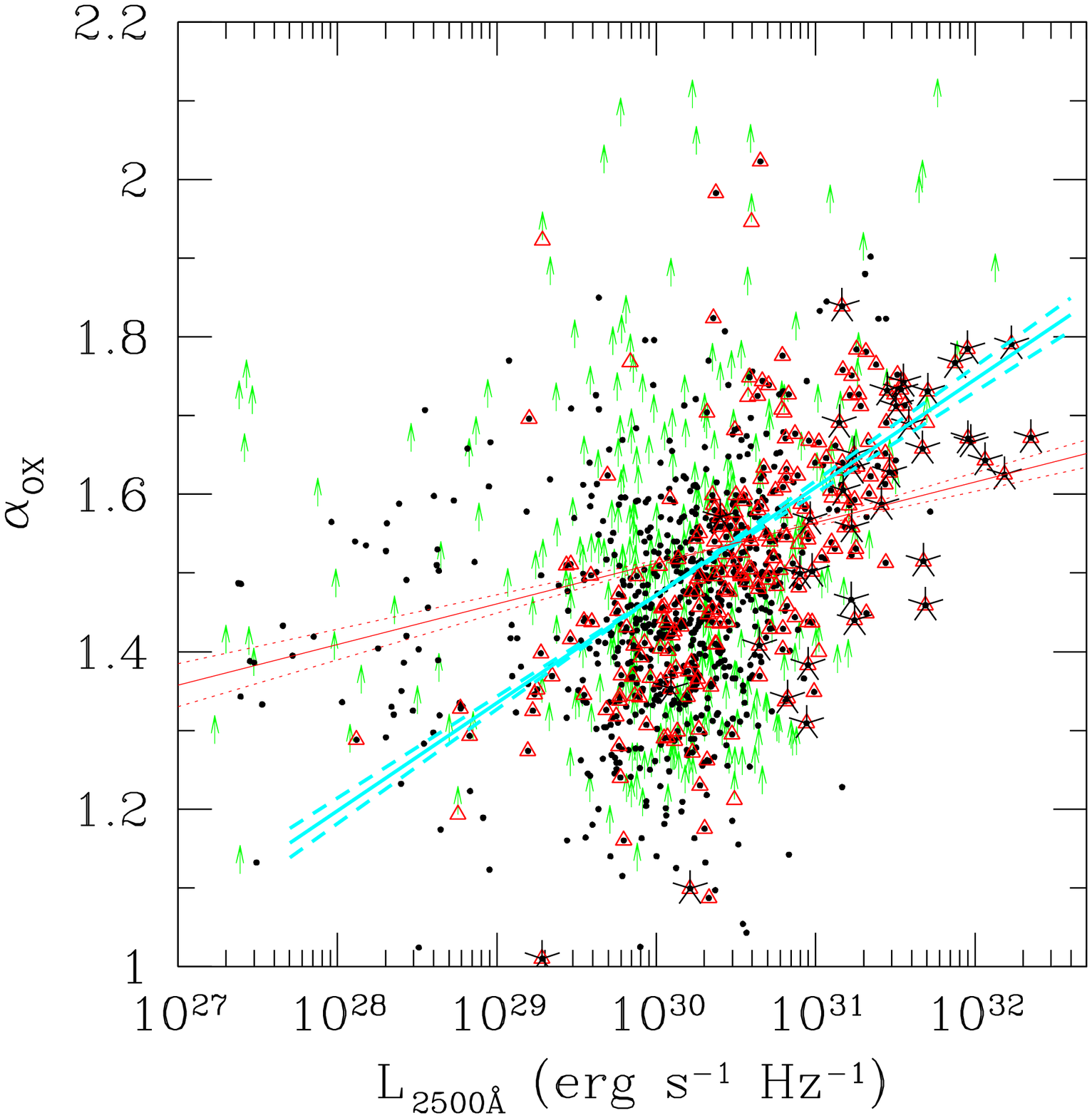}
\caption{\aox\, vs. optical 2500\AA\, log luminosity
for the Main~sample (LEFT) and the D2LNoRB~sample (RIGHT).  See
Fig~\ref{imagx} for symbol types.  The best-fit OLS $Y(X)$
regression for each ChaMP sample is shown as a red line with errors.
The best-fit relation from \citet{Steffen06} is shown as a solid cyan
line.}    
%\vskip-0.2cm
\label{opteml_aox}
\end{figure*}

Fig~\ref{opteml_aox} (left) shows the trend of \aox\,
with \opteml\, for the Main~sample.  While the detected QSOs appear as a
large `blob', this is due to a combination of the SDSS depth,
the QSO optical luminosity function, and the highly efficient QSO
selection in the corresponding redshift range 1$<$$z$$<$2.5 (see
Fig~\ref{rawdethistos}).  There is a highly 
significant correlation: $P<10^{-4}$ by Cox Proportional Hazard,
Kendall's $\tau$ or Spearman's $\rho$ for any of the samples
considered (with or without limits). Our best-fit regression line
minimizing residuals only in \aox\,  (e.g., a $Y(X)$ regression) for
the Main~sample is  

$$\aox\, = 
	(0.061\pm 0.009)\,{\mathrm log}(l_{\mathrm 2500\AA})
	- (0.319\pm 0.258) $$

\noindent This is significantly flatter than the results of S06
($0.137\pm0.008$), consistent with the more closely linear
relationship we find above between X-ray and optical luminosity.
As can be seen from Fig~\ref{opteml_aox} (bottom) and
Table~\ref{treg_oaox}, keeping targets but removing RL and BAL QSOs
(the D2LNoRB~sample) yields similarly flat slopes. 

The classical Seyfert~1/QSO dividing line at $M_{B}\sim-23$
corresponds here to \\
${\mathrm log}\,(l_{\mathrm 2500\AA})\sim$29.8
(or log\,($\nu_{\mathrm 2500}l_{\mathrm 2500\AA})\sim$44.9). While this  
conventional partition is essentially arbitrary, it does represent a
sharp discontinuity in the luminosity histogram of the current sample:
six times as many objects have ``QSO-like'' optical luminosity
as ``Seyfert-like''.  The hiLo~sample, restricted to luminous QSOs as above,
yields a slope for the \aox(\opteml) relation of $0.128\pm 0.007$, most
similar to S06 and previous work. The SDSS/ChaMP sample does
boast a larger number of low optical luminosity objects than most
previous analyses, in large part because of the sensitivity of the
\Chandra\, observations, and these lower luminosity objects may be
responsible in part for our different results across the full
\opteml\, range.  Analysis of the (much smaller, with $N_{det}$=176
and $N_{tot}$=260) the D2LSy1~sample with Seyfert-like luminosities, 
suggests that no
significant correlation exists ($P>0.6$ for all tests).  This
highlights that the trend of \aox(\opteml) may not be linear, 
as also found by S06 and \citet{Kelly07a}, or may only apply for high
luminosities. 

Note that we have investigated whether the details of our luminosity
and \aox\, calculations affect the results.  No significantly
different scientific conclusions result from our use of best-fit
$\Gamma$ (which affects the X-ray $K$-correction) 
to calculate the X-ray luminosity. For example, the the MainDet~sample 
\aox(\lo) regression that results from instead using fixed
$\Gamma$=1.9 and best-fit \nhintr\, has slope $0.077$ and intercept
$-0.872$.

Inasmuch as \aox\, probes intrinsic accretion processes, it probably
samples the balance between optical/UV blackbody thermal emission from
a geometrically thin but optically thick accretion disk, against X-ray
emission from a hot, optically thin corona that upscatters the seed UV
photons from the disk.  Various physical models can explain an
\aox\,-\opteml\, correlation with plausible parameters, e.g.,
a disk truncation radius that increases with luminosity
\citep{Sobolewska04}.  We also know that \aox\, measurements can be
affected by intrinsic absorption.  This is proven by the extreme
example of the BALQSOs here and elsewhere \citep{Green01,Gallagher02}.
Further evidence may come from a recent sample of AGN host galaxies of
5 clusters observed by \Chandra and the ACS onboard {\em HST}, where
\citet{Martel07} found that the X-ray to optical flux ratio of the AGN
correlates with the inclination angle of the host
galaxies\footnote{The X-ray/optical vs. inclination correlation holds
for late- but not early-type galaxies, so may not apply directly to
Type~1 QSOs if they are mostly in elliptical hosts.}  If the
observed trends are not dominated by selection effects (e.g.,
\citealt{Tang07}), it seems likely that instrinsic absorption acts
to increase the dispersion of an intrinsic relationship which
is dominated by accretion physics. 

\section{Evolution of \aox}

In a sample with a strong \aox(\opteml) correlation,
\aox\, will naturally correlate strongly with redshift
as well, due to the powerful redshift-luminosity trend 
shown in Fig~\ref{lz}.  To determine whether any redshift
evolution of \aox\, occurs independent of its luminosity dependence,
we use two methods.  

First, we examine a subsample with a narrow range in \opteml\, but
a reasonably broad range in redshift.  the zBox~sample
(Table~\ref{treg_oaox}) contains all the MainDet~sample objects with 
30.25$<$log\,(\opteml)$<$31 and 0.5$<z<$2.5. This sample shows no
significant correlation between \aox\, and redshift ($P_{max}=19\%$).
Accordingly, the nominal best-fit regression has a slope consistent
with zero (-0.001$\pm$0.014).  

\begin{figure}
% \plotfiddle{PSFILE}{VSIZE}{ROTANG}{HSCALE}{VSCALE}{HTRANS}{VTRANS}
%\epsscale{2}
\plottwo{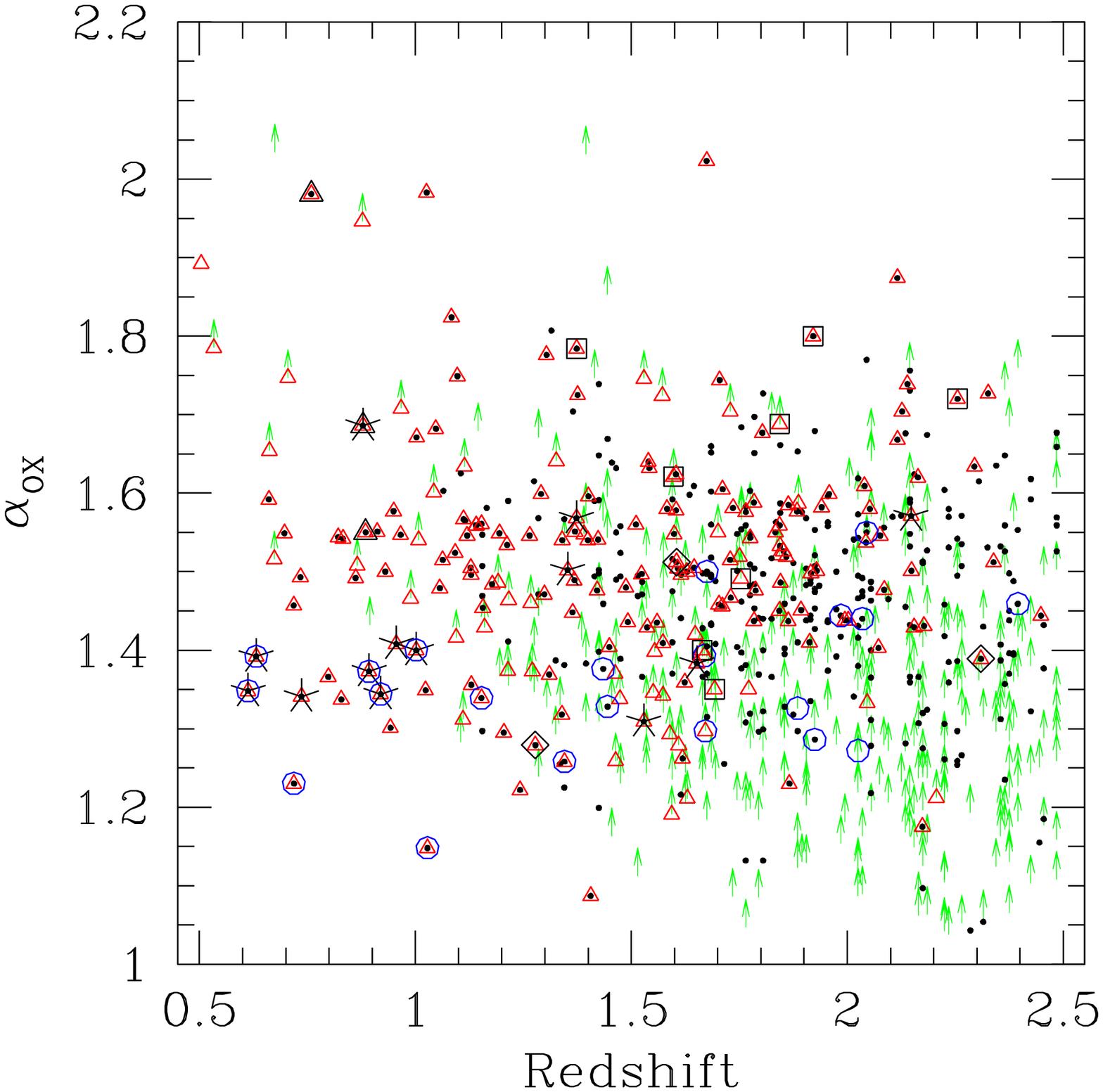}{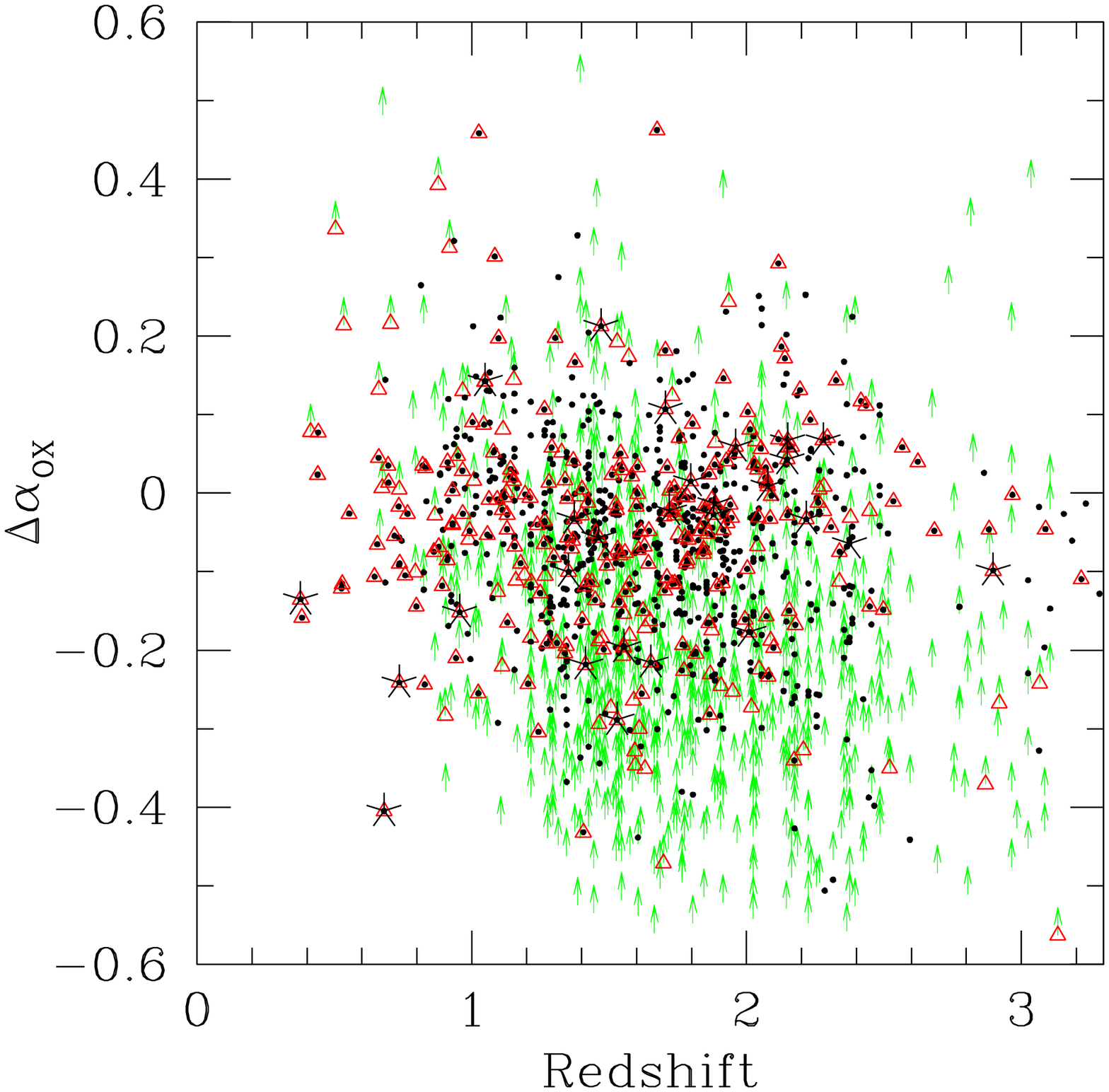}
\caption{LEFT: \aox\, vs. redshift for the zBox~sample,
restricted in log\,\opteml\, to the range 30.25 -- 31.
No significant trend exists. See Fig~\ref{imagx} for
symbol types.  RIGHT:  $\Delta$\aox\, vs. 
redshift for the hiLo~sample  (log\,(\opteml)$>$29.8)
after subtraction of its best-fit \aox(\opteml) relation in
Table~\ref{treg_oaox}. Although the redshift
range remains wide for this subsample, no trend is apparent.
% The best-fit line to the full hiLo~sample (blue line with dashed
% error bars) shows graphically the effect of a significant number of
% upper limits on the line intercept.
}    
\label{z_aox}
\end{figure}

Next, we can subtract the best-fit \aox(\opteml) regression 
to the more luminous hiLo~sample (log\,$(l_{\mathrm 2500\AA})>$29.8),  
where a simple linear fit seems applicable, and look for any residual
\aox($z$) dependence (evolution).  The expected \aox($z$) 
trend is significant for the hiLo~sample, induced by the \aox(\opteml) and
\opteml($z$) relationships.  We then subtract the best-fit
regression trend from Table~\ref{treg_oaox}.  The residual $\Delta\aox$
shows no trend with redshift in Fig~\ref{z_aox} (bottom)
or in correlation tests ($P_{max}=0.80$).

The large subsample sizes afforded by the SDSS/ChaMP QSO sample
allow us to conclude, without resort to more elaborate 
statistical analyses, that any apparent evolutionary trend can be
accounted for by the \opteml($z$) correlation in our sample,
and that such evolution is at best very weak in the range
$0.5<z<2.5$.

\subsection{\aox\, vs. X-ray Luminosity}
\label{aox_lx}

A weaker trend has also been noted in the correlation between
\aox\, and \xeml\, \citep{Green95,Steffen06}.  For the Main~sample, we find

$$\aox\, = 
	(0.003\pm 0.010)\,{\mathrm log}(l_{\mathrm 2\,keV})
	+ (1.384\pm 0.261) $$

\noindent which (while the correlation is significant) is consistent
with zero slope.  With a higher detection fraction,
the the D2L~sample yields slope $-0.0280\pm 0.0087$,
so that objects more luminous in X-rays are also X-ray brighter
(relative to $\opteml$).  Further removing RL and BAL QSOs (the
D2LNoRB~sample) does not change the best-fit parameters.
While the effects of different samples on the measured regression
is significant, the \aox(\xeml) relationship is particularly affected 
by limits, since they affect both axes.  

An apparent luminosity dependence of \aox\, is generated artificially if
the intrinsic dispersion in optical luminosity $\sigma_o$ exceeds that
for X-rays $\sigma_x$ \citep{Yuan98,Tang07}.\footnote{Any correlation
between a dependent variable $B$ which is derived via $B\propto
A^{-1}$ from an independent variable $A$ will be similarly affected in
samples with large dispersion.} The
significance of the induced correlation is proportional to
$\sigma_o^2/\Delta\,l_o^2$ where $\Delta\,l_o$ is the optical/UV
luminosity range of the sample. The magnitude of the biases also
depends on the luminosity function and sample flux limits.  Given
these effects, the apparently strong correlations so far published are
all consistent with no intrinsic dependence \citep{Tang07}.  The most
convincing remedy is likely to be a volume-limited sample that spans a
large range of both redshift and luminosity with high detection
fraction.  This requires careful treatment of large combined samples,
similar to \citet{Silverman05b,Silverman08}, including detailed
pixel-by-pixel flux (and consequently volume) limits.  While the
groundwork has been laid by the ChaMP's {\tt xskycover} analysis, such
a project is well beyond the scope of this paper.

\subsection{\aox\, vs. $\Gamma$}
\label{aox_gamma}

\begin{figure}
% \plotfiddle{PSFILE}{VSIZE}{ROTANG}{HSCALE}{VSCALE}{HTRANS}{VTRANS}
\plotone{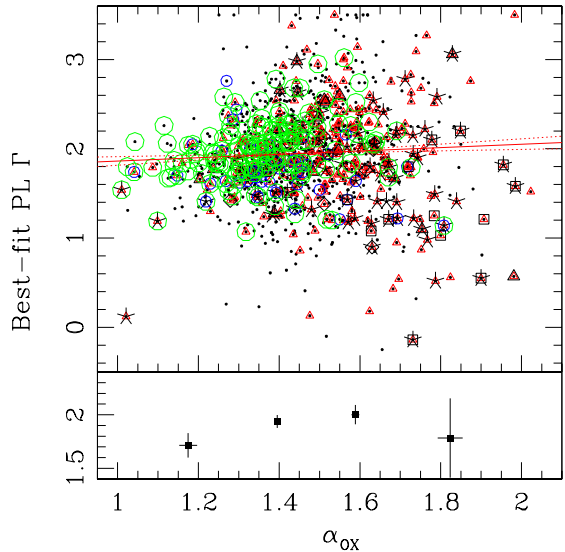}
\caption{TOP: Best-fit X-ray spectral power-law slope $\Gamma$
vs. \aox\, for the MainDet~sample.  The best-fit OLS regression
relation for this sample is shown with a red line, and associated
errors in dashed lines. Large open green circles show QSOs
with more than 200 (0.5-8\,keV) counts and simultaneous $\Gamma$-$\nhintr$
fits.  See Fig~\ref{imagx} for other symbol types.  
BOTTOM: Mean $\Gamma$ values are shown 
at the mean redshift for QSOs in 4 redshift bins of width
$\Delta\,\aox=0.25$ for the MainDet~sample. Errorbars show the error
in the means for both axes.}
%\vskip-0.2cm
\label{fgamma_aox}
\end{figure}

Fig~\ref{fgamma_aox} shows best-fit $\Gamma$ plotted against
\aox\, for the MainDet~sample. We detect for the first time a
significant but shallow correlation between $\Gamma$ and \aox.
Quasars that are relatively X-ray weak (larger \aox) tend to have
softer continuum slopes (larger $\Gamma$).  Fig~\ref{fgamma_aox}
shows the best-fit regression relation for the MainDet~sample,
$$\Gamma\, = 
	(0.188\pm 0.106)\,\aox + (1.676\pm 0.153).$$

The measured slope of the correlation is likely somewhat 

suppressed by the warm absorbers commonly found even in Type~1 AGN.
In Fig~\ref{fgamma_aox}, for the hard/weak region bounded by
\aox$>$1.6 and $\Gamma<1$, we find a BAL (indicated
by open black squares) is visible for every object for which a
spectrum exists covering blueward of restframe CIV\,$\lambda$1550. It
is likely that most if not  all objects in this region are BALs, as
could be determined  e.g., from restframe UV spectroscopy.  Again,
these objects are probably not intrinsically flat, but rather have a
hard best-fit $\Gamma$ due to undetected intrinsic absorption. 

Other samples shown in Table~\ref{treg_gaox} omit targets, and RL or
possibly absorbed QSOs all show steeper slopes.  The steepest
slope shown is for the HiCtNoTRB sample, which also includes only
objects with counts$>$200, where \nhintr\, is fit independently
from $\Gamma$.  This further supports that absorption
if anything flattens the apparent relation compared to the intrinsic
relation. 

\section{Discussion}

In this study, we have presented the largest homogeneous study 
to date of optically selected broad line quasars (from the SDSS) with
sensitive X-ray flux limits (from \Chandra; mode 2$\times
10^{-14}$\fcgs; 0.5-8\,keV). Our large sample highlights the large
dispersion in quasar properties that is unveiled whenever sensitive
limits and wide sky areas are combined. 

We confirm and extend several well-known multiwavelength relations.
The relationship between \aox\, and 2500\AA\, luminosity 
is confirmed for high luminosities ($M_B\lax -23$, or
${\mathrm log}\,(l_{\mathrm 2500\AA})\gax$29.8 (log\,($\nu_{\mathrm 2500}l_{\mathrm 2500\AA})\gax $44.9) with slopes similar to those found
previously (e.g., S06).  Including a wider luminosity range
inevitably produces a flatter relation across a range of subsamples
which test the effects of excluding \Chandra\, PI targets, 
RL QSOs, and QSOs with BALs, or NALs as well as NLS1s.  
No significant $\aox$($\opteml$) correlation exists for objects 
(68\% detected) at lower luminosities. Another possibility is that 
the relation simply flattens at low luminosities, or has a 
higher order luminosity dependence (e.g., \citealt{Kelly07a}).

We find for the first time a significant, robust and rather steep
dependence of X-ray continuum slope $\Gamma$ on X-ray luminosity
\xeml\, in the sense that the spectrum hardens with increasing \xeml.
A trend in the opposite sense has been reported recently for AGN in
the \Chandra\, Deep Fields \citep{Saez08}, but may be dominated by
differences between Type~1 and Type~2 AGN.  

We also report a shallow but significant trend that $\Gamma$
is harder for relatively X-ray bright (low \aox) QSOs. 
We note that X-ray bright QSOs include the RL QSOs, and that
RL QSOs also have predominantly flatter (harder) $\Gamma$.
We thus speculate that the overall trends of $\Gamma$(\aox)
and $\Gamma$(\xeml) both reflect an increase in the non-thermal 
emission fraction toward higher X-ray luminosities.  Not all
QSOs with a strong non-thermal X-ray emission component are 
necessarily radio loud - detectable radio loudness may pertain 
only to a fraction of these objects.  Radio bright phases
may be short compared to the overall QSO lifetime
and/or episodic.

The black hole masses in AGN are 5--8 orders of magnitude larger
than those in Galactic black hole X-ray binaries (GBH).  Since 
for a given $L/L_{Edd}$ the disc temperature scales with mass as
$M^{-1/4}$, AGN disks are cooler than in GBH.  The thermal 
accretion disk emission component that dominates the soft X-ray
emission ($\sim$2\,keV) in GBH corresponds to optical/UV
($\sim$2500\AA) emission in AGN.   Similarly, the non-thermal emission
(probably from Comptonized emission from the disk's corona) 
sampled as X-ray emission in AGN comes from harder ($\sim$20\,keV)
X-rays in GBH.  Type~1 QSOs may be analogous to Galactic black hole
binaries (GBH) in the high/soft state \citep{Sobolewska08} where a
similar trend is seen in $\Gamma$ vs. a disc/Comptonization index
$\alpha_{GBH}$ (analogous to \aox\, for QSOs) as we report here.  
\cite{Jester06} also compare the disk vs. non-thermal emission
fraction of GBH and AGN, and find that AGN segregate by radio loudness 
similarly to GBH in regions where luminosity and/or non-thermal 
fraction are high.  While the correlations we find lend
significant support to these interpretations, the scatter is large,
and could be significantly reduced if extrinsic effects can be
identified and corrected for.

To better understand the intrinsic physics of accretion
requires identifying and quantifying
extrinsic effects such as absorption and non-thermal processes.
Absorption may occur close to the SMBH gravitational radius, in the
BLR, a molecular torus, surrounding starforming regions, the outer
host galaxy or at intervening redshifts.  The intrinsic absorption may
be orientation-dependent, may evolve with redshift, and may be a
function of luminosity as well.  Contributions from non-thermal
processes certainly play a role, whether or not it is reflected in
detectable radio emission, and that role may change with SMBH mass,
spin, and environment, so consequently with lookback time as well.

Let's face it - quasars are complicated, but on average they don't
change much.  If quasar SED changes with luminosity were large
compared to the dispersion in the population and relatively
immune to selection effects, we would have been using
them for standard candles a long time ago.  Large samples,
uniformly observed and analyzed, offer the greatest hope
to disentangle the intertwining mysteries.

\acknowledgements

Support for this work was provided by the National Aeronautics and
Space Administration  through \Chandra\, Award Number AR4-5017X and
AR6-7020X issued by the \Chandra\, X-ray Observatory Center,  which is
operated by the Smithsonian Astrophysical Observatory for and on
behalf of the National  Aeronautics Space Administration under
contract NAS8-03060. 

GTR was supported in part by an Alfred P. Sloan Research Fellowship
and NASA Grant 07-ADP-7-0035. DH is supported by a NASA
Harriett G. Jenkins Predoctoral Fellowship.

{\em Facilities:} \facility{CXO, Mayall (MOSAIC-1 wide-field camera),
  Blanco (MOSAIC-2 wide-field camera), FLWO:1.5m (FAST spectrograph),
  Magellan:Baade (LDSS2 imaging spectrograph),  
Magellan:Clay (IMACS), WIYN (Hydra)}

We acknowledge use of the NASA/IPAC Extragalactic Database (NED),
operated by the Jet Propulsion Laboratory, California Institute of
Technology, under contract with the National Aeronautics and Space
Administration.

Funding for the SDSS and SDSS-II has been provided by the Alfred
P. Sloan Foundation, the Participating Institutions, the National
Science Foundation, the U.S. Department of Energy, the National
Aeronautics and Space Administration, the Japanese Monbukagakusho, the
Max Planck Society, and the Higher Education Funding Council for
England. The SDSS Web Site is http://www.sdss.org/. The SDSS is
managed by the Astrophysical Research Consortium for the Participating
Institutions. The Participating Institutions are the American Museum
of Natural History, Astrophysical Institute Potsdam, University of
Basel, Cambridge University, Case Western Reserve University,
University of Chicago, Drexel University, Fermilab, the Institute for
Advanced Study, the Japan Participation Group, Johns Hopkins
University, the Joint Institute for Nuclear Astrophysics, the Kavli
Institute for Particle Astrophysics and Cosmology, the Korean
Scientist Group, the Chinese Academy of Sciences (LAMOST), Los Alamos
National Laboratory, the Max Planck Institute for Astronomy (MPIA),
the Max Planck Institute for Astrophysics (MPA), New Mexico State
University, Ohio State University, University of Pittsburgh,
University of Portsmouth, Princeton University, the United States
Naval Observatory, and the University of Washington. 

\appendix
% Appendices appear at the end of the text, following the
% Acknowledgments, but before the Reference list. Multiple appendices
% are lettered with capital letters: Appendix A, Appendix B, Appendix
% C, etc., and sections within an appendix are numbered A.1, A.2,
% A.2.1, etc. If there is only one Appendix it is referred to simply
% as "the Appendix." 

 The ChaMP has developed and implemented an {\tt xskycover} pipeline
which creates sensitivity maps for all ChaMP sky regions imaged by
ACIS.  This allows (1) identification of imaged-but-undetected objects
(2) counts limits for 50\% and 90\% detection completeness (3) flux
sensitivity vs. sky coverage for any subset of obsids, necessary for
\lognlogs\, and luminosity function calculations and (4) flux
upper-limits at any sky position.  The basic recipe is as follows.  We
use the {\tt wavdetect} detection algorithm in CIAO \citep{Freeman02} to
generate threshold maps at each {\tt wavdetect} kernel scale actually run
(1, 2, 4, 8, 16, and 32 pixels).  The threshold maps, computed from the
local background intensity, determine the magnitude of the source
counts necessary for a detection at each pixel with a detection
threshold of $P=10^{-6}$ (corresponding to 1 false source
per $10^6$ pixels).  Thus, when a source is {\sl not} detected 
at a given location, the threshold map value serves
as an upper limit to the source counts.  Nominally, a source with
true intensity equal to this counts limit would be detected in
approximately half the instances that the source is observed under
the same conditions. To retain fidelity yet create a reasonably-sized
and easily-sampled sensitivity table covering the full ChaMP, we 
first average the threshold map values in sky pixels, each
$10\times10$ arcsec, whose boundaries are chosen to match a regular
commensurate grid across the sky. The final value of this
counts limit in any given sky pixel is interpolated from the two
threshold maps computed at wavelet scales that bracket the size of the
PSF at that location.\footnote{The 39\% radius is determined using the
PSF enclosed counts fraction, which corresponds to the 1$\sigma$
2-dimensional PSF size, from the library generated by the CXC
Calibration group as documented at
http://cxc.harvard.edu/cal/Hrma/psf.}

Note that the identification of the threshold map value one-to-one
with the counts limit is valid only for a specific shape of the PSF
(see Equation 6 of Freeman et al.\ 2002).  In particular, the strength
of the putative source can vary widely based on the correspondence
between the PSF size and the wavelet scale, as well as the shape
of the PSF.  For non-Gaussian PSF shapes (such as are found with
\Chandra\, at larger off-axis angles), the threshold values must be
corrected before a source counts limit is determined.  We calibrate
this correction factor by comparing the detection threshold map values
with simulated source retrieval experiments conducted on a subsample
of ChaMP fields - the 130 Cycle\,1--2 obsids studied by \citet{MKim07b}.
While the threshold values give us a reliable map
of variation on the sky, we must find a normalization from these
simulations.  Summing over a large number ($\sim$50,000) of simulated
sources, we compare the actual detection fraction from the simulations 
to the ratio of input (simulated) counts to derived (threshold)
counts.  From the fine binning available in these data, we
interpolate to find the normalization that yields the correct 
counts values for 50\% and 90\% completeness, across a range
of exposures, background levels, chip types, and off-axis angles.
Our experiments show that the only significant dependence 
of the correction factor (normalization) is on
off-axis angle, and that dependence warrants only a linear
correction dependence with best fit $N_{50}=1.32 + 0.198\times$OAA
($N_{90}=2.08 + 0.331\times$OAA) for 50\% (90\%) source
detection probability.  

Our method has been verified recently by \citet{Aldcroft08} using the
\Chandra\, Deep Field South (CDF-S) data available from the CXC. 
These data include the full 1.8\,Msec from 2000 \citep{Giacconi02},
and Director's Discretionary Time observations in 2007.  
The full 1.8 Msec exposure was used to generate a source list which
extended to faint fluxes.  The 21 individual obsids comprising the
full exposure were then treated as realizations of an experiment to
detect these sources and the ensemble of detection statistics were
analyzed by the same method as used for the $\sim$50,000 simulated sources.
 We found excellent agreement for the 50\% detection threshold and a
slight disparity (~10\%) for the 90\% detection threshold which could be
explained by source variability.

{}

\clearpage
\input{tab1}   % tsample
\clearpage
\pagestyle{empty}
\setlength{\voffset}{21mm}

\input{tab2empty}

\begin{figure}
%\epsscale{1}
%\plotone{tab2a.ps}
%\plotfiddle{PSFILE}{VSIZE}{ROTANG}{HSCALE}{VSCALE}{HTRANS}{VTRANS}
\plotfiddle{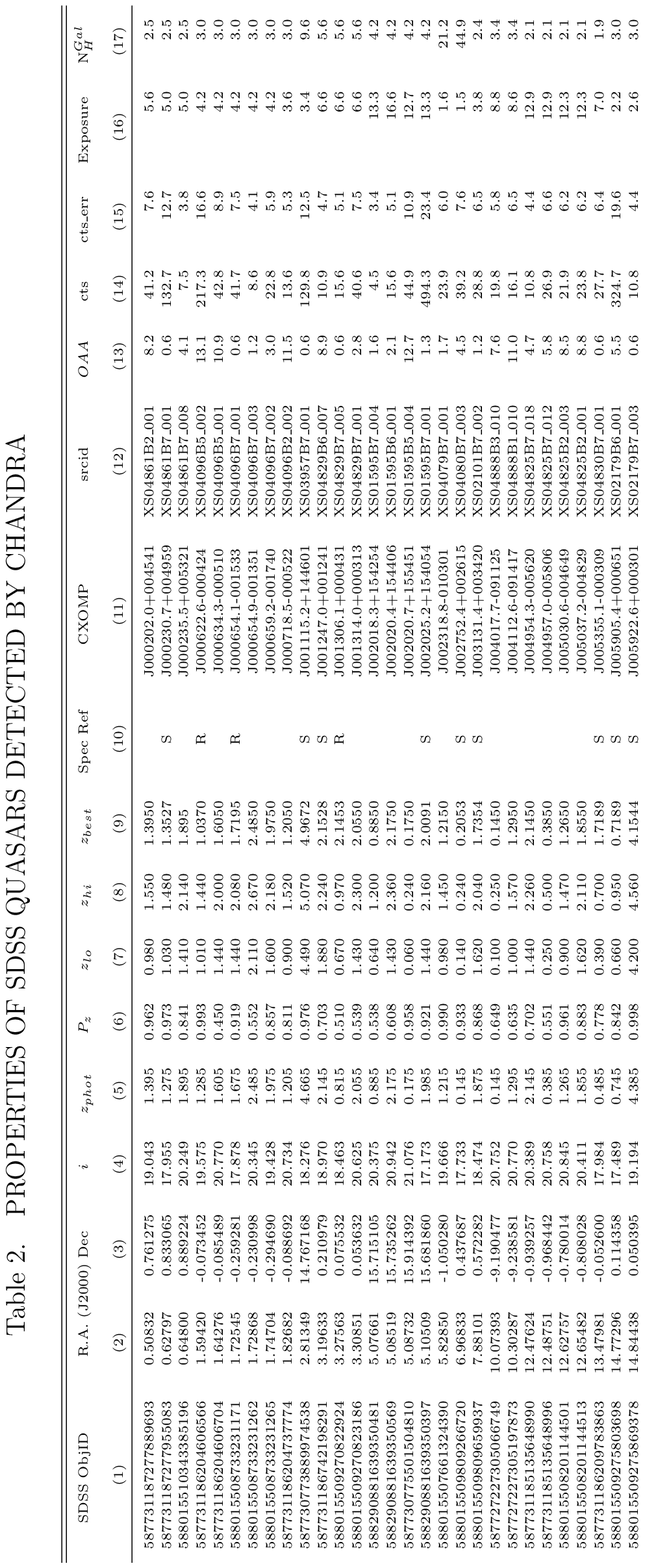}{6in}{0.}{90}{90}{-340}{-10} 
\label{ftab2a}
\end{figure}

\clearpage
\begin{figure}
\plotfiddle{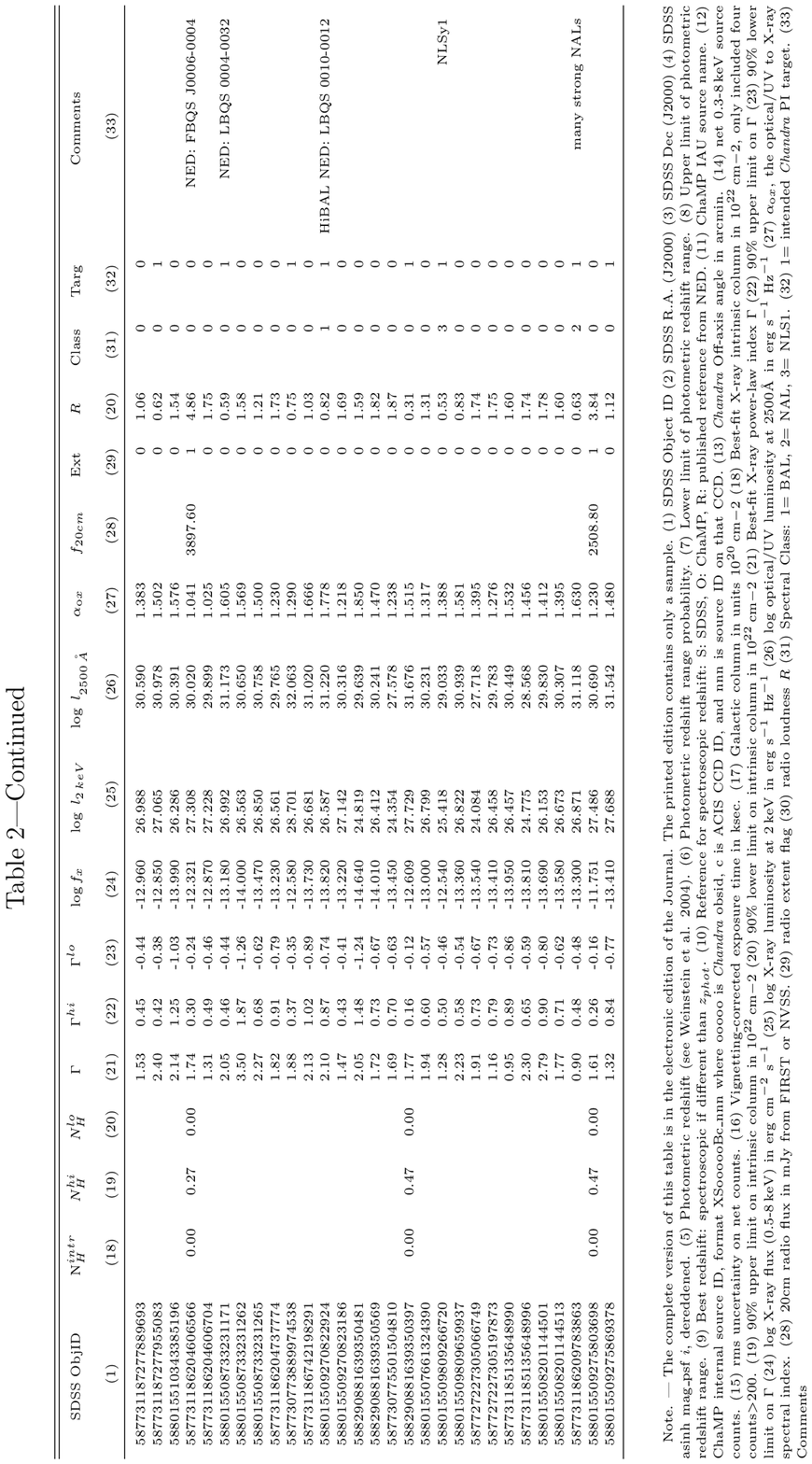}{6in}{0.}{100}{100}{-340}{-180} 
\label{ftab2b}
\end{figure}

\begin{figure}
\plotfiddle{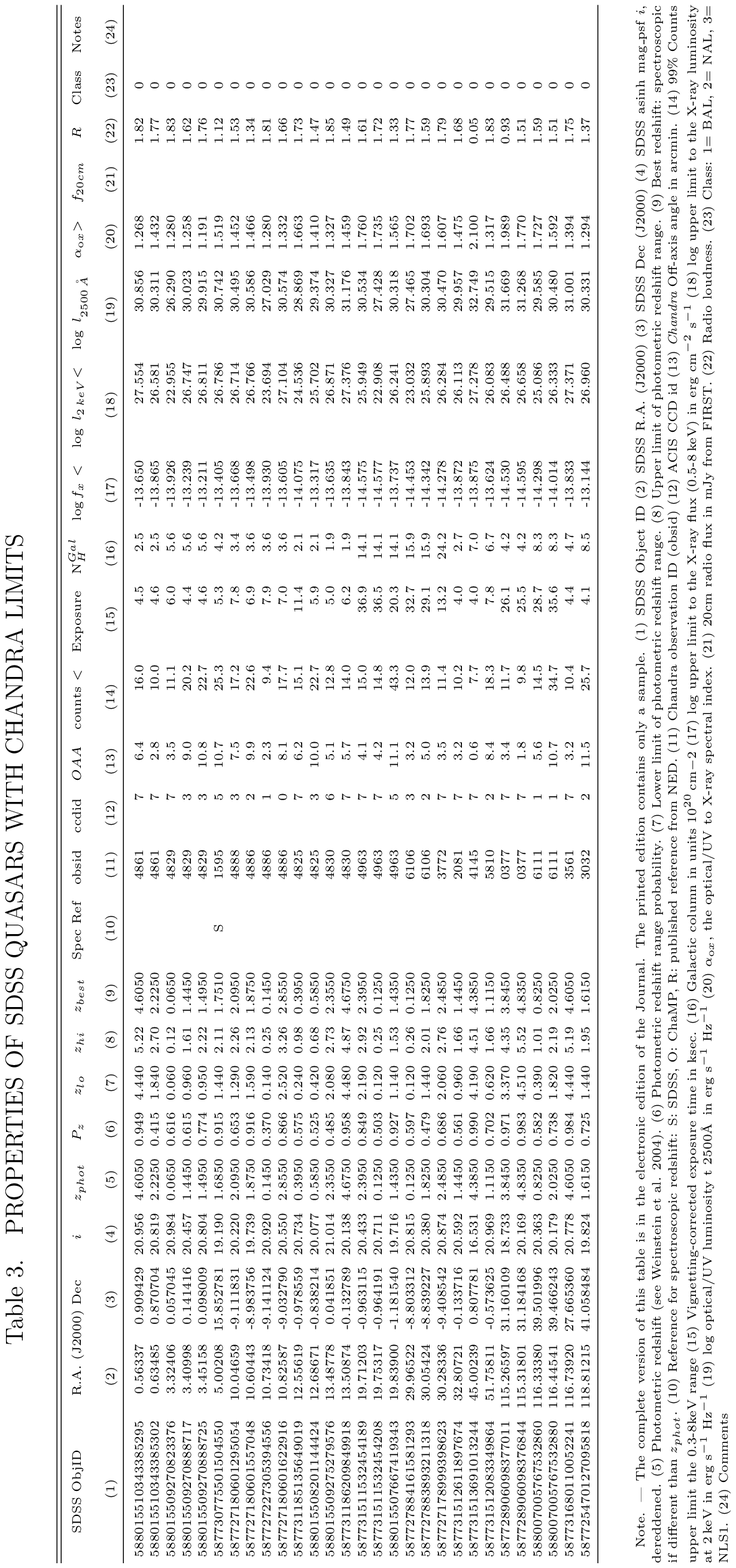}{5in}{0.}{85}{85}{-340}{-220} 
\label{ftab3}
\end{figure}
\input{tab3empty}

%\tabletypesize{\tiny}
%\clearpage
%  \LongTables % optionally
%    \input{stub.tab2}   % tdets
%    \clearpage

%\clearpage
%  \LongTables % optionally
%    \input{stub.tab3}   % tlims
%    \clearpage
%\tabletypesize{\normalsize}

\clearpage
\input{tab4}   % tuni
\clearpage
\input{tab5}   % treg_gamma
\input{tab6}   % treg_oaox
\clearpage
\input{tab7}   % treg_gaox

\end{document}

%% file: tab1.tex
\begin{deluxetable}{lccccccrrrr}
\tablecaption{Quasar Sample Definitions
\label{tsamples} }
\tabletypesize{\small}
\tablewidth{0pt}
\tablehead{
 \colhead{Sample} &
 \colhead{Lims \tablenotemark{a}} &
 \colhead{Targets} 	&
 \colhead{RL} 	&
 \colhead{Abs\tablenotemark{b}} &
 \colhead{$T_{min}$} 	&
 \colhead{OAA} 	&
 \colhead{$N_{det}$} 	&
 \colhead{$N_{lim}$} 	&
 \colhead{$N_{total}$}  &
 \colhead{\%\,Det}  	
}
\startdata
	 \multicolumn{11}{c}{}  \\
	 \multicolumn{11}{c}{Primary Samples}  \\
%SAMPLE LIMS TARG RL BAL >ksec <OAA      Ndet  Nlim Ntot  DetFrac 
Main       &  y &  . &   . & . & . &   . &    1135 & 1173 & 2308 & 49  \\ % 8
MainDet   &  . &  . &   . & . & . &   . &    1135 & 0    & 1135 & 100 \\ % 0
noTDet    &  . &  n &   . & . & . &   . &    1053 & 0    & 1053 & 100 \\ % 1
D2L        &  y &  . &   . & . & 4 &  12 &    922  & 347  & 1269 & 72  \\ % 6
D2LNoRB    &  y &  . &   n & n & 4 &  12 &    866  & 338  & 1204 & 71 \\ % 12
hiLo       &  y &  . &   n & n & . &   . &    847  & 961  & 1808 & 46 \\ % 10
HiCtNoTRB  &  . &  n &   n & . & . &   . &    129  & 0    & 129  & 100\\ % 21
	 \multicolumn{11}{c}{}  \\ 
	 \multicolumn{11}{c}{Other Samples}  \\ 
NoRB     &  y &  . &   n & n &   . &   . &    1054 & 1144 & 2198 & 47  \\ % 11       
NoRBDet &  . &  . &   n & n &   . &   . &    1054 & 0    & 1054 & 100 \\ % 2
D2LNoTRB &  y &  n &   n & n &   4 &  12 &    828  & 338  & 1166 & 71  \\ % 13
hiLoLx   &  . &  . &   n & n &   . &   . &    801  & 0    & 801  & 100 \\ % 22
zLxBox   &  . &  . &   . & . &   . &   . &    817  & 0    & 817  & 100 \\ % 5           
LoBox    &  . &  . &   . & . &   . &   . &    530  & 0    & 530  & 100 \\ % 4           
zBox     &  y &  . &   . & . &   . &   . &    420  & 360  & 780  & 53  \\ % 14      
zBoxDet &  . &  . &   . & . &   . &   . &    420  & 0    & 420  & 100 \\ % 3           
D2LSy1   &  y &  . &   n & n &   . &   . &    176  & 84   & 260  & 68  \\ % 17     
HiCt     &  . &  . &   . & . &   . &   . &    157  & 0    & 157  & 100 \\ % 20
HiCtNoTRB&  . &  n &   n & n &   . &   . &    129  & 0    & 129  & 100 \\ % 20
%  7 &  y &  n &   . & . &   4 &  12 &    863  & 347  & 1210 & 71 & high $f_{det}$ no Targ \\
%  9 &  y &  n &   n & n &   . &   . &    1006 & 1144 & 2150 & 46 & Cleaner w/ Lims \\
% 15 &  y &  . &   . & . &   . &   . &    530  & 511  & 1041 & 50 & $L_o$Box      \\
% 16 &  y &  . &   n & n &   4 &  12 &         &      &      &    & Sample~12-\aox\, regr  \\
% 18 &  y &  . &   n & n &   . &   . &         &      &      &    & Sample~10-\aox\, regr \\
% 19 &  y &  . &   n & n &   4 &  12 &    690  & 254  & 944  & 73 & high $f_{det}$ QSO Box     \\
\enddata
\tablenotetext{a}{If `y', sample includes  X-ray non-detections.}
\tablenotetext{b}{If `n', sample excludes QSOs with evident BALs or
  NALs and also NLS1s.} 
\end{deluxetable}

%% file: tab2empty.tex
\begin{table*}
\caption{  }
\label{tdets} 
\small
\begin{tabular}[t]{cc}
	 \multicolumn{2}{c}{}  
\end{tabular}
\end{table*}

%% file: tab3empty.tex
\begin{table*}
\caption{  }
\label{tilms} 
\small
\begin{tabular}[t]{cc}
	 \multicolumn{2}{c}{}  
\end{tabular}
\end{table*}

%% file: tab4.tex
\begin{deluxetable}{lrcccc}
\tablecaption{Quasar Sample Univariate Results
\label{tuni} }
\tabletypesize{\small}
\tablewidth{0pt}
\tablehead{
 \colhead{Sample\tablenotemark{a}} &
 \colhead{$N$} &
 \colhead{Mean} &
 \colhead{error\tablenotemark{b}} &
 \colhead{Median} &
 \colhead{$P_{max}\tablenotemark{c} (\%)$} 
}
\startdata
	 \multicolumn{5}{c}{$\Gamma$\, Distributions\tablenotemark{d}}  \\
MainDet & 1135& 1.94 & 0.02 & 1.93 & \ldots \\
RQ      & 704 & 1.91 & 0.02 & 1.86 & 0.2  \\
RL      & 43  & 1.73 & 0.05 & 1.65 & 0.2 \\
BAL     & 15  & 1.35 & 0.15 & 1.30 & 0.05  \\
NAL     & 9   & 1.68 & 0.13 & 1.75 & 19   \\
NLS1    & 19  & 2.01 & 0.15 & 1.95 & 35 \\
$z>3$   & 56  & 1.80 & 0.07 & 1.76 & 21 \\
	 \multicolumn{5}{c}{}  \\
	 \multicolumn{5}{c}{\aox\, Distributions}  \\
D2L   & 1269 & 1.421 & 0.005 & 1.365 & \ldots \\
RQ    & 680  & 1.527 & 0.008 & 1.452 & 0.0 \\
RL    & 31   & 1.382 & 0.030 & 1.392 & 0.0 \\
BAL   & 23   & 1.717 & 0.028 & 1.664 & 0.0\\
NAL   & 8    & 1.463 & 0.056 & 1.500 & 49  \\
NLS1  & 19   & 1.540 & 0.080 & 1.433 & 24 \\
$z>3$ & 47   & 1.817 & 0.077 & 1.786 & 55 \\
\enddata
\tablenotetext{a}{For each parameter tested, the numeric sample at the
top is the parent for comparison subsamples below.  The RL subsamples
are tested against non-RL QSOs from the parent sample. The remaining 3
QSO subsamples for each parameter are tested against the parent sample
excluding all 4 QSO subtypes.}
\tablenotetext{b}{Error in the mean from the  Kaplan-Meier estimator
  as implemented in ASURV.  An estimate of the dispersion can be
  obtained by multiplying this by $\sqrt{N-1}$
}
\tablenotetext{c}{The maximum probability for the null hypothesis
(of indistinguishable samples) from 3 tests described in \S~\ref{uni}.
Only for $P_{max}<5$ do we consider the distributions significantly
different.  RL and RQ samples are contrasted to each other. 
Other samples are compared to their parent sample (MainDet or
D2L)-$X$, where $X$=BALs+NALs+NLS1s, except for BALs, whose parent
sample is RQ QSOs only.} 
\tablenotetext{d}{These are distributions of ''best-PL'' measurements, 
best-fit $\Gamma$, which always includes \nhgal, and also includes
\nhintr\, for 0.5-8\,keV counts $>$200.}
\end{deluxetable}

%% file: tab5.tex
\begin{deluxetable}{lllrl}
\tablecaption{Quasar Sample Bivariate Regression Results: 
$\Gamma$(\xeml)  OLS
\label{treg_gamma} }
\tabletypesize{\small}
\tablewidth{0pt}
\tablehead{
 \colhead{Sample} &
 \colhead{Slope} &
 \colhead{error} &
 \colhead{Intercept} &
 \colhead{error} 
}
\startdata
%	 \multicolumn{5}{c}{}  \\
MainDet   & --0.1465 & 0.0199 & 5.8575 & 0.5276 \\ %  0
noTDet    & --0.1424 & 0.0188 & 5.7579 & 0.4959 \\ %  1
HiCtNoTRBt & --0.1213 & 0.0497 & 5.2473 & 1.3333 \\ % 21
NoRBDet   & --0.1528 & 0.0262 & 6.0343 & 0.6980 \\ %  2
hiLoLx     & --0.2336 & 0.0388 & 8.1937 & 1.0337 \\ % 22
HiCt       & --0.1454 & 0.0472 & 5.8540 & 1.2679 \\ % 20
\enddata
\tablecomments{Samples tested are arranged in the same order as in
  Table~\ref{tsamples}. 
OLS refers to the ordinary least squares regression.}
\end{deluxetable}

\begin{deluxetable}{lllrl}
\tablecaption{Quasar Sample Bivariate Regression Results: 
\xeml(\opteml) OLS Bisector
\label{treg_ox} }
\tabletypesize{\small}
\tablewidth{0pt}
\tablehead{
 \colhead{Sample} &
 \colhead{Slope} &
 \colhead{error} &
 \colhead{Intercept} &
 \colhead{error} 
}
\startdata
	 \multicolumn{5}{c}{}  \\
	 \multicolumn{5}{c}{Primary Samples}  \\
Main & 1.1171 & 0.0170 & --7.5929 & 0.6365 \\ %    8
MainDet & 0.9372 & 0.0266 & --1.9178 & 0.8797 \\ %    0 
NoTDet  & 0.9719 & 0.0259 & --2.9508 & 0.8418 \\ %    1
D2L & 1.1350 & 0.0209 & --8.1162 & 0.6329 \\ %    6
D2LNoRB & 1.1667 & 0.0238 & --9.0641 & 0.7199 \\ %   12
hiLo & 1.2976 & 0.0340 & --13.1316 & 1.0356 \\ %  10
	 \multicolumn{5}{c}{}  \\
	 \multicolumn{5}{c}{Other Samples}  \\
NoRB & 1.1502 & 0.0146 & --8.5922 & 0.6125 \\ %   11
NoRBDet & 0.9359 & 0.0253 & --1.8745 & 0.8903 \\ %    2
zLxBox & 0.8421 & 0.0173 & 1.0177 & 0.6695 \\ %      5
LoBox & 0.9907 & 0.0310 & --3.5191 & 0.9357 \\ %    4
zBox & 1.6658 & 0.1125 & --24.3936 & 3.4355 \\ %  14
zBoxDet & 1.4088 & 0.0803 & --16.3512 & 2.4564 \\ %   3
D2LSy1 & 1.2299 & 0.0504 & --10.6975 & 1.4746 \\ %  17
D2LNoTRB & 1.1908 & 0.0196 & --9.7838 & 0.5921 \\ %   13
	 \multicolumn{5}{c}{}  \\
S06 & 0.72  & 0.01   &   4.53~~     & 0.69 \\ % 
% & 1.1665 & 0.0159 & --9.0564 & 0.5722 \\ %    7
% & 1.1550 & 0.0152 & --8.7369 & 0.5979 \\ %    9
% & 1.4947 & 0.0462 & --18.9987 & 1.4128 \\ %  15
% & 1.0629 & 0.0319 & --5.9073 & 0.9693 \\ %   19
\enddata
\tablecomments{Samples tested are arranged in the same order as in
  Table~\ref{tsamples}. 
OLS refers to the ordinary least squares regression.
S06 are results from \citet{Steffen06} for comparison.}
\end{deluxetable}

%% file: tab6.tex
\begin{deluxetable}{llllrl}
\tablecaption{Quasar Sample Bivariate Regression Results: 
\aox(\opteml) OLS
\label{treg_oaox} }
\tabletypesize{\small}
\tablewidth{0pt}
\tablehead{
 \colhead{Sample} &
 \colhead{Slope} &
 \colhead{error} &
 \colhead{Intercept} &
 \colhead{error} 
}
\startdata
%  Name  slope err Int err Notes          REDONE 4 April
	 \multicolumn{5}{c}{}  \\
	 \multicolumn{5}{c}{Primary Samples}  \\
Main & 0.0598 & 0.0066 & --0.2776 & 0.1988 \\ %  8
MainDet & 0.0826 & 0.0066 & --1.0331 & 0.1978 \\ %  0 
NoTDet & 0.0732 & 0.0089 & --0.7496 & 0.2681 \\ %  1
D2L & 0.0610 & 0.0085 & --0.3189 & 0.2580 \\ %  6
D2LNoRB & 0.0516 & 0.0078 & --0.0358 & 0.2377 \\ % 12
hiLo & 0.1284 & 0.0070 & --2.3754 & 0.2151 \\ % 10
	 \multicolumn{5}{c}{}  \\
	 \multicolumn{5}{c}{Other Samples}  \\
NoRB & 0.0513 & 0.0073 & --0.0239 & 0.2217 \\ % 11
NoRBDet & 0.0804 & 0.0085 & --0.9667 & 0.2546 \\ %  2
D2LNoTRB & 0.0482 & 0.0104 & ~~0.0652 & 0.3146 \\ %   13
zLxBox & 0.1943 & 0.0067 & --4.4484 & 0.2033 \\ %  5
LoBox & 0.1895 & 0.0105 & --4.2956 & 0.3176 \\ %  4
zBox & 0.1019 & 0.0242 & --1.5709 & 0.7404 \\ % 14
zBoxDet & 0.1119 & 0.0267 & --1.9335 & 0.8172 \\ %  3
D2LSy1 & 0.0058 & 0.0175 & ~~1.3253 & 0.5160 \\ %   17
	 \multicolumn{5}{c}{}  \\
S06 & 0.137 & 0.008  & --2.638~  & 0.240~ \\ % 
%  7 & 0.0553 & 0.0091 & --0.1463 & 0.2768 \\ % 
%  9 & 0.0503 & 0.0072 & 0.0078 & 0.2156 \\ %  
% 15 & 0.1255 & 0.0158 & --2.2938 & 0.4807 \\ % 
% 19 & 0.1124 & 0.0125 & --1.8549 & 0.3824 \\ % 
% 23 & 0.1417 & 0.0131 & --2.7853 & 0.3965 \\ % 
% 24 & 0.1493 & 0.0120 & --3.0290 & 0.3645 \\ % 
% 25 & 0.1428 & 0.0084 & --2.8417 & 0.2584 \\ % 
% 26 & 0.1325 & 0.0069 & --2.5057 & 0.2116 \\ % 
\enddata
\tablecomments{Samples tested are arranged in the same order as in
  Table~\ref{tsamples}. OLS refers to the ordinary least squares
  regression. S06 are results from \citet{Steffen06} for comparison.}
\end{deluxetable}

%% file: tab7.tex
\begin{deluxetable}{llllrl}
\tablecaption{Quasar Sample Bivariate Regression Results: 
$\Gamma$(\aox) OLS
\label{treg_gaox} }
\tabletypesize{\small}
\tablewidth{0pt}
\tablehead{
 \colhead{Sample} &
 \colhead{Slope} &
 \colhead{error} &
 \colhead{Intercept} &
 \colhead{error} 
}
\startdata
%  Name  slope err Int err Notes          REDONE 4 April
	 \multicolumn{5}{c}{}  \\
MainDet   & 0.188 & 0.106 & 1.676 & 0.153 \\ %   0
NoTDet    & 0.274 & 0.108 & 1.566 & 0.154 \\ %   1
NoRBDet   & 0.340 & 0.109 & 1.466 & 0.154 \\ %   2
HiCt      & 0.342 & 0.202 & 1.507 & 0.259 \\ %  20
HiCtNoTRB & 0.358 & 0.179 & 1.529 & 0.225 \\ %  21
\enddata
\tablecomments{Samples tested are arranged in the same order as in
  Table~\ref{tsamples}. OLS refers to the ordinary least squares
  regression. }
\end{deluxetable}